# Microwave-assisted synthesis and modification of nanocarbons and hybrids in liquid and solid states: Strategies and results

Nagaraj Nandihalli*

Critical Materials Innovation Hub, Ames National Laboratory, U.S. Department of Energy, Iowa State University, Ames, Iowa, USA

* Nagaraj Nandihalli ✉ nagaraj.nandi001@umb.edu

*Abstract*

Over the past 20 years, nanocarbons have become more significant as nanostructured fillers in composites and, more recently, as functional elements in a brand-new class of hybrid materials. Microwave-assisted synthesis and processing is a burgeoning subject matter in materials research with significant strides in the realm of nanocarbon during the last decade. The review examines recent approaches to producing various nanocarbons using microwaves as energy sources, the characterization of such materials for various applications, and their results. The underlying factors supporting the increased performance of such materials or their composites are analyzed and reaction mechanisms are presented wherever necessary. In particular, the recently developed and implanted approaches to produce porous carbon materials, CNTs and fibers, carbon nanospheres, carbon dots, reduced graphene oxide, corresponding nanocarbon hybrid materials, and the purification and modification of CNTs are discussed. Finally, the reduction of graphene oxide and the preparation of graphene derivative hybrids using solid-state and liquid-state routes such as polyopl, mixed solvents, ionic liquids, and microwave-assisted hydrothermal/solvothermal methods are analyzed in detail. In addition, the principles of microwave heating in liquid and solid states, the use of metals or their particles as arcing agents or catalysts, and carbonaceous materials as internal or external susceptors during synthesis and modifications are presented in detail.

## 1 Introduction

With the ability to exist in zero, one, or two dimensions, nanocarbons enable the construction of three-dimensional nanocarbon frameworks and a comprehensive understanding of carbon atom bonding and arrangement. Unlike nanocomposites, which typically blend the inherent qualities of component substances, nanocarbon hybrids provide access to a vast surface area required for gas/liquid-solid interactions as well as a larger interface, enabling processes related to charge and energy transfer to produce synergistic effects that yield distinctive characteristics and improved functionality. Applications for nanocarbons and their hybrids include heterogeneous catalysis, electrocatalysis, photocatalysis, supercapacitors, photovoltaics, health, and sensors.[1-6] Carbon dots outperform fluorescent semiconductor quantum dots in terms of strong fluorescence, broad excitation spectra, and narrow, tunable emission spectra. They are biocompatible, very small in size, low in molecular weight, and less toxic, making them preferable to quantum dots.[4] Similarly, carbon nanofibers[7], carbon nanospheres[8,9] mesoporous carbon[10] have energy storage, adsorption and various technological applications. The primary issue in creating such carbon hybrid designs is controlling the decorating fraction, formalizing the foreign nanostructures, and precisely arranging them on the supporting materials. It is important to consider the critical fabrication processes involved in nanocarbon hybridization or heterostructuring with external noble metal, ceramic, or semiconductor nanoparticles (NPs) or nanowires (NWs). By using such materials, different device types can perform better and have more functionality.

Nanocarbon hybrids have the greatest potential for use in energy storage technologies.[11] As anodes in Li-ion batteries (LIBs), metals and metalloids (Ge, Sn, Si, and, Sb), metal oxides, and metal-phosphides/sulfides/nitrides (P, N, S) have all been used. However, because of their low conductivity,[12] poor cyclability, rapid capacity fading from high volume expansion and shrinking that results in electrode pulverization, and severe electrode collapse during cycle operations, these materials have low rate capability. Carbon is the most widely utilized anode material. This is both the most practical and widely used anode material in terms of economics. Moreover, carbon in the anode increases mechanical strength, surface area, and electronic conductivity, all of which contribute to efficient charge carrier transport and rate capability.[13] Graphene-based materials have been found to have up to 1000 mAh/g of reversible capacity.[14,15] Mesoporous graphene (MPG), carbon tubular nanostructures (CTN), and hollow carbon nanoboxes (HCB) are attractive candidates for LIB anodes. High capacities of around 1100 mAh/g, 600 mAh/g, and 500 mAh/g at 0.1 A/g are reported for MPG, CTN, and HCB, respectively. MPG, CTN, and HCB anodes maintain capacities of 181 mAh/g, 141 mAh/g, and 139 mAh/g at 4 A/g, respectively.[16] High capacity and rate capability are provided by nanocarbons with graphene layers that undergo the lithiation/delithiation process, a high ratio of graphite edge structure, and a large specific surface area (abbreviated as *SSA* hereafter) that permits capacitive behavior. Graphene offers improved ion intercalation through its high in-plane conductivity and vertical interlayer spacing, which increases capacity and rate capability. Electron transport is facilitated by the carbon shells in carbon nanospheres, which permit volume expansion and contraction during charge storage. The confluence of these two aspects improves cycling stability and rate performance. Nanocarbon, having 1D tube architecture, is also suitable for Li-ion transport due to its high conductivity. Carbon protects the Li-ion against dendritic growth while recharging.[17] Synthesizing the optimal porous structure for nanocarbons that helps reduce transport routes would provide the structure and chemistry necessary for Li-ion storage while maintaining accessibility to electrolyte ions.[18] Reduced graphene oxide (rGO), a graphene derivative, has a wide range of potential uses. With significant electromagnetic wave absorption across broad optical bandwidths and over the MW regime, rGO composites have been adopted as radar-absorbing elements that offer good MW shielding.[19-21]

The strong photon energy absorption in the solar spectrum band and high carrier mobility, has enabled innovative rGO-based solar devices and photodetectors.[6, 22] Patterned rGO can be used as transparent conductive electrodes for LEDs[23] and solar cells[24] by lowering the thickness of the material to reduce light absorption.

Among the two types of supercapacitors, namely electrical double layer capacitors (EDL) and pseudocapacitors, EDL capacitors feature carbon-based materials (CNT,[25] activated carbon,[26] graphene,[27] etc.) as the electrode material, while metal oxides[28] and conducting polymers[29] are the electrode materials for pseudocapacitors. One approach for improving supercapacitor performance is to add pseudocapacitive materials[30] (e.g., $RuO_2$,[31] $MnO_2$,[32] conductive polymers,[33] etc.) to graphene, which may minimize graphene sheet aggregation and restacking,[34, 35] resulting in a higher specific capacitance ($C_{sp}$) for the composite.

Microwaves (MWs) as a source of heating and for facilitating reactions have various advantages over traditional (firing or resistive) heating, the majority of which stem from how energy is delivered. MWs can transport energy directly to target materials, bypassing the main energy transfer processes in conventional heating, namely conduction and convection. MW heating is more energy efficient and faster than conventional heating, and selective heating is possible without contact between the microwaves and their environment. Although the synthesis of nanocarbons and their hybrids is advancing at an increasingly rapid pace, developing approaches aimed at the above-mentioned applications offer significant opportunities.

## 1.1 Microwave-assisted heat generation in liquid and solid media

MWs range in frequency from 300 MHz ($3 \times 10^8$ cycles/s) to 300 GHz ($3 \times 10^{11}$ cycles/s) (Figure 1a). The quantum energies of MW photons at 0.3 GHz, 2.45 GHz, and 30 GHz are $1.2 \times 10^{-6}$ eV, $1.0 \times 10^{-5}$ eV, and $1.2 \times 10^{-3}$ eV, respectively. These energies are a million times lower than those of X-rays. They cannot, therefore, ionize or break chemical bonds in situations where the quantum energy is at least a thousand times greater.[36] Nonetheless, they can only cause molecular rotations (Figure 1b). Material processing typically involves MWs at 915 MHz, 2.45 GHz, 5.8 GHz, and 24.124 GHz frequencies.[37] 2.45 GHz ($\lambda$ of ~12.24 cm) is generally used in laboratory and household MW ovens.

Ionic conduction and dipolar polarization are the two main ways that MW radiation heats a material or liquid. While charged particles (typically ions) engage in ionic conduction, the dipoles in the reaction mixture (polar solvent molecules or reagents) participate in the dipolar polarization effect.[38, 39] When exposed to MWs, dipoles/molecules reorient themselves to align with the oscillating electric field ($E$- field) (Figure 1b). Since the $E$-field component oscillates, dipoles attempt to return to a random state in the absence of an $E$-field, which does not happen quickly. Inertial, elastic, frictional, and molecular interactions prevent frequent changes in dipole/molecule orientations that increase molecular kinetic energy and result in volumetric heating. The MW energy loss induced by this dipole/molecular friction process is known as dipolar loss,[39, 40] whereas heating employing the $E$-field component of high-frequency MW radiation is known as dielectric heating. The magnitude of the heat generaated during this process is directly proportional to the dipoles' capacity to align with the frequency of the applied field, which is dictated by the molecule's functional groups and volume. When the $E$-field component oscillates at $10^{12}$ times/sec, even the smallest molecules can no longer rotate substantially. Similarly, the MW field causes charged particles in a reaction medium (often ions) to oscillate backward and forward, and in this process, charged particles tumble and knock against their adjacent molecules, or atoms, causing impacts that generate motion and heat. Ionic conductivity channels generate significantly more heat than the dipolar rotation process.[39,41]

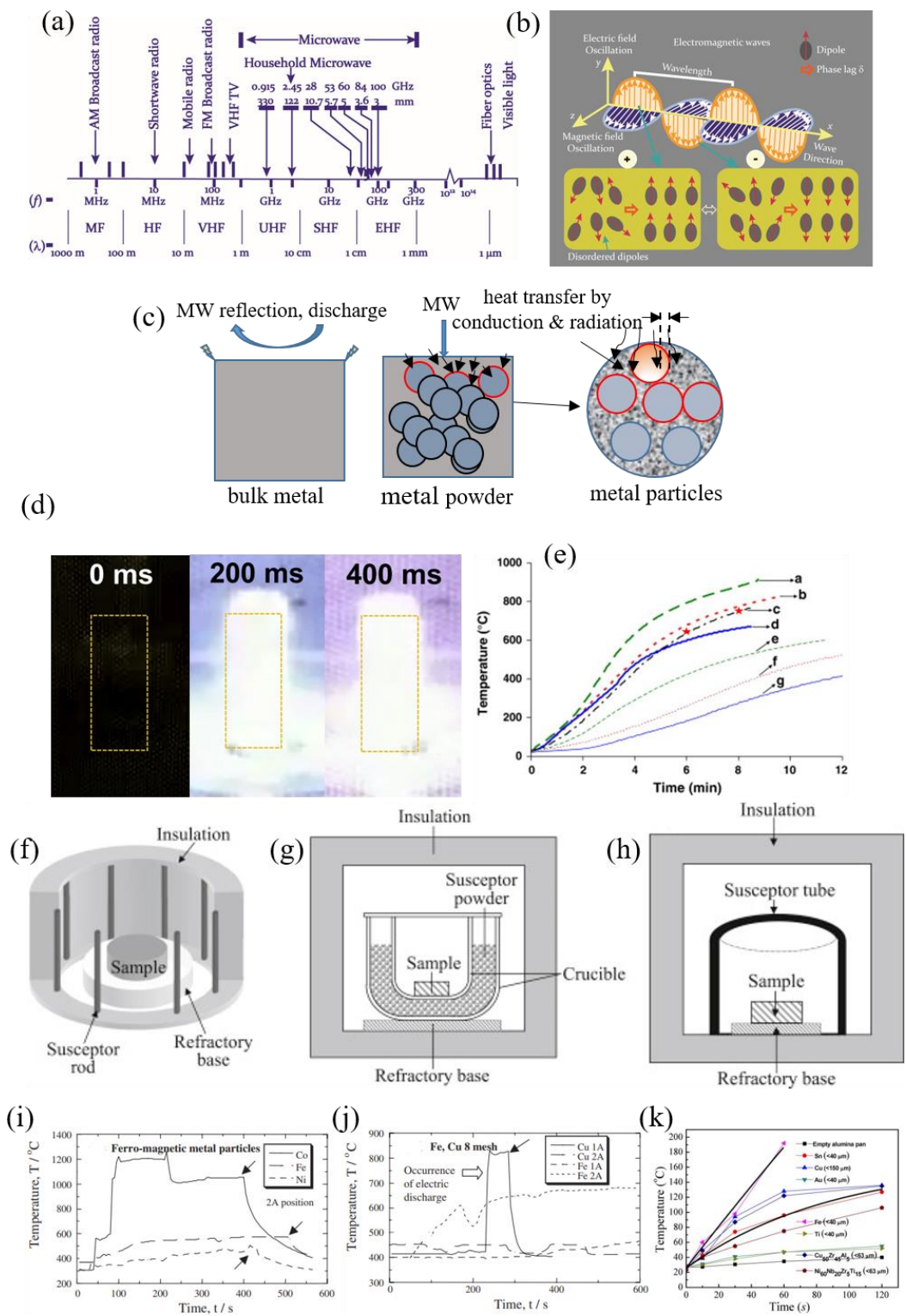

Figure 1: (a) MW region of the EM spectrum; (b) Heating mechanism for dielectric insulator materials. The oscillating *E*-field agitates the molecular dipoles; (c) Schematic representation of the MW heating of metal particles[42]; (d) magnitude of the light intensity with time during MW heating of metal acetate/GO[43]; (e) Temperature rise during MW heating of a mixture of graphite with various additives in a wt. ratio of 1:1. (a) graphite (50%) + magnetite ($Fe_3O_4$) (50%); (b) graphite (50%) + SiC (50%);

(c) graphite (100%); (d) graphite (50%) + manganese dioxide (50%); (e) $Fe_3O_4$ (100%); (f) SiC (100%); (g) manganese dioxide (100%).[44] The ★ on the graphite curve (c) represent 640 and 750 ºC after 6 and 8 min of MW exposure, respectively. (f-h): Most commonly utilized hybrid heating arrangements involving susceptors; (f) rod-type (picket fence arrangement);[45] (g) powdered (in the two-crucible set-up) and (h) tubular susceptors.[46] (i) The process of heating several ferro-magnetic materials (particle sizes, 70 µm); (j) Heating of large Fe & Cu particles at two spots (Cu, Fe particles of ∅ of >2 mm (8 mesh)). In 2A: Cu not heated & Fe heated well. The arrows indicate the time when the power was turned off. Adapted with ref. [42] (k) Metal powder samples heated in a MW oven. The black curves reflect fitting according to the theoretical model presented below in the plot. The heating curves for metallic glassy particles are also provided for comparison. Adapted with ref. [47]

The loss tangent (also known as the energy dissipation factor), determines a material's or solvent's capacity to convert MW energy into heat at a specific frequency and temperature and is given by

$$tan\delta = \frac{\varepsilon''}{\varepsilon'}$$  *Equation 1*

Where $\varepsilon''$ is the dielectric loss factor (a measure of how well electromagnetic radiation is converted to heat), $\varepsilon'$ is the relative permittivity (describes the polarizability of molecules in the *E*-field), and $\delta = 90º - \phi$ is the phase lag or loss angle between the applied *E*-field and the materials polarization. A MW reactor's standard working frequency (2.45 GHz) necessitates a high tan$\delta$ reaction medium for effective absorption and heating.[38] Materials having a low tan$\delta$ value are practically transparent to MWs (at that particular frequency) and will not be significantly heated. Selective heating occurs in mixtures of solvents based on the constituents' relative dielectric characteristics.[38]

If suitable MW-absorbing reagents are not available for material production, a susceptor can be used, which can also serve as a heat source. A susceptor is a substance with a high dielectric tan$\delta$. This substance absorbs electromagnetic energy and converts it into heat. A susceptor could have direct contact with the precursors. For instance, when a strongly MW absorbing solute molecule is added to a MW non-absorbing solvent, the heat absorbed by the entire solution is caused by the MW-heat absorbing solute's convective heating of the solvent medium. Alternatively, a sample can be stored separately by embedding the reaction vessel/test tube in a container made of susceptor material or wrapping it with susceptor material. Susceptors typically employed in synthesis include carbon (graphite or amorphous carbon), SiC, and copper (II) oxide.[38] Figure 1f-h depicts the most commonly utilized hybrid heating arrangements involving susceptors. Combining susceptors with reagents can lead to product contamination, necessitating an additional separation step.

*Table 1* lists the tan$\delta$ values of commonly used solvents.

*Table 1:* Loss tangent (tan$\delta$) values for different solvents.[40] Data acquired at 2.45 GHz of MW radiation and 20 ºC. [bmim]$PF_6$ = 1-butyl-3- methylimidazolium hexafluorophosphate. NMP = N-methyl-2-pyrrolidone; BP = boiling point

| Solvent | BP (ºC) | tan$\delta$ | Solvent | BP (ºC) | tan$\delta$ |
|---|---|---|---|---|---|
| ethylene glycol (EG) | 198 | 1.350 | DMF | 153 | 0.161 |
| Ethanol | 78 | 0.941 | 1,2-dichloroethane | 84 | 0.127 |
| DMSO | 189 | 0.825 | Water | 100 | 0.123 |
| 2-propanol | 82 | 0.799 | Chlorobenzene | 131 | 0.101 |
| Formic acid | 100.7 | 0.722 | Acetonitrile | 82 | 0.062 |
| Methanol | 65 | 0.659 | Acetone | 56-57 | 0.054 |
| 1,2-dichlorobenzene | 180.5 | 0.280 | Tetrahydrofuran | 66 | 0.047 |
| NMP | 202 | 0.275 | Dichloromethane | 39.8 | 0.042 |
| [bmim]PF$_6$ | - | 0.185 | Toluene | 111 | 0.040 |
| acetic acid | 118-119 | 0.174 | Hexane | 68-69 | 0.020 |
| Glycerol | 290 | 0.651 | 1,3-Propanediol | 211-217 | 1.30 (at 25 °C) |
| 1,4-Butanediol | 230 | 0.783 | 1,5-pentanediol | 242 | 0.456 |
| Nitrobenzene | 202 | 0.589 | Benzaldehyde | 178.1 | 0.337 |
| 1-2-propanol | - | 0.757 | Acetic acid | 113 | 0.174 |
| 1-Butanol | 118 | 0.571 | 2-Butanol | 100 | 0.447 |
| 2-Methoxyethanol | 124 | 0.410 | Chloroform | 61 | 0.091 |
| Nitromethane | 101 | 0.064 | o-Xylene | 144 | 0.018 |
| Ethyl acetate | 77 | 0.059 | 2-Butanone (MEK) | 80 | 0.079 |
| Tetrahydrofuran | 66 | 0.047 | Isobutanol | 108 | 0.522 |
| Dimethylformamide | | 0.161 | Dimethylsulfoxide | | 0.825 |
| Methylene chloride | | 0.042 | | | |

Metals/semiconductors (with free charge carriers and ions) interact with MWs in a distinctive and complex manner that differs from water and polar solvents. During MW irradiation, the material can couple with the MWs' $E$ or $H$-fields, or both. The principal heating mechanisms in the industrial high-frequency heating range ($10^7$ - $3 \times 10^9$ Hz), which encompasses radio frequency and MWs, include dipolar polarization, conduction, and interfacial polarization.

The conduction process relies on the presence of a limited number of free charges within the material matrix, for example graphite. At low MW frequencies, this process is typically fairly consistent; however, as the frequency increases, it diminishes and reaches a point of decline at about 100 MHz. The material has a low electrical conductivity ($\sigma$), and because of electrical resistance, the movement of unbound charges produces thermal energy.

The dipolar polarization mechanism is caused by the presence of molecules with a dipole moment in the dielectric material. When exposed to a fluctuating electric field in the presence of MW radiation, the dipoles within the dielectric oscillate, increasing the dielectric's internal energy. Frictional dissipation of internal energy may cause the material to heat up. The accumulation of charged particles along the interfaces of heterogeneous dielectric materials is known as interfacial polarization (or Maxwell-Wagner polarization).[48] The complex permittivity ($\varepsilon^*$) can help explain how a dielectric substance absorbs MWs and converts them into thermal energy. This means that the dielectric constant ($\varepsilon'$) is reflected in the real part of the complex permittivity in this case, and the dielectric loss ($\varepsilon''$), in the imaginary part, i.e., $\varepsilon^* = \varepsilon' - j\varepsilon''$. It should be noted that not all materials interact with an *E*-field or *H*-field in the same manner. As a result, it is critical to examine the material's dielectric properties under various processing conditions prior to developing MW equipment.

Under the MW exposure, the power dissipation/unit volume in a dielectric medium is expressed as follows:[49]

$$P = \sigma|E|^2 = (2\pi f \epsilon_o \epsilon_r' tan\delta)|E|^2 \qquad \text{Equation 2}$$

where $\sigma$ is the electrical conductivity, *f* is the frequency of the MW, *E* is the electric field, $tan\delta$ is loss tangent, $\epsilon_o$ is the permittivity of free space, and $\epsilon_r'$ is the relative dielectric constant.[38] The preceding equation shows that power dissipation depends linearly on $tan\delta$. The heating rate of a MW-irradiated material can be expressed as:[49]

$$\frac{\Delta T}{t} = \frac{\sigma|E|^2}{\rho c} \qquad \text{Equation 3}$$

Where $\rho$ denotes density, *c* the material's specific heat capacity, $\Delta T$ the temperature change, and *t* the duration. At room temperature (RT), graphite's high thermal conductivity and modest specific heat capacity allow for rapid heating under MWs.

Table 2 lists various minerals and inorganic chemicals that can absorb MWs at normal temperatures.[50,49] The temperatures reached by these materials as well as the corresponding exposure times when exposed to MW radiation in typical household MW ovens are also listed in the table. As displayed in Table 2, most types of carbon interact with MWs in their powder form. In particular, MWs (2.45 GHz) are quickly absorbed by amorphous carbon powder, which reaches 1550 K in less than a min in a domestic MW oven operating at 1 kW.[49] The observed permittivity values of carbon black, coal, and artificial graphite in the MW frequency range of 1-10 GHz[51] show that graphite's dielectric loss tangent is extremely high at 2.45 GHz.

Graphite and graphene are dielectric materials with a sp$^2$-bonded carbon network and carriers of a large number of delocalized $\pi$ electrons. These electrons, under the influence of the MW field, begin to move in the direction of the applied field generating an electric current.[52] Because these electrons are unable to keep in phase with a highly fluctuating alternating *E*-field, their energy is lost, culminating in heat generation in the material. GO and rGO possess polar oxygen functional groups that can engage with MWs due to differences in electron uptake ability between carbon and oxygen atoms, leading to electric dipole polarization.[53] When exposed to an electromagnetic field, these groups can cause vibrations that stretch and bend.[54] To produce nanoparticles, cobalt acetate tetrahydrate + thiourea for CoS and chloride precursor solutions of others (Ru, Pd, and Ir) were combined with GO and irradiated with MWs (1200 W). At 600 °C, GO reduced to rGO and the sample reached 1600 K in a 100 ms.[43] Figure 1d depicts the magnitude of the emitted light intensity over time. The rGO's defects are critical to attaining this high MW-induced temperature.[43] Figure 1e shows the temperature raise during MW heating of a mix of graphite with other additives with a 1:1 wt. ratio.[44] Therefore, rGO may be an effective susceptor for a wide range of MW solid-state processes (without any liquids). Susceptors, particularly carbonaceous materials (Table 2), as internal or exterior energy absorbing agents, can be adopted in MW solid state synthesis and modification of nanocarbons and nanocarbon composites, much as they are in liquid state MW-assisted techniques.[55] (Section 2.4.2).

Table 2: MW-active elements, natural minerals, and compounds.[50,49,56]

| Element/mineral/compound | time of MW exposure (min) | $T$(K) | Element/mineral/compound | time of MW exposure (min) | $T$(K) |
|---|---|---|---|---|---|
| Al | 6 | 850 | NiO | 6.25 | 1578 |
| C (amorphous, < 1 mm) | 1 | 1556 | $V_2O_5$ | 11 | 987 |
| C (graphite, 200 mesh) | 6 | 1053 | $WO_3$ | 6 | 1543 |
| C (graphite, < 1 μm) | 1.75 | 1346 | $Ag_2S$ | 5.25 | 925 |
| Co | 3 | 970 | $Cu_2S$ (chalcocite) | 7 | 1019 |
| Fe | 7 | 1041 | $CuFeS_2$ (chalcopyrite) | 1 | 1193 |
| Mo | 4 | 933 | $Fe_{1-x}S$ (phyrrhotite) | 1.75 | 1159 |
| V | 1 | 830 | $FeS_2$ (pyrite) | 6.75 | 1292 |
| W | 6.25 | 963 | $MoS_2$ | 7 | 1379 |
| Zn | 3 | 854 | PbS | 1.25 | 1297 |
| $TiB_2$ | 7 | 1116 | PbS (galena) | 7 | 956 |
| $Co_2O_3$ | 3 | 1563 | CuBr | 11 | 995 |
| CuO | 6.25 | 1285 | CuCl | 13 | 892 |
| $Fe_3O_4$ (magnetite) | 2.75 | 1531 | $ZnBr_2$ | 7 | 847 |
| $MnO_2$ | 6 | 1560 | $ZnCl_2$ | 7 | 882 |
| CaO | 4 | 389 | $Al_2O_3$ | 4.5 | 351 |
| Cu | 7 | 501 | Ni | 1 | 657 |
| Zr | 6 | 735 | Pb | 7 | 550 |
| Loss tangents of various carbon materials at 2.45 GHz MWs and RT ||||||
| Material | tan$\delta$ | Ref. | Material | tan$\delta$ | Ref. |
| CNT | 0.25-1.14 | 57, 58 | Carbon black | 0.35-0.83 | 59, 60 |
| Activated carbon (AC) | 0.57-0.80 | 61, 62 | Charcoal | 0.11-0.29 | 63, 62 |
| Graphitized carbon powder (60-80 mesh) | 0.4-1 | 59 | Carbon foam | 0.05-0.20 | 64 |
| Coal | 0.02-0.08 | 65 | Activated (at 400 K) carbon | 0.22-2.95 | 66 |
| tan$\delta_e$ (tan$\delta$ for $E$-field) of MW transparent materials at 2.45 GHz[67, 68,69] ||||||
| Material | tan$\delta_e$ | | Material | tan$\delta_e$ | |
| Alumina | 0.001 | | Borosilicate glass | 0.0012 | |
| PVC | 0.0056 | | Fused quartz | 0.0003 | |
| Silicon | <0.012 | | Polystyrene | 0.0003 | |
| Teflon | 0.00048 | | Silicon Nitride | <0.001 | |

MWs are generally reflected by bulk metals and alloys, and MW penetration is low.
In these situations, the MW penetration is frequently expressed using a metric known as skin depth, which is defined as:[67,70]

$$Ds = \frac{1}{\sqrt{\pi \mu \nu \sigma}} = 0.029(\rho \lambda_o)^{0.5} \qquad \textit{Equation 4}$$

Where $\sigma$ is the electrical conductivity (S/m), $\rho$ is electrical resistivity (Ω-m), and $\lambda_o$ (m) is the incident wavelength, $\nu$ is the MW frequency, and $\mu$ is the permeability of free space.

Therefore, it is clear that the energy absorbed during MW irradiation is dictated by the target material's electromagnetic characteristics as well as its thickness. The skin depth in bulk metallic materials is extremely small. When the diameters of the metallic particles are on par with the skin depth, a section of the surface area of the metal powder that couples with MW energy is sufficient to participate uniform MW heating (Figure 1c). Therefore, MW absorption and volumetric heating have been observed in various submicron and nanoscale metal powders.[71]

Investigation on the heating behavior of Fe, Co, Ni, Cu, and nonmagnetic nickel oxide (NiO) particles under the $E$ and $H$-fields using a single-mode applicator with all samples exposed to the same MW power input (0.2 kW) under Ar,[42] revealed that Fe particles (~70 mm in size) were heated well in the $H$-field, and the ferro-magnetic metal particles (with a higher Curie point) were heated even better. It was also possible to heat Fe bulk particles (~3 mm) in an $H$-field in the absence of an electric discharge (Figure 1i-j). Among Ni particles (45-150 mm in size), particles of smaller size were heated better. Diamagnetic Cu particles with diameter (⌀) > 2 mm did not absorb MW energy in 2A (the largest $H$-field position) and were heated only by the episodic recurrences of electric discharge in 1A (largest $E$-field position, Figure 1j). NiO particles (⌀ of 7 μm) were found to be heated well only under high $E$-field. NiO is a nonmagnetic (anti-ferro magnetic) ceramic material and has low $\sigma$ (<1 × 10$^{-12}$ S/cm) below 50 °C. As a result, it was heated solely by dielectric loss, rather than magnetic or eddy current loss mechanisms.

The magnitude of MW absorption by several types of metallic materials using a multimode 2.45 GHz MW applicator has also been examined.[47] Fe powder was reported to be the most efficiently heated material because of the eddy current loss (in oscillating $H$-field) and magnetic reversal loss (in oscillating $E$-field) mechanisms that ensued in Fe. Diamagnetic metals Cu and Sn were heated better in comparison to paramagnetic Ti, while Au was only mildly heated. Cu- and Ni-based metallic glassy particles were slightly heated. (Figure 1k).[47] It is shown that FeO and $Fe_2O_3$ get heated with the $E$-field of MWs, while $Fe_3O_4$ can be heated in both the $E$ and $H$ fields (see Figure 1e).[72] Conducting polymers are also known to absorb MWs effectively. For example, in the MW synthesis of nanostructured carbon from conducting polypyrrole nanospheres, nanofibers, or NTs (prepared using chemical polymerization of pyrrole using $V_2O_5$ nanofibers as the structure-directing agent), after 3-5 min of irradiation, the temperatures of the sample (20-50 mg) surpassed 800 °C.[73] In another case of PPy nanofibers serving as a MW energy absorbing agent, a mixture of 50 mg PPy nanofibers and 50 mg $Mo(CO)_6$ or $W(CO)_6$ in a glass via was irradiated with 1250 W MWs for 60 sec. Intense sparking was observed within the first 3 sec, and the temperature of the mixture reached ~1000 °C.[74]

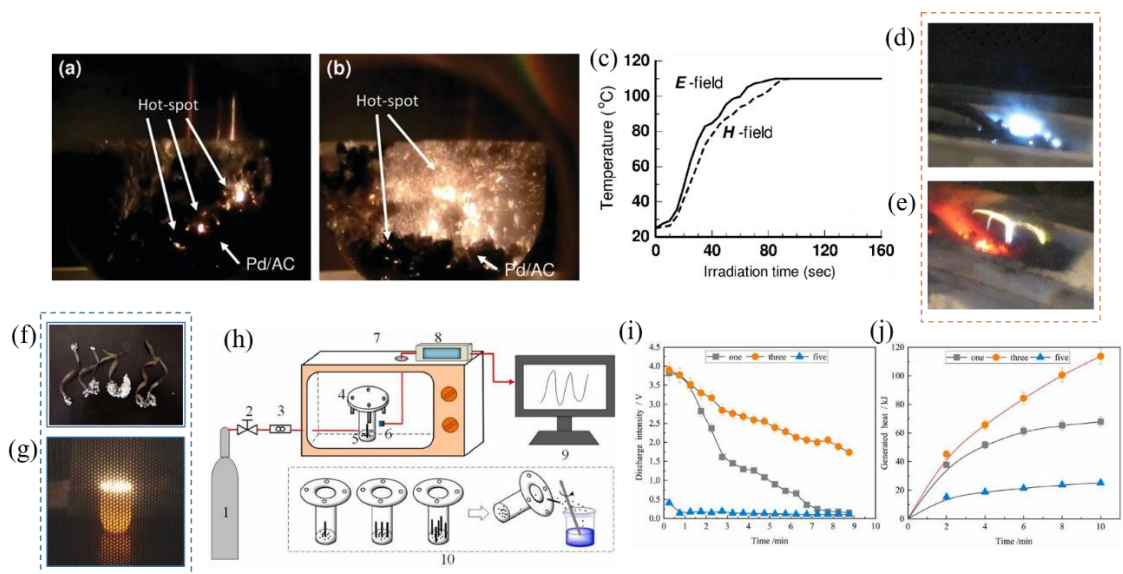

Figure 2: (a-c) formation of hotspots and their effect. Electric discharges occurring on the Pd/AC catalyst surface under *E*-field: (a) 40 sec of irradiation and (b) 120 sec of irradiation; (c) Temperature raise vs. time profiles in the synthesis of 4-methylbiphenyl with 70-W MWs (under *E*-field and *H*-field conditions).[75] (d) Ball lightning plasma; (e) arc discharge plasmas.[76] (f) Appearance of the metal strips following discharge & (g) Discharge phenomena observed when stainless-steel strips in quartz sand were exposed to MWs.[77] (h-j) Dynamic characterization of discharged intensity & generated heat. (h) experimental system: 1. Gas cylinder; 2. relief valve; 3. flow meter; 4. quartz reactor; 5. needle-shape metal; 6. photo-resistance; 7. MW oven; 8. micro control unit; 9. CPU; 10. hydro-thermal exchange system; (i) Effect of number of metal wires on the discharge performance & generated heat (j) where metal: Fe wire; MW power: 480 W; wire length: 4 cm; wire ⌀: 1 mm were employed.[78]

  On carbon material surfaces, the dielectric properties may not be homogenous. The localized concentration of dipole/π-electrons in certain regions may be different, thus differences in their ability to absorb MWs. This effect gives rise to what is called hot-spots (temperature gradients),[79] where the temperature can exceed several 100 °C as compared to the surrounding regions. The hot-spot effect is thought to be responsible for some of the non-thermal effects caused by MW heating. In a MW-assisted polyol approach (300 W) for synthesizing Ag NWs in the presence of polyvinylpyrrolidone, a hot spot effect at the wire ends promoted NW growth.[80] Schatz investigated the polarizability of noble metal truncated tetrahedral nanoparticles after activation of the material's plasmon resonance and concluded that this results in a more than two orders of magnitude rise in the local electric field at the vertices.[81] Because the NW endpoints are evolving sites with defects, they may attain greater temperatures than the middle part and accrue charge as a result of MW field polarization. Similarly, Hu et al. used the MW-assisted polyol technique (using polar diethylene glycol, permittivity $\varepsilon \approx 32$, and BP = 244-245 °C)[82-84] to produce ZnO colloidal nanocrystal clusters (57 to 274 nm).[85] Because of polar ZnO's outstanding MW-absorbing properties, hot-spots on solid ZnO may have formed under MW irradiation, accelerating nanocrystal growth and subsequent cluster formation. In an applied alternating *E*- field, an increased driving force for mass transport and directed crystallographic fusion of nanocrystals can be generated on the hot surface of ZnO by localized ionic currents caused by the high conductivity and polarization of the diethylene glycol solvent and ZnO. Because the MW field may directly heat the freshly generated ZnO nuclei and nanocrystals, the ZnO crystals reach significantly greater temperatures than the remainder of the bulk mixture.[85]

Chen et al. demonstrated that in SiC particles/paraffin oil, hot spots can cause temperature differences between surrounding substances of several 100 °C, or even more in certain instances.[86] The SiC particles had a much greater temperature profile than the surrounding paraffin oil. The temperature variations are greater for larger SiC particles. The magnitude of hot-spot effects is heavily influenced by the amount and speed with which heat created by the strong MW-absorbing medium is transferred to the weakly absorbing medium. Investigation on the formation of hotspots and their effect on the heterogeneous Suzuki-Miyaura coupling reaction during the synthesis of 4-methylbiphenyl in toluene solvent in the presence of Pd/activated carbon (AC)[75] shows that under high $E$-field conditions, hotspots formed easily on the surface of the AC catalyst support. Chemical yields of 4-methylbiphenyl were twice as high under the MWs' $H$-field as under the $E$-field. The excessive creation of hotspots has a negative impact on this coupling reaction. Figure 2a shows whitish-orange arcing on the catalytic surface after approximately 40 sec of irradiation. With additional MW irradiation, the frequency of electric discharges continued for up to 120 seconds (Figure 2b). Figure 2c shows that the rate of heat increase after 60 sec under $E$-field conditions (1.5 °C/min) was similar to that under $H$-field heating (1.4 °C/min). To prevent the formation of hot spots, Petricci et al. have proposed a non-toxic biomass-derived $\gamma$-valerolactone (BP of 208 °C) as a potential green substitute for toluene in MW-assisted Pd/C catalyzed processes.[87]

## 1.2 Microwave arcing phenomena

When a mixture of organic solutions and metal wires or powders is irradiated with MWs, electric arcing occurs intensely between the tips of metal wires (or sharp edges of metal particles), resulting in the carbonization of organic liquids and the formation of amorphous carbons.[88] The applied $E$-field causes an electric current in the conducting material, resulting in the accumulation of +ve and -ve charges on opposite sides. On sharp edges and submicroscopic irregularities, the surface charge density can be extremely high, which can lead to an electrical breakdown in the medium and the creation of an electrical spark or arc. The MW arcing phenomenon is substantially more vigorous in nonpolar solvents than in polar ones. During MW arcing, it produces a plasma environment, and the temperature of the immediate vicinity surpasses 1100 °C, resulting in the carbonization of organic molecules and the creation of amorphous carbon structures. Arcing process in metal-solvent systems under MWs is determined by a variety of parameters: the number, morphology, size, metal particles' surface characteristics, ionization energy, purity and $\tan\delta$ of the solvent, and the MW power.[88] In the past, metal particles were purposefully introduced in polar organic solvent systems to aid MW absorption and expedite solution heating for rapid chemical synthesis.

Aside from liquid-phase synthesis applications in synthetic chemistry, arcing phenomena hold significance in material science and nanomaterial research. Core/shell metal/carbon NPs,[89] $Fe_3O_4$/carbon nanocomposites,[90] $ZrB_2$/metal oxide nanostructures,[91, 92] rGO, partially rolled graphene,[93] Cu and Ni NPs and CuS nanorods,[93] $ZnF_2$,[93] and $NiF_2$,[93] CNTs,[94] and metal fiber/polymer composite materials have all been produced through arcing effects involving highly conductive metals or metal oxides.[95] MW triggered electrical discharges (corona, spark or arc discharges) need not to occur in liquid media. Under specific atmospheric conditions, arc discharge between graphite or coal rod electrodes can produce a wide range of effects: doping, exfoliation, intercalation, and reduction, thus aiding in the fabrication of advanced carbon products.[96, 97] The same is true in the case of MW-generated arc discharge, where various atmospheres, such as $CH_4/N_2$, Ar, $CH_4/CO_2$, and air, were introduced to obtain different nanocarbons.[98-100] Using carbon or graphite-type materials that show electric discharges at MW conditions, high-temperature pyrolysis reactions[101] and the rapid synthesis of graphene sheets[102] and nanoscrolls[103] have been conducted. When activated carbon, biochar, black coke, semi-coke, and metals (W, Fe, Cu, and Al) are exposed to MW radiation, MW discharge occurs in the form of tiny sparks or electric arcs.[104-106] Two types of plasma—ball lightning plasma and arc discharge plasma—have been observed while MW heating of carbons (Figure 2d & e).[76] Included in the experiment were graphite, metallurgical coke, anthracite, activated carbon, and biomass pyrolysis char. Temperatures below 400 °C were more likely to produce ball lightning plasmas, while temperatures between 400 °C and 700 °C favored arc discharges. Plasma regions ignited earlier than other carbon sections because they experienced faster temperature increases than the rest of the carbon.[76] MW-induced metal discharge has been used in the decomposition of $N_2O$,[107] decomposition of biomass tar,[108] pyrolysis of waste tire[109, 110] and the metal recovery from the e-wastes[111, 112] Figure 2f illustrates how the heat generated by the discharge might melt the stainless-steel strip terminals.[77, 113] Figure 2g shows the occurrence of discharge phenomena when stainless-steel strips embedded in quartz sand were irradiated with MWs.[77]

The dynamic characterization of MW-induced metal discharge with plasma and thermal effects and its progression is complex.[114-116] High-speed video technology can provide photographic records of the MW discharge, while spectral analysis equipment can reveal the composition of plasma generated by the MW discharge.[117] To comprehend the dynamic properties of MW discharge, photo-voltage conversion approach, which relies on the use of photosensitive resistance is employed.[118, 119] A study of the dynamic characteristics of discharge intensity and generated heat using light-to-voltage conversion and hydrothermal exchange methods applied to Fe, Cu, and Al generated discharges found that Fe had better discharge efficiency than Al and Cu.[78] The discharge intensity of Fe was 1.66 times that of Al and 1.22 times that of Cu. The increase in MW power from 320 to 480 W resulted in an extended discharge duration (5.75 to 7.25 min). Despite a significant decline in discharge intensity with time, MW discharge could increase the efficiency of electric energy conversion into heat energy by 48.41% when using one wire of Fe with a length of 4 cm and a ⌀ of 1 mm at 480 W. In the experiment, a photoresistor was utilized to determine the intensity of MW discharge, and the generated heat in both the discharge and non-discharge stages was evaluated using the hydrothermal exchange method (Figure 2h). Figure 2i-j shows that utilizing three Fe wires resulted in an 80% higher avg. intensity of MW discharge than using a single Fe wire. There are two possible explanations for this pattern. First, when three Fe wires were exposed to MW radiation, the discharge occurred either simultaneously or alternately, greatly increasing the frequency of MW discharge.[78] Another explanation for this is that, unlike using a single Fe wire for discharge, inserting three Fe wires would result in a coupling effect that would increase the electromotive force and potential difference.[117] According to Figure 2j, the avg. heat produced from one wire of Fe during the experiment was ~5 kJ (with five wires inserted simultaneously). Recent studies on the synthesis of a variety of nanocarbons using the arc effect are discussed in the following sections.

This review examines recent findings on the synthesis of nanocarbons and their composite materials obtained by various MW-assisted routes, as well as their properties. Since carbon allotropes are good MW absorbers, they are used in the modification of nanocarbon materials. The article discusses the latest results of such studies.

## 2 Preparation of nanocarbons and their hybrids

### 2.1 Fabrication of porous carbon materials

In addition to NPs, nanospheres, nanosheets (NSs), CNTs, fibers, and so on, carbon nanomaterials encompass nanostructured bulk materials: porous materials that are useful in different applications due to their high *SSA*. Heat-treating solid carbon materials to produce activated carbons is one possible preparation process.

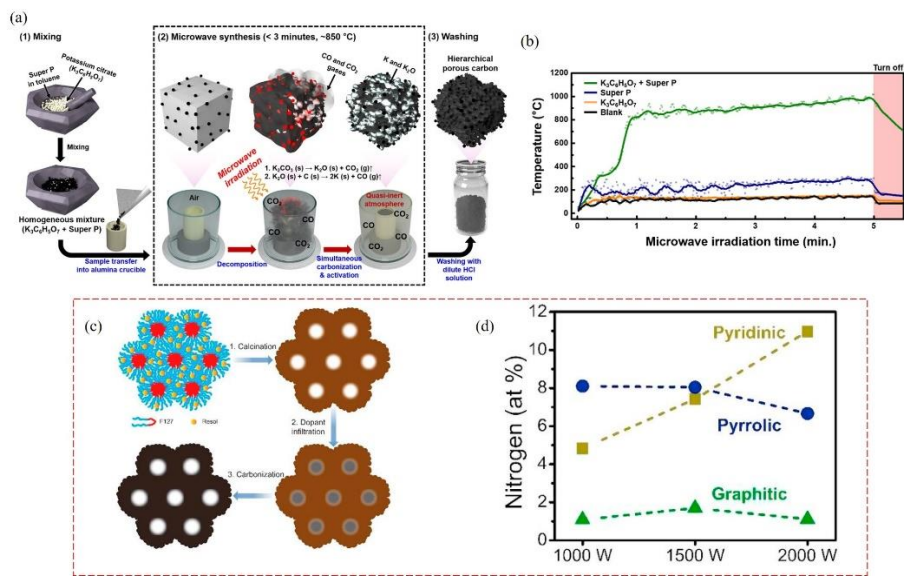

Figure 3: (a-b): HPAC synthesis. (a) Depiction of synthesis of HPAC from a uniform mixture of potassium citrate and super P; (b) Temperature variations in the samples in the first 5 min of MW irradiation. $K_3C_6H_5O_7$ (2000 mg), super P (20 mg), and a combination of the two were independently synthesized in $Al_2O_3$ crucibles, and the temperature variations were monitored with an IR thermometer (yellow, blue, and green lines, resp.). The temperature of an empty $Al_2O_3$ crucible was recorded under the identical conditions (shown by the black line).[120] (c-d): N-doped ordered mesoporous carbon. (c) Fabrication of N-doped ordered mesoporous carbons on Ni foam involves (1) cooperative assembly to form micelle-templated crosslinked resol, (2) calcination at 350 °C to decompose F127 template to yield a mesoporous framework, (3) permeation of melamine to the mesopores as a N source, and (4) carbonization to produce mesoporous N-doped carbons; (d) Influence of MW power on the condition of N bonding inside the carbon framework, as revealed by XPS. The N content is displayed using an absolute scale, derived from XPS data.[121]

Traditionally, activation is carried out using traditional heating processes, with treatment temperatures spanning 400-950 °C and activation durations ranging from 0.5-5 hours.[122] Based on the variations in their pore sizes, porous carbons are classified as microporous (pore size ≤ 2 nm), macroporous (pore size > 50 nm), mesoporous (2 nm ≤ pore size ≤ 50 nm), and hierarchically porous (micro-mesomacroporous) carbons with ordered/disordered porosities and amorphous/graphitic textures. In recent times, MW heating has developed as a viable activation technique because of the volumetric heating properties of MWs.

A small quantity of carbon black (super P) was introduced as an internal susceptor in a fast carbonization/activation process of potassium citrate ($K_3C_6H_5O_7$) for producing hierarchical porous activated carbon (HPAC) using a one-step MW heating technique (700 W, 2.45 GHz, exposure duration 1 to 4 min).[120] The graphitic structure of the susceptor (super P) allows for effective Joule heating by absorbing MW energy through delocalized π-electrons,[55] which aids in volumetric heating of the homogenous mixture. Both carbonization and activation processes were carried out concurrently and completed quickly under MWs after encasing the samples to prevent abrupt changes in environmental conditions from air to $CO/CO_2$ gases. This resulted in the formation of hierarchical porous carbon structures with a large *SSA* of 1280.4 m$^2$/g. Subsequently, MW-induced hierarchical porous carbon was employed as the electrode material in electrical double-layered capacitors (EDLCs), which demonstrated a $C_{sp}$ of 318 F/g at a current density of 1 A/g in a 3-electrode system. The MHPC materials were produced as shown schematically in Figure 3a. Figure 3b depicts the sweeping temperature variations of the samples exposed to MWs over the first 5 minutes. The reaction mechanism states that the super P powder dissolved in the organic salts effectively absorbs the MW energy, causing the sample to quickly reach ~300 °C, the temperature at which the $K_3C_6H_5O_7$ organic moieties are pyrolyzed. At this point, the breakdown of $K_3C_6H_5O_7$ results in the simultaneous formation of $K_2CO_3$ and the bulk carbon matrix. Because of this endothermic process, the temperature is sustained for a while. The temperature then rises quickly once more as a result of the additional MW energy that is absorbed by super P and the bulk carbon matrix. As a result, sequential activation and more carbonization happen on its own. In the process, thick CO and $CO_2$ gases were vigorously produced from the organic moieties of $K_3C_6H_5O_7$ within 1 min, completely filling the glass chamber during the MW exposure.

The preparation of activated carbon from agricultural by-products using conventional heating processes and MW treatment reveals that the optimum conditions are 400-800 °C (1-3 h duration) for conventional heating and 350-700 W (5-15 min duration) for MW-based heating.[123] For example, using MW-assisted chemical activation technique (600 W, 6 min), activated carbon was obtained from pineapple peel using KOH and $K_2CO_3$ (with char:KOH of 1:1.25 wt.%).[1] The KOH activated sample had a suitable development of pore structure, with *SSA* (1006 m$^2$/g), total pore volume (0.59 m$^3$/g), and avg. pore size (23.44 Å), respectively (Figure 4a). Alternatively, activated carbon can be obtained by preparing a ternary compound of potassium-nongraphitic carbon and tetrahydrofuran (THF) and heating it with MWs for 2 minutes. This expands the interstices between adjoining graphene layers in the ternary compound, resulting in a microporous structure with a large number of ultramicropores (ø of < 0.7 nm).[124] After MW irradiation of nongraphitic carbon, the constituent stacked graphene layers were fractured, yielding a *SSA* of 563 m$^2$/g.

Xerogels are dried gels that retain at least some of their porous nature after drying at RT.[125] Xerogels have higher porosity, a larger *SSA*, and extremely small pore diameters, making them suitable as precursors for the synthesis of porous carbon structures. For comparison, conventional and MW heating procedures were used to produce resorcinol-formaldehyde carbon xerogels with various starting pH values. The effect of the precursor solution's pH and the synthesis process used on the final materials' textural and chemical qualities was investigated.[126] High pH results in just microporous carbon xerogels, but low pH leads the materials to turn into micro-mesoporous and subsequently micro-macroporous carbon xerogels. Apart from the synthesis duration (~5 h for MW-assisted synthesis vs. several days for conventional methods), the primary distinction between the two heating methods is the meso-macroporosity of the obtained materials. This is because MW radiation primarily yields mesoporous carbon xerogels with a particular mesopore size across a broader pH range than conventional synthesis. For instance, mesoporous MW samples have a pH range of 4.5-6.5, whereas conventionally synthesized similar samples need an initial pH of 5.8-6.5.

The MW-assisted glycerol-mediated hydrothermal (HT) method followed by annealing is used to create composites with ultrafine $Na_3V_2(PO_4)_2F_3$ nanocrystals (< 20 nm) implanted in porous carbon nanospheres.[127] As a cathode material for SIB, the product shows a charge/discharge capacity of 106 mAh/g at 0.5 C after 100 cycles with a retainment of almost 100%. The intensity ratio ($I_D/I_G$) of the D band to the G band correlates the degree of disorder and avg. size of the $sp^2$ domains.[128] The product's $I_D/I_G$ value of 0.86 in the Raman spectrum suggests the presence of amorphous carbon. Xia et al. presented a method for producing ordered, high *SSA* (258 m$^2$/g) mesoporous carbons with a regulated nitrogen content directly on a Ni foam framework using MWs.[121] Melamine serves as the source of nitrogen, and cooperative assembly of phenolic resin (resol) and Pluronic F127 coated on the Ni foam creates a pathway to hierarchically organized composite structure to facilitate effective transport. Four steps were involved in producing nitrogen-doped mesoporous carbon: permeating the precursor solution into the Ni foam, heating to crosslink the resol and decompose the Pluronic template, melting melamine into the mesopores for nitrogen sources, and carbonization to generate the mesoporous carbon (1000-2000 W under $N_2$ for 3 min) (Figure 3c). As MW power increases, so does the total nitrogen absorbed into the framework, with pyridinic nitrogen taking precedence (Figure 3d). As a result, increased MW power increases both total nitrogen doping and the oxygen reduction reaction (ORR) catalytically active pyridinic-N doped content.[129, 130]

Heteroatom doping (donor or acceptor atoms) is helpful in carbonaceous materials to regulate the ion transport, electronic transfer, and active site for electrochemical energy storage.[11, 131-134] Doping carbon with N and B can lead to higher conductivity and pseudocapacitance.[135-137] Furthermore, both B and N atoms can achieve a high doping concentration because of their similar atomic radii. Using polyacrylamide as a carbon source and MgO as a template, a heteroatom-doped (N and O) hierarchical porous carbon (HPC) was synthesized without any activation via a conventional carbonization under $N_2$ atmosphere at 700 °C for 120 min.[138] The produced samples have a large *SSA* (989.836 m$^2$/g), a 3D linked hierarchical porous structure, and a high concentration of oxygen (10.3 at%) and nitrogen (10.97 at%) functional groups. With a carbonization temperature of 700 °C and $N_2$ serving as the carbonization atmosphere (Figure 4b), the optimized electrode sample (NHPC-700) exhibits excellent rate characteristics, cycling performance of 100.76% capacitance maintenance after 10000 cycles, and a $C_{sp}$ of 295.3 F/g at 0.5 A/g. Moreover, in a 1 M $Na_2SO_4$ aqueous electrolyte, the constructed NHPC-700//NHPC-700 symmetrical supercapacitor offers an *energy density* of 15.16 Wh/kg.

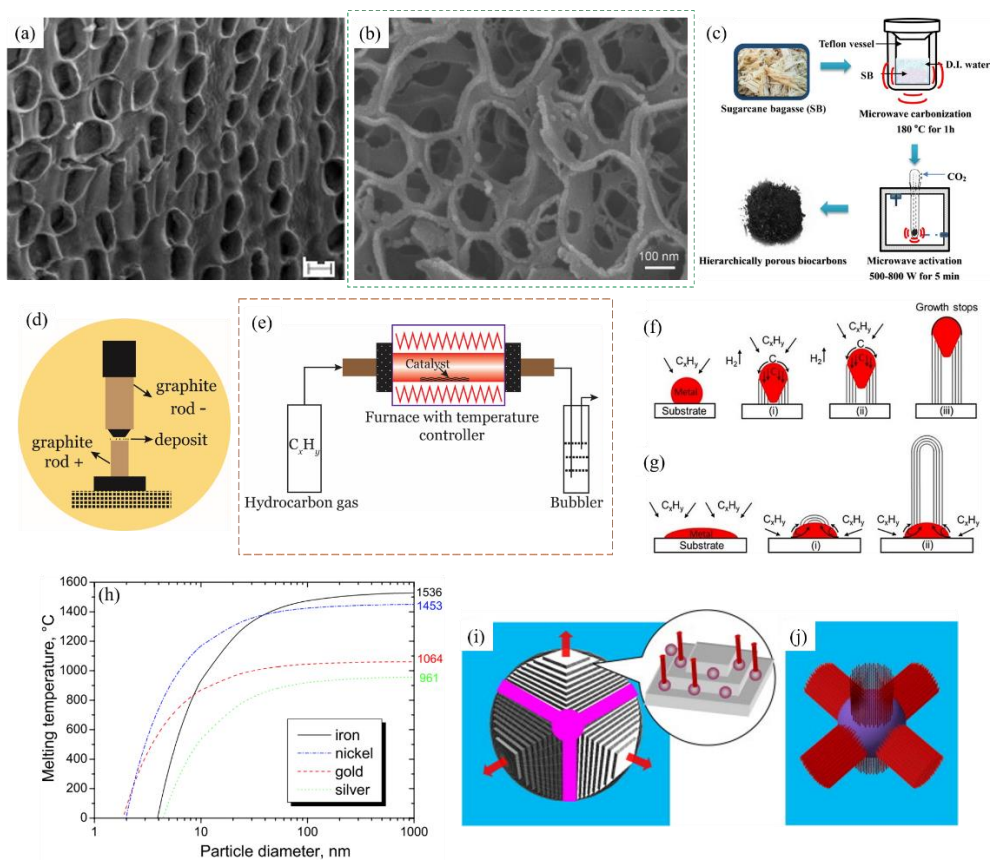

Figure 4: (a) Activated carbon obtained by KOH chemical activation at 600 W for 6 min, scale bar: 10 μm.{Foo, 2012 #503} (b) SEM images of a sample with carbonization temperature of 700 °C and $N_2$ as carbonization atmosphere.[138] (c) Steps for producing hierarchically porous biocarbons from SB. The SB was first treated with DI water, dehydrated at 105 °C for a day before being crumpled and ground into a fine powder.[139] (d) Experimental setup for synthesizing CNTs using the traditional arc discharge method. (e) Schematic of the simplest CVD system. (f-g) Generally accepted CNT growth mechanisms. (f) tip-growth model; (g) base-growth model; (h) Melting temperatures for some metals vs. particle size.[141]; (i-j): Orthogonal growth of CNTs on $Al_2O_3$ substrate.[142]

One emerging technique that economically removes ions from salt water is called capacitive deionization (CDI), which uses electrosorption at low voltages.[143, 144] Biocarbons were selected as one of the best materials for CDI electrodes due to their high pore volume, large *SSA*, durability, and low cost. Sugarcane bagasse (SB) biowastes are utilized as the starting materials to prepare hierarchically porous carbons using MW carbonization and activation with KOH in the presence of $CO_2$ (Figure 4c).[139] By regulating the MW power and the physical activation agent, $CO_2$, it is possible to tune *SSA*s and ratios of $V_{meso}/V_{total}$. The biocarbons (SB-CO$_2$-700) that are activated at 700 W of MW irradiation in a $CO_2$ atmosphere had a *SSA* of 764 m$^2$/g and a higher mesopore to total pore volume (~64.1%). The electrosorption capacity of the SB-CO$_2$–700 is measured to be 11.4 mg/g in 10 mM NaCl solution at 1.2 V, whereas the $C_{sp}$ of the same sample is found to be around 123 F/g at 5 mV/s. Further, defects may be noticed, as the $I_D/I_G$ for biocarbons MW treated at 700 W is greater than those activated at 500 and 800 W.

## 2.2 Synthesis of CNT, carbon nanofiber (CNF), CNT purification and modification and CNT hybrids

### 2.2.1 CNT and CNF

There are three primary methods for producing carbon nanotubes (CNTs): chemical vapor deposition (CVD),[145] arc discharge,[146] and laser ablation[147, 148]. These procedures typically entail vacuum or inert gas protection, or high atmospheric pressure, and high temperatures. Figure 4d shows the experimental setup for synthesizing CNTs using the traditional arc discharge approach. A DC arc discharge between two rods of pure graphite consumes the anode and produces nanocarbon. CNTs were produced in large quantities by arc discharge in a helium atmosphere at 60 kPa.[149] Arc discharge and laser ablation methods require a solid target and extremely high temperatures to evaporate it, and the CNTs produced by both methods are tangled. CNTs are detected in the arc discharge method with a high concentration of amorphous carbon, and they are not aligned. Further, the complicated techniques and instruments required for CNT synthesis keep the cost of as-produced CNTs high. At the industrial level, CVD is a widely used method for synthesizing CNT. CVD encompasses different sub techniques such as catalytic CVD (CCVD)[150](commonly used technique), water-assisted CVD,[151] thermal or plasma-enhanced (PE) oxygen assisted CVD,[152] laser-assisted CVD,[153] radio frequency CVD (RF-CVD),[154] microwave plasma CVD (MPECVD),[155] and hot-filament CVD (HFCVD)[156, 157]. Figure 4e depicts the commonly used experimental setup for CNT growth by CVD. A hydrocarbon vapor (~15-60 minutes) is circulated through a tubular reactor that has a substrate or catalyst support present at a temperature high enough (600-1200 °C) to break down the hydrocarbon. The CNTs are grown on the catalyst and collected after cooling the system to RT. If the hydrocarbons are in liquid state (benzene, alcohol, etc.), the liquid is heated in a flask, and the flask is purged through an inert gas, which carries the hydrocarbon vapor to the reactor zone. Volatile CNT precursors: camphor, ferrocence, naphthalene, etc. directly turn into vapor and pass over the catalyst placed in the high temperature zone. For utilizing a solid hydrocarbon as a CNT precursor, it can be positioned in the low-temperature area of the reaction tube.

Achieving hydrocarbon breakdown solely on the catalyst surface and forbidding airborne pyrolysis are essential to obtaining pure CNTs. In general, the diameter of the CNT is determined by the size of the catalyst particles. As a result, metal nanoparticles of regulated size, produced using dependable procedures, can be used to generate CNTs of controlled diameter. Because carbon is highly soluble in Fe, Co, and Ni metals at high temperatures and has a high carbon diffusion rate in these metals, these metal catalysts are the most widely utilized.[158] Compared to other transition metals, Co, Fe, and Ni adhere more strongly to the growing CNTs, making them more effective in the formation of high-curvature (low ⌀) CNTs.[159] Other transition metals, including Cu, Au, Ag, Pt, and Pd, were discovered to stimulate CNT development from diverse hydrocarbons.[141] A range of metals: Fe, Co, Ni, Cu, Pd, Pt, Mn, Cr, Mo, Sn, Mg, Au, and Al have been favorably used for horizontally aligned SWCNT growth on $SiO_2$ substrates.[160] High densities of horizontally aligned SWNT arrays were produced by the breakdown of methane and ethanol on Cu nanoparticles supported on Si wafers at temperatures between 825 and 925 °C.[161] Methane decomposition on Re catalyst[162] is also been reported. It has been shown that transition metals are also effective catalysts in arc-discharge and laser vaporization processes. Some studies have indicated that the catalysts need not to be particles. Catalyst thin films applied to a variety of substrates have also been shown to be effective in producing consistent CNT deposition.[163] In an analogous manner, high-yield CNT growth has been obtained by acetylene decomposition on a stainless steel sheet at 730 °C.[164] Alloys have been proven to have higher catalytic activity than pure metals. For example, Fe-Co catalyst on zeolite support,[165] and Co-Mo and Ni-Mo on MgO support.[166] The following hypothesis for CNT growth mechanism is a well-accepted. A hydrocarbon vapor first decomposes into carbon and hydrogen species when it comes into contact with hot metal nanoparticles. The hydrogen sails away, and the carbon dissolves into the metal. After exceeding the carbon solubility limit in the metal at this temperature, the dissolved carbon precipitates and crystallizes in the form of a cylindrical network without dangling bonds and is therefore energetically stable. The exothermic nature of hydrocarbon decomposition causes some heat to be released into the metal's exposed zone. The endothermic process of carbon crystallization, on the other hand, takes heat from the metal's precipitation zone. The process is kept sustainable by the exact temperature gradient inside the metal particle.[158] In the CNT formation, there are two broad scenarios. If there is insufficient catalyst-substrate interaction, hydrocarbon breaks down on the metal's top surface, carbon diffuses down through the metal, and CNT precipitates out over the metal bottom, pushing the entire metal particle away from the substrate (Figure 4f(i)). As long as the metal's top is open to new hydrocarbon breakdown, the concentration gradient persists, allowing carbon diffusion and CNT to grow longer and longer (Figure 4f(ii)). Excess deposition of carbon on the metal inhibits catalytic activity and CNT development (Figure 4f(iii)). This type of growth process is called tip-growth model. In other scenario (Figure 4g), when the catalyst-substrate interaction is strong, initial hydrocarbon decomposition and carbon diffusion occur similarly to tip-growth, but the CNT precipitation does not to push the metal particle up, forcing the precipitation to emerge from the metal's apex. Carbon initially crystallizes as a hemispherical dome, which then spreads upwards as a smooth graphitic cylinder. Subsequent breakdown of hydrocarbons occurs on the metal's lower peripheral surface, and as-dissolved carbon diffuses upward. As a result, CNT develops with the catalyst particle rooted on its base, and hence, this growth process is dubbed the base-growth model. Nonetheless, it remains unclear whether the metal is in a solid or liquid condition during CNT growth. Some studies suggest the active catalyst being in the liquid phase.[167, 168] Figure 4h shows that the melting point of nanoparticles of some metals (< 10 nm) decreases rapidly. At around 800 °C, an 8-nm Fe and Au particle (or 4-nm Ni particle) could melt. The typical temperature range for CNT growth is 700-900 °C. This means that the catalyst metal may be liquid at higher temperatures (>800 °C) or solid at lower temperatures (< 800 °C). For detailed analysis of the physical state of the catalyst, mode of carbon diffusion, and chemical state of the catalyst, readers are directed to an article by M. Kumar.[158]

The catalyst-substrate interaction plays an important role in achieving effective CNT development. Commonly employed catalysts supports (substrates) are silicon, alumina, quartz, silica, $CaCO_3$, magnesium oxide, and alumino-silicate (zeolite).[158] Catalysts in the nanopores of zeolite substrates have produced very high yields of CNTs with a narrow range of ⌀s.[169] High yields of aligned CNTs with a high aspect ratio have been observed when Fe catalyst films are deposited on alumina substrate (10 nm thick).[170] It has been suggested that the oxide substrate may be involved chemically in the formation of CNTs when it serves as a physical support for the metal catalyst.[171] The direction of CNT growth is determined by the exposed substrate surface's crystallographic orientation. For example, CNTs grow more preferentially at 90° to the surface of Si(100) but at 60° to the surface of Si (111).[172] MgO (001) single-crystal substrate exhibits preferential CNT growth along [110] direction.[173] On micro-sized alumina particles, an orthogonal CNT growth has been seen through CVD of a ferrocene-xylene mixture at 600 °C (Figure 4i-j).[142]

Common precursors for CNTs include ethylene, methane, acetylene,[174] xylene, benzene, and carbon monoxide. A broad range of precursors have been employed to prepare CNT structures including, cyclohexane,[175] Fullerene,[176] alcohol CVD on Fe-Co-impregnated zeolite support,[177] amino-dichloro-s-triazine, pyrolyzed on cobalt-patterned $SiO_2$ substrates,[178] camphor,[179] kerosene,[180] liquefied petroleum gas,[181] and coal gas[182]. Because of the numerous advantages of MW-assisted synthesis over traditional approaches, producing CNT (or similar structures) with MWs as an energy source is gaining more attention.

Rapid synthesis of CNTs can be achieved by directly MW irradiating catalyst (Fe) particles on the surface of a solid carbon source, such as activated carbon fiber (ACF).[183] The impregnated metal catalysts rapidly absorb energy from the MWs via the dielectric heating method.[38, 184] Furthermore, the solid carbon source itself is a good absorber of MW energy (Table 2), causing self-annealing and facilitating the creation of CNTs. As indicated in Table 2, Fe is a good absorber of MW energy as well. Direct MW irradiation may cause structural changes in ACF, accompanied by the generation of carbon-containing gases. These gases could be employed as carbon feedstock in the synthesis of CNTs. In this investigation, an ACF specimen (1.5 cm × 2 cm, surface area of 1785 $m^2$/g) was impregnated with catalyst NPs via a catalytic solution.[185] The dried ACFs were placed in a quartz tube reactor (0.3048 m long, 0.0254 m in ⌀) and irradiated with MWs directly via a custom-built MW irradiation apparatus (2.45 GHz; 2 kW) under Ar flow. Figure 5a-c illustrate the morphology of the synthesized CNTs when the irradiation duration tuned from 30-120 sec with 400 W of MW power. It was discovered that there is an appropriate window during the irradiation period for the MW-assisted growth of CNT. At a brief irradiation duration, carbon nanostructures as small as a few tens of nm formed an aggregate (Figure 5b). However, as the irradiation period rose, fibrous carbons predominated, and the result became nearly particle-free. Large carbon particles and very thick fibrous carbons emerged after an extended exposure time. Furthermore, a sample without catalyst impregnation was irradiated for 90 sec (Figure 5c), and no CNTs were detected. The ratio, $I_D/I_G$ dropped from 0.91 to 0.41 following MW irradiation, indicating that the specimen's crystallinity had increased.

A single-mode MW reactor was utilized to grow MWCNTs and CNFs on Si, quartz glass, and mica substrates with identical dimensions.[186] The catalyst layers were made from various Fe, Ni, and Co salts and applied to Si substrates by dropcasting ethanolic metal salt solutions and then evaporating the solvent. The dried substrates were subjected to MW irradiation (200 W, 5 min) in the presence of several carbon sources (ethanol, 2-propanol, butanol, and EG), while recording temperature and pressure (Figure 5d-g). The best results were obtained with Ni-based catalyst materials (⌀ of ~14 nm; 2 min of irradiation) and ethanol as the carbon source in terms of size and homogeneity of the produced CNTs. Individual CNTs and Y-junction CNTs were produced without the use of an additional catalyst via MW breakdown of methane.[187] The synthesis was performed in a vertical quartz tube heated in a domestic MW oven (2.45 GHz, 6 kW, Figure 5h). Carbon felt was used as a MW susceptor without a catalyst and exposed to MWs. The experiment involves flushing the quartz tube with $N_2$ to remove oxygen from the reaction chamber before heating to the desired temperature. The reaction was carried out at 1100 °C for 60 min using a mixture of $CH_4$ and $N_2$ gas (1:4 vol, 75 sccm). A significant number of individual CNTs can also be synthesized when the Ar flow rate is reduced to 15 sccm. According to the author's hypothesized growth mechanism for Y-junction CNTs via MW CVD, many free radical species are liberated following methane decomposition and present in the form of clusters due to MW polarization. These active carbon clusters may act as nuclei for the formation of CNTs. With low gas flow rate, individual CNTs with few defects can be generated. With an increase in the gas flow rate the number of defects, including heptagons increases. As a result, these active sites would absorb superfluous carbon atoms, forming new branches. The curved graphene sheets and junctions of branched CNTs reveal a notable concentration of lattice edges or plane defects, as indicated by the Y-junction CNTs' $I_D/I_G$ value of 0.92.

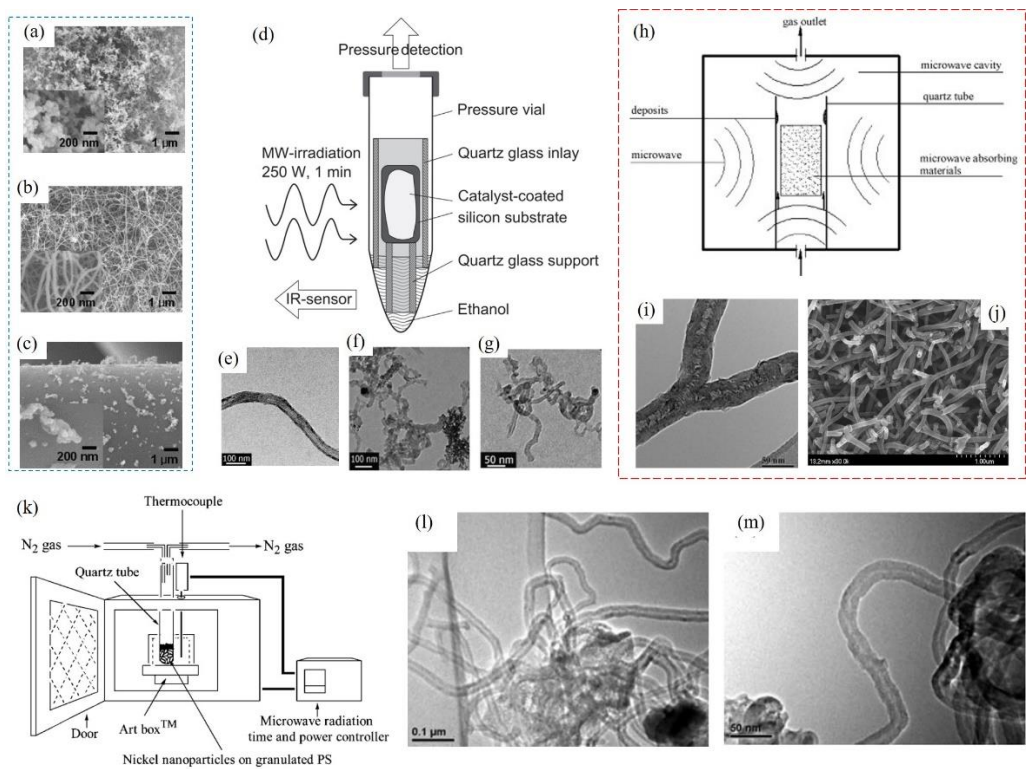

Figure 5: (a-c): CNT synthesis using Fe catalyst and ACF. Post MW irradiation changes on the surface of ACF specimen. Irradiation times, (a) 30 sec; (b) 90 sec; (c) ACF specimen without Fe catalyst impregnation was irradiated for 90 sec.[183] (d-g): MWCNTs and CNFs synthesis on different substrate. (d) The substrate is mounted in a sealed pressure vial on a quartz support over a reservoir of liquid ethanol, and irradiation ensues; (e) TEM image of CNFs synthesized on substrates coated with $Fe_2O_3$; (f) TEM image of CNF grown on substrates coated with $CoC_2O_4$; (g) TEM image of CNF grown on substrates coated with $Ni(NO_3)_2$.[188] (h-j): Y-junction CNTs. (h) Experimental set up; (i) TEM image of a single Y-junction CNT with bamboo-like structure; (j) FESEM image of Y-junction CNTs.[187] (k-m): CNTs from polystyrene particles and Ni NPs. (k) Customized domestic MW oven and an Art Box to synthesize CNTs; (l & m) TEM images of the products calcined at 700 °C and 900 °C respectively.[94]

Mendez et al. prepared CNTs by vaporizing pulverized graphite or graphite blends with sucrose or boric acid under vacuum.[189] Nanotubes were generated in 60 min using a domestic MW-800 W oven. The inclusion of calcined sucrose reduces the number and size of nanotubes. In the presence of boric acid, samples contain a high concentration of nanotubes and bamboo-like structures. Additionally, aligned multi-layer CNTs were produced (50-400 nm in sizes). Vivas-Castro et al. produced MWCNTs by heating the mixture of graphite and iron acetate in a quartz ampoule with 1000 W irradiation from a household MW oven.[190] The powder mixture absorbs MW, causing iron acetate to pyrolyze, and its decomposition yields metallic Fe NPs that act as catalysts in the formation of nanotubes and other carbon nanostructures. The study compared direct irradiation of samples in vacuum-sealed quartz ampoules against attenuated irradiation achieved by partially immersing the ampoules in water. By irradiating quartz ampoules that are partially immersed in water, the material temperature rises very slowly and it takes longer for reactions to occur during synthesis. With a 1000 W power, the material temperature reaches 1000 °C in 30 min. After 30 min of irradiation, CNT grows disordered on graphite particles' surfaces. CNTs are more prevalent in samples after 60 min of exposure. Curled and disordered MWNTs were the most common structure type in the 90 min exposure sample. After 120 min of exposure, major portion of the initial graphite powder has reacted with Fe catalytic particles, resulting in MWNT. Direct irradiation causes the temperature of the combined particles to rise dramatically in a matter of seconds. After 10 min of exposure to MWs, the material forms 7-10 µm-thick blocks of aligned MWNTs. For the formation of MWNTs, 7 min of irradiation is adequate, and the powders react almost completely. A TEM analysis of these samples revealed several Fe nanoparticle-filled MWNTs as well as MWNT tips coated with Fe. Ampoules immersed in water are dynamically exposed to MWs, with temperature gradients varying frequently due to their rocking motion and materials produced in this scenario are curled MWNTs after prolonged exposure.

Takagi et al. proposed a Ni nanoparticle approach to synthese CNTs by heating a mixture of commercially available pulverized polystyrene (PS) and Ni NPs at 800 ºC for 10 min in a nitrogen atmosphere.[94] An Art Box, which has a SiC coating inside, is a small furnace inside this MW oven that is heated by MW radiation. Granulated polystyrene and Ni NPs (avg. size of 100 nm) were consecutively added to a quartz test tube. The Art Box was placed in a household oven and irradiated with MWs until the temperature in the box reached the appropriate level (600, 700, 800, or 900 ºC). Subsequently, the test tube was placed in the Art Box. PS was calcined for a set time (5, 10, 15, or 20 min) under $N_2$ gas (Figure 5k). It was discovered that the yield of CNTs swiftly increases when the Ni wt. ratio is 0.20 and the catalytic ability of Ni NPs became stagnant above 0.20 of the Ni wt. ratio. These CNTs appeared in different ⌀s (25-100 nm), and their maximum length was 2 µm. The amount of CNTs increased exponentially when the mixture's total amount was increased twice and five times while maintaining a constant nickel wt. ratio of 0.30 (a Ni ratio of 30 wt%). This was explained by better atomic carbon absorption into the Ni NPs in the test tube's narrower open space. Three other precursors were also examined. Sugar did not yield CNTs. PVA produced extremely few CNTs, but polyethylene produced a reasonable amount of CNTs. Therefore, precursors containing oxygen atoms or hydroxyl groups were unsuitable for CNT synthesis. Precursors with a higher weight content of carbon are preferred.[94]

The MW arcing method of nanocarbon synthesis offers various advantages, including the need for a typical household MW oven, the use of common organic compounds, energy efficiency, and effectiveness in synthesis, but it also has certain downsides, particularly in terms of safety and end product recovery. Cu, Mo, and Fe wires, and steel fiber were used as arcing agents to start the pyrolysis of ferrocene ($C_{10}H_{10}Fe$), triggering CNT development in a one-step MW arcing and heating process.[191] In this strategy, reaction triggers included commercially available Cu wire (⌀ of 0.15 mm), Mo wire (⌀ of 0.57 mm), Fe wire (⌀ of 0.20 mm), and steel fiber (SF, ⌀ of 0.20 mm). Four quartz crucibles were filled with comparable ferrocene (80-100 mg each). Then, 2-3 pieces of Cu, Mo, Fe, and SF wires measuring 2-3 cm each were vertically placed into the ferrocene powder (Figure 6I). Subsequently, specimens containing quartz crucibles were heated in a MW oven (2.45 GHz, 800 W) for 20-40 sec in air. Upon MW irradiation, the metal tips amplified the electric field, resulting in strong arcing and a rapid temperature increase of more than 1100 °C.[192, 193] Ferrocene decomposed at high temperatures into atomic Fe and cyclopentadienyl ($C_5H_5$). The atomic Fe's condensation and agglomeration enabled the formation of Fe NPs, which served as catalysts to commence the pyrolysis of cyclopentadienyl, resulting in the synthesis of black CNT powder. The orange-yellow ferrocene powder, irrespective of the metal used as a MW trigger, turned fluffy black after 20-40 sec of MW irradiation. The residuary metal triggers in the crucible were easily extracted from the CNT powder. The product was treated with acid to obtain high-purity CNTs (Figure 6II).

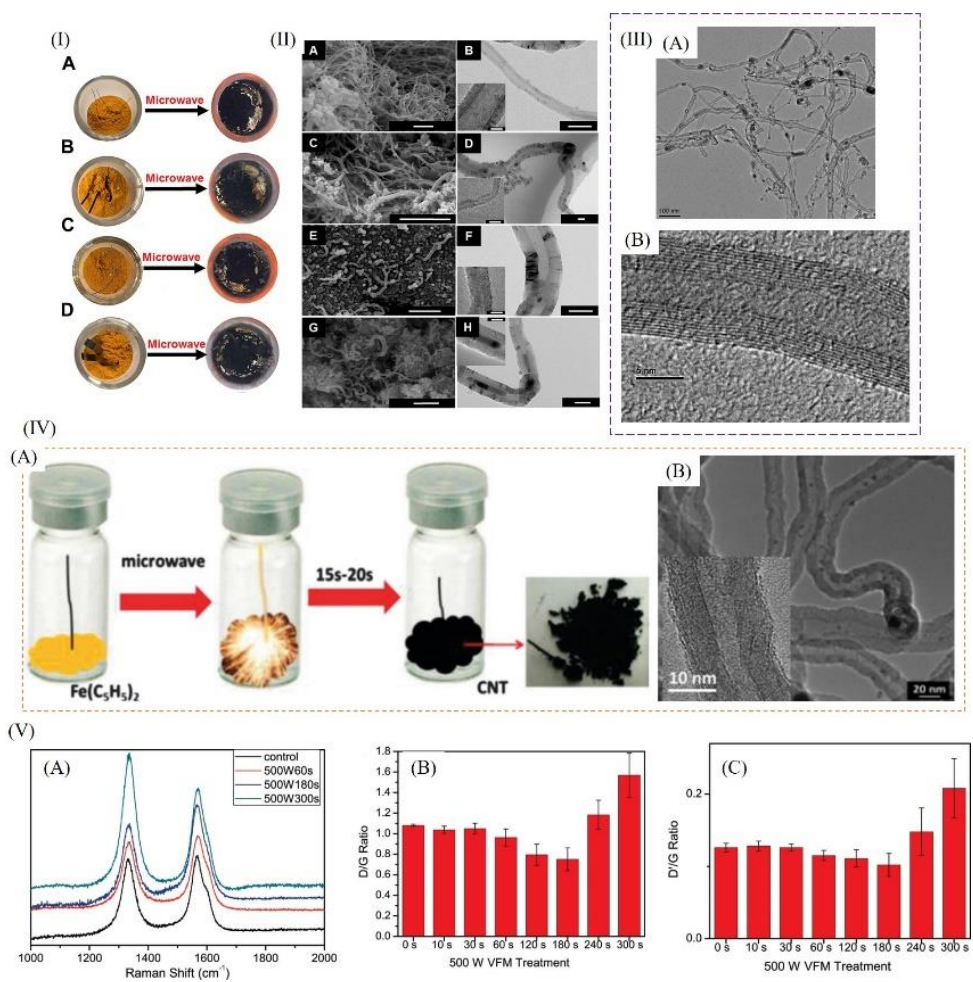

Figure 6: (I-II): Pyrolysis of ferrocene using metal wires to obtain CNTs. (I) Schematics of the CNTs growth directly from ferrocene triggered by metals (A Cu-wire, B Mo-wire, C Fe-wire, D steel wire)

via MW heating; (II) SEM, TEM, & HRTEM images of CNTs prepared by microwaving of ferrocene and metals: (A & B) Cu-wire; (C & D) Mo-wire; (E & F) Fe-wire; (G & H) steel fiber, inset: HRTEM images. (Scale bars: 1 μm for SEM; 50 nm for TEM; 20 nm for HRTEM).[191] (IIIA-B): Ferrocene to CNT. TEM images of MWCNT obtained by MW heating of the graphite, C-fiber, ferrocene mix for 5 sec (1800 W power).[193] (IVA): CNT synthesis using single C-fiber arcing; (IVB) TEM image of the multi-wall shape of CNTs with ⌀ of ~ 20 nm.[192] (V) Defects healing in VACNT. V(A) A domain of the Raman spectra of the pristine VACNT sample and the VACNT samples processed with VFM (500 W) at different durations; V(B-C) $I_D/I_G$ (B) and $I_D/I_G$ (C) of the reference (control) VACNT sample and the VACNT samples processed by VFM (500 W) at various duration.[194]

A one-step approach to fabricating CNTs using a single carbon fiber (C-fiber) to initiate ferrocene pyrolysis in the presence of a MW field has been reported.[192] A single C-fiber released a significant amount of heat, and the solid residue it produced formed an active center for the formation of MWCNTs. This led to a rapid reaction rate, which was comparable to the creation of radical polymer chains. Other than ferrocene and a single fiber, this approach did not require any chemical reagents or preparation steps. As shown in Figure 6(IV)A, a single C-fiber was inserted in a vial containing ferrocene and irradiated with MW. Arcing occurred at the C-fiber tips, resulting in temperatures above 1000 °C, which were sufficient to breakdown the ferrocene and carbonize the cyclopentadienyl. The solid product then absorbed MW energy, heating the remaining ferrocene. As a result, in a matter of sec, the temperature of the mixed powders rose dramatically, resulting in a quick reaction and the generation of a black CNT powder. Rapid synthesis of large quantities of CNTs is possible by substituting commercial CNTs or other MW receptors for the single fiber. When the same amount of CNTs was homogeneously mixed with ferrocene, however, no reaction occurred. C-fiber powder was also employed as a receptor. In this case, the CNT yield was extremely low after 20 sec of irradiation. Typically, a 3-4 cm length of C-fiber is sufficient to sustain the pyrolysis of 80-100 mg ferrocene. TEM scans revealed the multi-wall shape of CNTs (~ 20 nm in ⌀) (Figure 6(IV)B).

In another case of MW arcing and heating, employing C-fibers to commence quick arcing (within 2-3 sec) and using precursor materials: graphite and ferrocene under ambient conditions resulted in rapid CNT growth (in 5 sec).[193] Graphite was used as a support material for ferrocene due to its high MW radiation absorbency (Table 2). The coarse surface of the C-fiber enhances the electric field, resulting in arcing along the fiber's tip and surface. Such arcing generates a confined plasma environment with a temperature of approximately 1000 °C.[76, 195] This arcing results in the fast spreading of absorbed MW energy through graphite and ferrocene. In this investigation, a commercial MW oven (1800 W, 2.45 GHz) was used to heat the quartz vial containing the precursors for 5 sec at 20%, 40%, 60%, 80%, and 100% of the 1800 W operating power. On avg. the conversion efficiency from ferrocene to CNT with 100% MW power is ~82 wt%. When the graphite quantity in the mixture is large, the yield declines while the carbon conversion efficiency increases. CNTs' capillary action causes catalyst particles to be enclosed inside the tubes at various positions throughout their length (Figure 6IIIA-B). Despite being an efficient MW absorber, graphite was shown to be incapable of producing CNTs within 5 sec in the absence of C-fiber. Graphite was not able to generate sufficient arcing inside the vial within 5 sec, and no CNTs were seen. This means that no reaction occurs within this time frame without the use of C-fiber. When the C-fiber was added, at all MW powers and graphite/ferrocene ratios (and with the same amount of C-fiber), intense arcing in < 5 sec was observed from the specimens.

Carbon nanostructures, such as CNTs, NPs, and graphitic flakes, are produced from medium- and high-ash Indian coals via MW-assisted pyrolysis with Fe as a susceptor/catalyst.[196] Fe proved to be a suitable susceptor for creating various nanostructures in coal char. CNTs and carbon NPs produced have avg. ⌀s ranging from 20 to 120 nm and 130 to 190 nm, respectively. Increased Fe mass yielded large diameter CNTs and NPs, but increasing the amount of coal in the mixture resulted in lower CNT and nanoparticle ⌀s (20-30 nm). The significant ash concentration of the coal matrix contributed to the low yield and quality of carbon nanostructures. When compared to coal chars from earlier studies, MW pyrolysis produces coal chars with a high *SSA* (130 m$^2$/g). The coal tars from both coals had high heating values (HHVs) of 35-37 MJ/kg. The most common noncondensable gases were $H_2$ (> 40 vol%), $CH_4$ (20-25 vol%), and CO (17-18 vol%). Evidently, unlike traditional techniques: electrical arc discharge, furnace treatment, and thermal plasma jet, MW pyrolysis produces carbon nanostructures in less time and with equal or higher yields. Pyrolysis tests were carried out in a multimode MW oven.

### 2.2.2 CNT purification and modification

CNTs must be of high quality and well-aligned in order to be used in microelectronic industries.[197, 198] Transition metal particles (e.g., Ni, Co, Fe) are known to act as catalysts in the vapor-grown CNT synthesis process, which uses hydrocarbons (e.g., $CH_4$, $C_6H_6$) as gas sources. Metal catalysts are typically required to stimulate CNT growth. CNTs may have limited utility in various applications because they contain a small proportion of metal catalysts and have defects along the walls of the graphene tubes. Defects in the MWCNTs would degrade their electrical and structural qualities. Clearly, it is essential to remove metal catalysts, and elimination of defects is the key to harness the salient features of individual CNT. Traditionally, reducing CNT defects requires a high-temperature thermal annealing (>1900 °C for MWNTs).[199, 200] Given the excellent MW absorption and heat generation capabilities of CNT,[201] there is increased interest in using MWs to cure defects in CNTs.

An effective approach to purify MWCNT prepared using electron cyclotron resonance chemical vapor deposition (ECRCVD) was proposed.[202] To dissolve metal catalysts, this technique employs a two-step acidic treatment in the MW digesting system. $HNO_3$ or HCl can swiftly absorb MW heat and energy while completely dissolving metals in CNTs without resulting in damage. The two-step MW-assisted and acid-treated methods for dissolving metal in MWCNTs take less than an hour. Following purification, the amount of residual catalyst metals in samples is determined by thermogravimetric analysis (TGA), which reveals a high yield of MWCANT with roughly 5% metal content. In this technique, a raw MWCNT sample was transferred to a 100 mL Pyrex digestion tube. The first digestion stage was performed at 210 °C for 20 min using a 1:1 combination of 5 M $HNO_3$ and 5 M HCl. The second digestion process lasted for 30 min at 210 °C (at 100 W MW power). The suspension from the second digestion was filtered over a 0.1 mm PTFE (poly(tetrafluoroethylene)) membrane in DI water. A thin, black layer of MWCNTs was produced after rinsing with alcohol and drying the sample. $HNO_3$ is widely known for its ability to dissolve metal particles, whereas HCl is effective in dissolving metal oxides. Further, $HNO_3$ is a strong oxidant that can eliminate amorphous carbon. In the MW system, however, inorganic acids: $HNO_3$ and HCl quickly absorbed MW energy without agitation, and metals dissolved quickly without damaging MWCNTs.

Harutyunyan et al. designed a solid-state MW irradiation approach in which SWCNTs were irradiated for 20 min (150 W in Ar flow) and then refluxed in 4 M HCl for 6 hours, yielding a residual catalyst level of < 0.2 wt%.[203] Vazquez et al. outlined a new approach based on the treatment of raw HIPCO (high-pressure CO conversion) nanotubes in a MW oven (80 W, 5 sec) under air and without solvent, resulting in a significant reduction in iron concentration in the soot. Following washing with strong HCl, an iron concentration of 16 wt% was observed, and the content decreased to 9 wt% with further treatment.[204] Particles from many metals are good MW absorbers. Therefore, they can be easily heated using MWs. MWs loosen or decompose the carbon passivation layer covering the metal catalyst, causing the metal to oxidize. At the same time, amorphous carbon byproducts are transformed into $CO_2$. As a result, the metal impurities can be quickly eliminated using a moderate acid, eliminating the need for protracted refluxing or ultrasonication. In this MW-assisted approach, the generation of high temperatures can be achieved locally, inducing no major damage to the CNT's general structure.

A MW annealing process was developed to minimize the defect density of vertically aligned CNTs (VACNTs).[194] Thermogravimetric and Raman tests revealed a significant decrease in defects for CNTs annealed in MWs for 3 min. In the tensile test, fibers spun from annealed CNTs showed an increase in tensile strength (~0.8 GPa) and modulus (~90 GPa) by ~35% and 65%, respectively, compared to pure CNTs. Additionally, $\sigma$ improved by ~20% (~80000 S/m). The employed variable-frequency MW (VFM) chamber (6.425 GHz frequency, 1.150 GHz bandwidth, and 0.1 sec sweep time) was a self-configured; VACNT sample containing glass chamber was Ar-filled. The VACNT samples emitted an intense light, heated up quickly, and outgassed when exposed to MW radiation. In this investigation, a 500 W MW output was sufficient to heat the VACNT samples above 400 °C in a few seconds. Figure 6V(A) compares the Raman spectra of pristine VACNT samples and those treated with VFM. Additionally, the ratio of the D′ band integrated intensity ($I_{D'}$) at about 1610/cm to $I_G$ was examined.[205] Figure 6V(B-C) shows the respective $I_D/I_G$ and $I_{D'}/I_G$. It is clear that $I_D/I_G$ and $I_{D'}/I_G$ fall with an extended duration of MW irradiation, showing a structural order refinement in the MW-annealed CNTs within < 180 seconds. The authors hypothesised that decreased $I_D/I_G$ and $I_{D'}/I_G$ resulted from the defective sites on the synthesized CNT repairing and forming graphitic structures during MW annealing. Given that extremely effective Ar protection is difficult to obtain in the VFM chamber, $I_D/I_G$ and $I_{D'}/I_G$ increment with a prolonged MW irradiation time is most likely attributed to CNT degeneration in MW or to CNT oxidation.[206]

Under MW irradiation, Pd-coated SWNTs, DWNTs, and MWNTs were catalytically unzipped in the presence of an oxygen-containing liquid media (water), yielding the few-layer graphene (FLG) sheets.[207] In this unzipping process, the Pd particles function as scissors, cutting the nanotube lengthwise. By significantly reducing the associated energy barrier, the Pd nanocatalysts and oxygen near vacancies facilitated the unzipping process to produce graphene from nanotubes. A quartz vessel containing the sample was irradiated with MWs (300 W, at 200 °C for 60-120 min) from a multi-mode commercial system. An ionic liquid, 1-ethyl-3-methyl imidazolium tetrafluoroborate (EMIM $BF_4$) was used to split MWCNT or SWCNTs to produce graphene nano-ribbons using MW irradiation.[208] MWCNT/EMIM $BF_4$ mixture was sonicated and subjected to MW radiation for 240 sec at 700 W. The authors describe the unzipping mechanism as akin to CNT fluorination. Semi-ionic bonds are formed between $BF_4^-$ radical fluorine ions and defect sites on MWCNT surfaces. Once the C-F bond formed, the unzipping reaction continues with electrophilic fluorination along the strain caused by the C-C bond angle. Aside from the fluorine moiety of the ionic liquid, EMIM BF4's imidazolium cation is acidic owing to the comparably high acidity of the hydrogen in the imidazolium nucleus. Imidiazolium cations are deprotonated under MW irradiation, resulting in carbine ligands that can help expedite the unzipping process.

### 2.2.3 CNT hybrids

Nanocarbons in diverse morphologies, when combined with various appropriate constituents, find utility in a wide range of technological applications.[209-211] It is crucial to develop effective methods for preparing materials with exceptional performance for a specific application.

Doping is a typical technique for modifying the electronic characteristics of semiconductor materials. For example, when CNTs are doped with N or B atoms, they become *n*-type or *p*-type.[212] As a result, combining doped CNT with graphene can produce beneficial results. Sridhar et al. proposed a fast MW approach for synthesizing N-doped CNTs anchored on graphene substrates (G-Fe@NCNT) using azobis(cyclohexanecarbonitrile) (ACN) as the N-doped-CNT precursor and iron as the catalyst. (Figure 7a-e).[213] Mesoporous hierarchical nanostructures were created by vertically anchoring μm-long, N-doped CNTs to graphene. XPS analysis showed that nitrogen moieties exist in the form of pyridinic and graphitic nitrogen. As an anode material in LIBs, the materials had a high capacity of 1342 mA h/g even after prolonged cycling, indicating that the 3D network can withstand significant volume changes that occur during the lithiation/delithiation process.

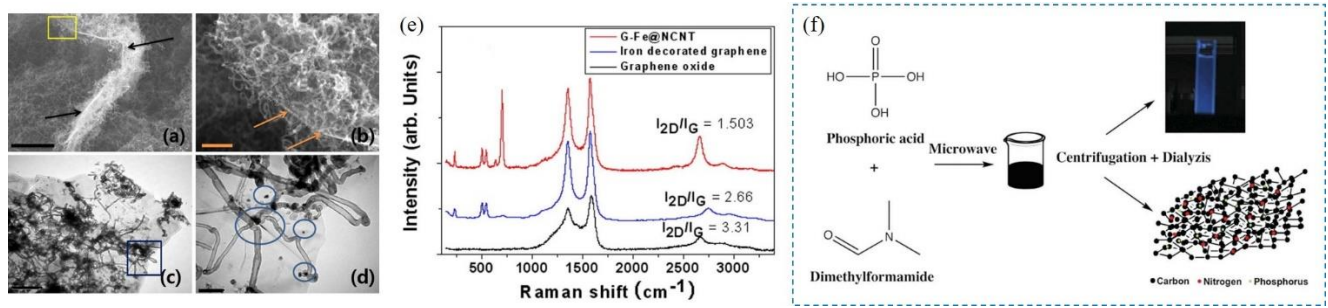

Figure 7: (a-e): G-Fe@NCNT. SEM (a & b) and TEM (c & d) images of N-doped CNT vertically docked on graphene substrate. Scale bars: 5 μm, 2 μm, 1 μm, and 200 nm in (a-d) respectively; (e) Raman spectra of different structures.[213] (f) Scheme 1. Formation of N-P doubly doped nanocarbon & photoluminescent carbon.[214]

As shown in Figure 7(a & b), dense CNT forests (several μm- several tens of μm in length) were seen vertically adhered to a porous graphene substrate. The Raman spectra of GO, Fe-decorated graphene, and G-Fe@NCNT are shown in Figure 7e. The systematic decrease in the ratio of intensities of 2D band to G band ($I_{2D}/I_G$ ratio) from 3.31 in GO to 1.503 in G-Fe@NCNT suggests that the defects on graphene caused by oxidation have been repaired due to Fe NPs decoration and subsequent CNT growth. In this study, to prepare G-Fe@NCNT hybrids, Fe-decorated graphene was mixed with ACN and microwaved at 700 W for 2 minutes.

In addition to $RuO_2$, another metal oxide, NiO is a potential material for charge storage (pseudo-capacitors) due to its high theoretical capacitance (~2500 F/g), low cost, high redox reversibility, and ease of preparation.[215] On the negative side, NiO exhibits weak electrical and ionic conductivity, larger volume changes during charge/discharge cycling, and dismal rate performance in energy storage devices. To overcome these challenges, NiO electrodes are assembled with CNT due to nanocarbon's superior chemical stability, favorable $\sigma$, and high *SSA*. Biochar was employed as a substrate to develop and synthesize biochar-CNT/NiO, which will serve as the anode in LIBs.[216] Ni acts as a catalyst for the synthesis of CNTs using MW-assisted CVD method. After 100 cycles, biochar-CNT-NiO demonstrated capacity of 674.6 mAh/g, better than biochar-NiO without CNTs (368.4 mAh/g) and biochar-CNT-Ni (140.9 mAh/g). Also, the biochar-CNT-NiO composite has a better discharge rate performance. The inclusion of CNTs in biochar-CNT-NiO reduced the volume change of NiO between cycles and accelerated electron transfer, resulting in better performance. In other report,[215] CNTs were decorated with NiO NPs via MW-assisted solution-based technique to obtain NiO/CNT nanocomposite (*Figure 8*a), which displayed a $C_{sp}$ of 258 F/g (at 1 A/g) as an electrode. After 2500 charge and discharge cycles, the composite material retained 86% of its initial capacitance.

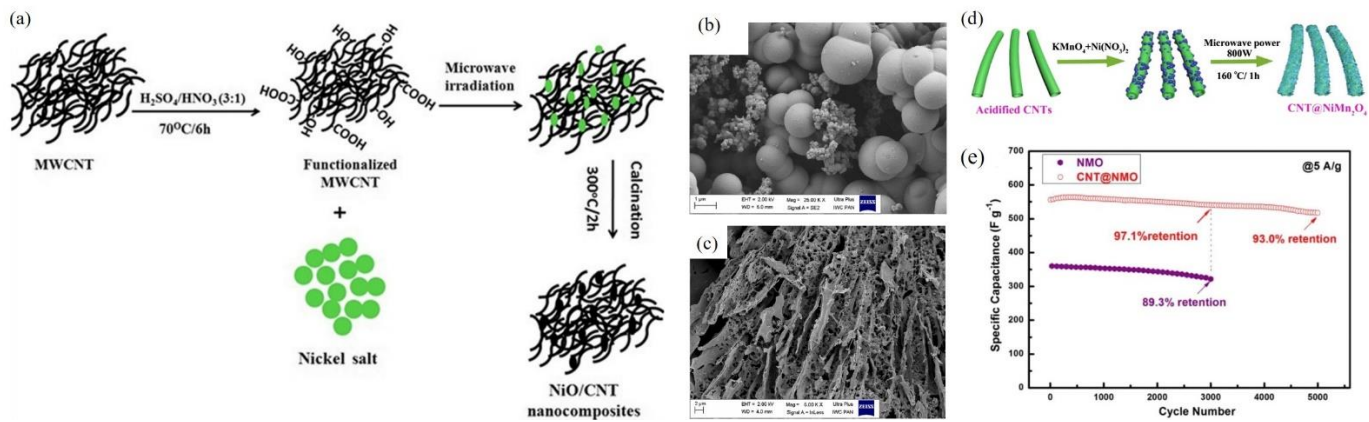

Figure 8: (a) Preparation of NiO/CNT nanocomposites.[215] (b-c): Carbon nanosphere. SEM images of the materials acquired under various experimental conditions. (b) 1 MPa pressure, 1000 W reactor power; 15 min reaction time; (c) 4 MPa pressure, 1000 W reactor power, 10 min reaction time.[217] (d-e): CNT@NiMn$_2$O$_4$. (d) Schematic image of the formation of CNT@NiMn$_2$O$_4$ nanocomposite; (e) Cycling stability of a CNT@NiMn$_2$O$_4$ nanocomposite and NiMn$_2$O$_4$ at 5 A/g.[218]

$MnO_2$ is a non-toxic, low-cost pseudocapacitive material with theoretical $C_{sp}$ of ~1370 F/g.[219, 220] Even so, it has low $\sigma$ ($10^{-5}$ to $10^{-6}$ S/cm), low ion mobility, and low structural stability.[220, 221] By combining it with materials made of carbon, the electrochemical performance can be improved. Amorphous $MnO_2$ was composited with CNTs in 60 sec by the solid-state MW method.[222] Varied ratios of CNT (10/20/30 mg) and $Mn(NO_3)_2 \cdot 4H_2O$ (50 mg) were evenly mixed, ground, placed in a corundum crucible, and irradiated with different MW powers (800, 600, and 400 W) and times (90, 60, and 30 sec). When used as supercapacitor electrodes, the interactions between $MnO_2$ and CNT increased the $C_{sp}$ to 1250 F/g, and after 7000 cycles, the cycling stability remained at 80% of the initial value.

Another transition-metal oxide material, $RuO_2$[223] based nanocomposite, $RuO_2$ NPs (1-2 nm in size)/MWCNTs has been prepared using $RuCl_3$ solution as precursor and $NH_3 \cdot H_2O$ as precipitator via a one-step MW-assisted method (domestic oven, ~700 W, 30 sec).[224] During the synthesis process, the chlorine ions might be readily eliminated during the synthesis step because $NH_4Cl$ decomposes at high temperatures.[225] The remaining $NH_3 \cdot H_2O$ was likewise easily removed during MW treatment. Further, the intermediate product $Ru(OH)_3$ may be immediately converted into $RuO_2$ NPs without additional annealing. It was discovered that $RuO_2$ NPs adhere to the sidewalls of MWCNTs. Some NPs appeared to fill the internal holes of MWCNTs. The Ru/O ratio is estimated to be close to 0.5.

A MW-assisted HT technique is used to create a core-shell nanocomposite of $CNT@NiMn_2O_4$ (Figure 8d).[218] An asymmetric supercapacitor (ASC) with $CNT@NiMn_2O_4$ as +ve electrode and activated carbon as -ve electrode was developed, which had an *energy density* of ~36.5 Wh/kg at a *power density* of ~800 W/kg and was found to be demonstrated good cycling stability with 82.8% retained capacitance after 10000 cycles at 5 A/g. In this study, $KMnO_4$, $Ni(NO_3)_2 \cdot 6H_2O$, $CO(NH_2)_2$, and $NH_4F$ were added to DI water. Acidified CNTs were added to the combined solution. The mixed solution in a 100-mL Teflon-lined container was subjected to MW-assisted HT treatment at 160 °C for 1 h with a power of 800 W. The product was annealed at 450 °C for 2 h in an Ar environment to obtain the final $CNT@NiMn_2O_4$ nanocomposite.

## 2.3 Carbon QDs, doped- QDs, carbon dots, carbon nanospheres, and hybrid materials

Carbon QDs have been synthesized using a variety of techniques, including arc discharge treatment,[226] electrochemical oxidation,[227] laser ablation,[228] nitric acid reflux treatment,[229] HT/solvothermal treatment,[228, 230] and ultrasonic processing.[231] However, most developed techniques have disadvantages, including high starting material costs, complex procedures, difficult synthesis conditions, long processing times, and energy-consuming instruments. Therefore, researchers have used MWs as an energy source to overcome these limitations. A one-pot MW-assisted technique is applied to generate metal-free N and P doubly doped, electrocatalytically active functionalized nanocarbon (FNC) and photoluminescent (PL) carbon nanodots (PCNDs).[214] Surface morphological and spectral studies show that P and N were incorporated into oxygen-rich PCNDs and FNC. FNC characterization studies reveal the presence of edge plane-like sites/defects, as well as excellent electrocatalytic activity. Further, FNC exhibits excellent characteristics as a metal-free oxygen reduction catalyst that is impervious to the effects of methanol crossover in alkaline media. PCNDs (5-10 nm), which fluoresce blue when exposed to UV light, were subsequently tested in bioimaging applications. PCND and FNC were synthesized via a one-pot, one-step MW synthesis (Scheme 1, Figure 7f) and then separated using a simple centrifugation process.

HT carbonization is one of the most attractive methods for producing functional carbonaceous materials from biomass. The reaction typically involves HT heating of biomass at mild temperatures (~200 ºC) under autogenous pressure in a closed container. As a result, a series of functional carbonaceous compounds have been synthesized from biomass using the HT carbonization technique, with potential applications in energy storage,[232, 233] hydrogen storage,[234] water purification,[235] and catalysis.[236] One of the main constituents of lignocellulosic biomass is cellulose, while the main structural component and byproduct of acid digestion of biomass is glucose; thus, both are often used as model systems in HT carbonization research. However, when MWs are introduced as energy sources and carbon-containing elements are used as MW absorbents, MWs accelerate the reaction and nanocarbons are produced very quickly and efficiently. Despite the fact HT carbonization of glucose or cellulose often yields well-distributed, μm-sized carbon spheres,[237, 238] these spheres are not conductive and must be further carbonized by annealing to make them conductive.

To make fluorescent carbon dots, candle soot, graphite oxide, conductive carbon black, and lamp black were heated in nitric acid under reflux.[239] The carbon dots made from graphite oxide had the highest quantum yield (QY) and narrowest emission of any developed. Carbon dots were also synthesized using MW-assisted processes. Compared to conventional heating under reflux, MW-assisted heating under reflux (120 °C, 8 h, 900 W) and a MW-assisted HT approach (15-60 min, 800 W) both reduced reaction times. Compared to carbon dots prepared by traditional reflux heating, those prepared using MW-assisted methods showed improved absorption, better QY, and longer fluorescence lifetime.

Using MW-assisted heating, glucose solutions were subjected to HT carbonization (HTC) at 210 °C for 15 minutes, producing uniformly sized spherical structures.[240] The avg. ⌀ varied based on the glucose solution's concentration, ranging from 201 nm to 650 nm. The carbon content of the dried microspheres was < 50% by atomic ratio, but it increased to >90% after a second MW irradiation (1 min) in a solvent-free environment. The N doping of the carbon particles was accomplished by adding urea to the glucose solution. The structures had 35-45% C and 5-10% N in atomic ratio. During the second MW irradiation of the dried particles, the C content increased to 60-80%, whereas the N content maintained 5-10% and the oxygen concentration declined to 0-3%. MW irradiation of polyethylene glycol at temperatures ranging from 160 to 220 °C for 40 min without a catalyst resulted in chains of graphitic carbon particles.[241] Chains were composed of individual particles (340-620 nm in size), which grew with the synthesis temperature; the $D/G$ ratio is reported to be 0.91, indicating a combination of amorphous and graphitic material. The chains were discovered to be an intermediary product that produces MWCNTs when heated further under HT conditions.

It has been shown that incorporating a minute quantity of graphene oxide to glucose (e.g., 1:800 wt. ratio) can considerably modify the morphology of the MW-assisted HTC product, yielding in more conductive carbon materials and a higher degree of carbonization.[242] HTC treatment produces distributed carbon platelets (tens of nm-thick) at low graphene oxide mass loading levels, and free-standing carbon monoliths at high mass loading levels. Comparative investigations with other carbon materials (graphite, CNTs, carbon black, and rGO) show that only graphene oxide has a major impact on HTC conversion, most likely due to its good water processability, amphiphilicity, and 2D structure, which may aid in the template of the initially carbonized materials.

As previously stated, one method for producing carbon spheres is to decompose organic compounds by heating solutions of high carbon content materials in steel or Teflon-lined autoclaves. This process can be used with a number of carbon sources, including sugars[243] and polymers.[244] Such procedures can use both conventional and MW-assisted heating. However, these techniques require well-calibrated synthesis parameters such as temperature, pressure, and time; thus, such parameters must be determined in order to produce the desired product. Using a MW-assisted solvothermal reactor, carbon spheres made from resorcinol and formaldehyde are produced and characterized.[217] A partial transition from spherical to polyhedral shape is observed with increasing reaction pressure; this is linked to a more ordered graphitic structure. If the reaction pressure was higher than 3 MPa, no carbon spheres were formed. When the reaction power was 2000 W and the reaction time was 10 min, the most effective $CO_2$ carbon adsorbents were generated. The material's graphitic structure content increased as reaction pressure increased, and the spherical shape collapsed (Figure *8*b-c).

N-doped carbon quantum dots (N-CQDs, size range: 4.44-13.34 nm; avg. ⌀ of 7.89 nm, Figure 9c) were prepared in a single step using an ammonia solution of xylan as the precursor at 200 °C for 10 min (commercial MW system, 200 W, Figure 9a).[245] The N-CQDs exhibited excellent crystal quality, favorable photoluminescence properties, and strong resistance to salts and metal ions. Furthermore, a sensing platform from the as-prepared N-CQDs demonstrated selectivity and sensitivity characteristics to tetracycline antibiotics due to an inner-filter effect, displaying a detection limit of 6.49 nM in a linear range of 0.05-20 µM under optimal conditions.

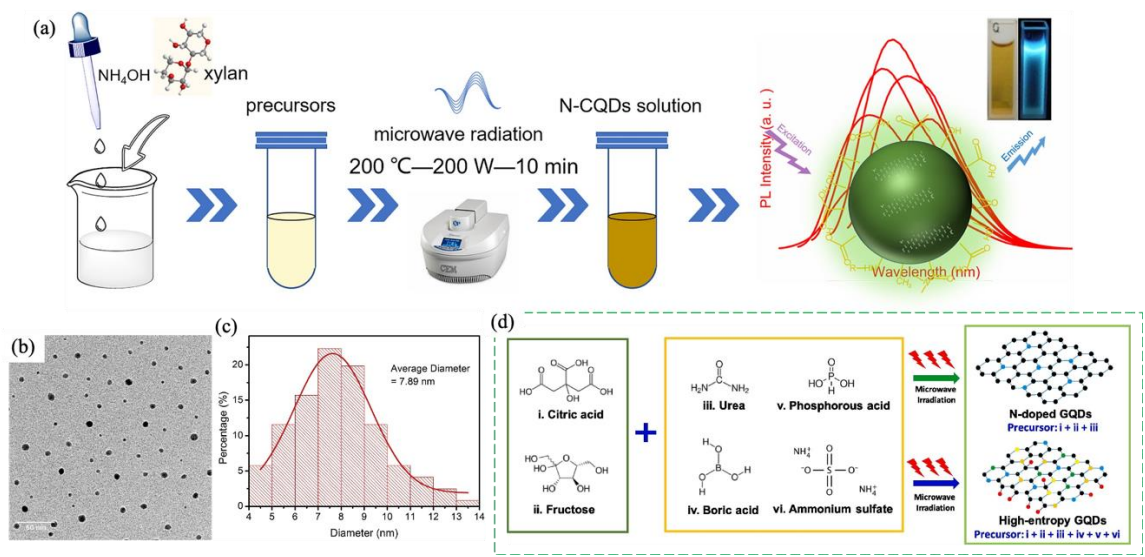

Figure 9: (a) Synthesis of N-CQDs using MW treatment; (b) TEM image of the N-CQDs; (c) N-CQDs size distribution with Gaussian fitting.[245] (d) A diagram exhibiting the material selection and solid-state MW synthesis process for N-doped and HE-GQD samples.[246]

In a similar vein, Liu et al. showed how to create CQDs from glycerol using a MW pyrolysis method in a matter of minutes.[247] Chen et al. prepared heteroatom-doped CQDs from polyols using MW pyrolysis in a one-step synthesis.[248] Xiao et al. and Gong et al. used MW-assisted techniques to obtain chitosan-derived CQDs.[249, 250] In a study by Liu et al., a solid-state MW method was used to synthesize high-entropy graphene QDs (HE-GQDs, 5 nm) and they were tested as metal-free electrocatalysts for the oxygen reduction reaction (ORR) in alkaline electrolytes.[246] The synthesis process combines pyrolysis of two carbon precursors and four dopant precursors with pulsed MW irradiation at 180 °C. The fabricated GQD catalyst consists of 6 elements arranged in graphite-like lattices with a variety of surface functionalities and dopants. This high-entropy GQD catalytic electrode had a large electrochemically active surface area and can withstand ORR tests. It demonstrated an explicit reduction peak at around -0.5 V and an onset potential at around -0.1 V compared to the saturated calomel electrode (SCE) in diluted KOH solution. Figure 9d shows the material combinations used to prepare N-doped and HE-GQD samples using a solid-state MW synthesis technique.

In a MW-assisted solid-state synthesis approach, carbon nanospheres were produced by combining carbon microparticles with 10 wt% ZnO microparticles (used as catalyst) and irradiating the mixture with MWs in a vacuum (340 W, 1-3.5 min).[251] The final product was reported to have a minimum level of amorphous carbon. HRTEM investigation revealed that after 3.5 min of interaction, carbon nanospheres with an avg. ⌀ of < 35 nm were formed, with ZnO NPs of 4 nm ⌀ inside. In the trials, 1 g of carbon powder was blended with 10% ZnO. The mixture was homogenized to produce particles with a ⌀ of <50 μm.

A theoretical and practical study was carried out on the generation of carbon nanostructures (multilayer graphene sheets and NPs) using MW plasma at atmospheric pressure.[252] The technique involves introducing $CH_4$ into a MW argon plasma atmosphere, where $CH_4$ molecules decompose and solid carbon is formed. The solid carbon nuclei are steered through the assembly zone by the plasma gas stream, where free-standing graphene sheets are synthesized selectively at 1 kW MW power, with Ar and $CH_4$ flow rates of 600 sccm and 2 sccm, respectively. $I_D/I_G$ and $I_{2D}/I_G$ are 0.62 and 0.8 respectively, indicating that multi-layer graphene was created under the optimized conditions. Temperature gradients, dwelling duration, and the concentration of carbon building units in the plasma reactor's assembly zone are the three essential parameters that decide which form of structure will be mostly produced. For lower temperatures in the assembly zone (7.5 sccm) and larger partial $CH_4$ fluxes, carbon NPs with a size range of 70-800 nm are formed. He et al. have shown that N-doped carbon dots (N-CDs) can be synthesized using acetylacetone and ammonia as starting reactants, where MWs (single-mode MW applicator) helped to provide a high-temperature, high-pressure environment.[253] The reactants were heated to various temperatures, such as 160 °C, 180 °C, and 200 °C, over a period of 5 min and held at these temperatures for 10 to 30 minutes. Organic fluorescent molecules with a high QY of 51.61% can be obtained more easily at low temperatures (160 °C). After dehydration and carbonization, N-CDs with perfect lattice defects may be synthesized at high temperatures (200 °C), and they exhibit the same strong photobleaching capabilities as QDs.

Graphitic carbon nitride (g-$C_3N_4$) has been proven to be an effective catalyst for the photodecomposition of water and is stable.[254, 255] The bulk g-$C_3N_4$ has a band gap of ~2.7 eV, and NSs of g-$C_3N_4$ may have wider band gaps, resulting in blue emission.[256] Therefore, g-$C_3N_4$ in nano size can be used to prepare luminescent composites. CDs@g-$C_3N_4$ composite phosphors have been synthesized in large quantity (20 g) using a two-stage MW-assisted liquid-state heating technique using citric acid and urea as precursors.[257] Carbon dots (CD) were uniformly decorated in situ and adhered to the g-$C_3N_4$ network, inhibiting CD agglomeration and triggering luminescence quenching. In comparison to manually mixed CDs and g-$C_3N_4$, CDs@g-$C_3N_4$ (in-situ deposition) phosphors displayed greater photoluminescence quantum yield (PLQY). The CDs@g-$C_3N_4$ composite loaded with CDs (1 wt%) emitted strong green light with a PLQY of 62% under 410 nm illumination and kept 92% and 80% of its original PL intensity after 1 h of continued UV light (1.6 W/cm$^2$) and heating to 160 °C, respectively. Utilizing CDs@g-$C_3N_4$ as a color conversion layer, a white light-emitting diode delivering cold white light with CIE color coordinates (0.29, 0.33) and a color temperature of 7557 K was devised having a power efficiency of 42 lm/W. While producing CDs@g-$C_3N_4$, MWs were used in two steps (750 W MWs for 3-4 min). In comparison with the PL spectra (with 360 nm excitation) of

CDs@g-C$_3$N$_4$ composites with varied CD loadings demonstrate that for loading amounts of CDs < 1 wt%, there are bands at 390 and 520 nm (Figure *10*a). The composite containing 1 wt% CDs produced the strongest emission. The band at 390 nm vanished as CD loading increased, while the band at 520 nm gradually redshifted and lost intensity.

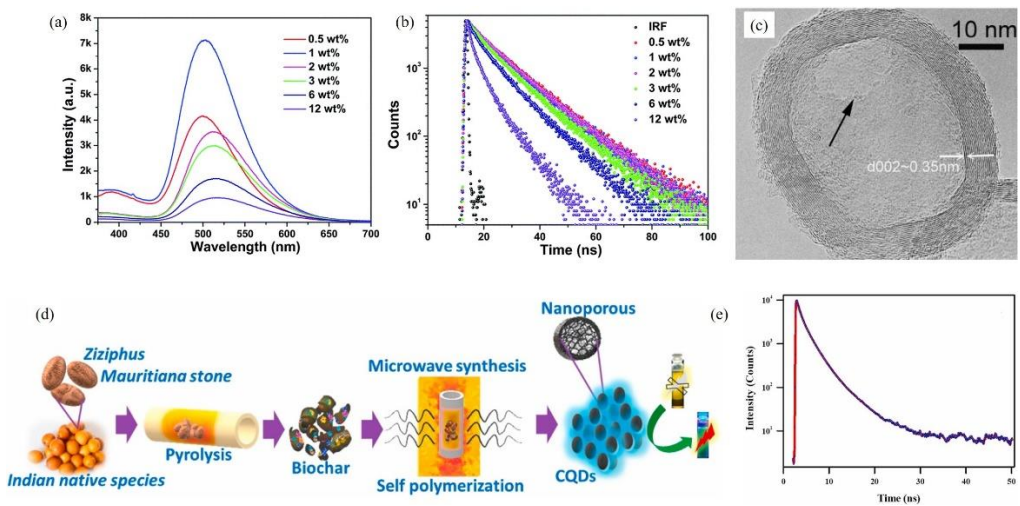

Figure 10: (a) PL spectra of CDs@g-C$_3$N$_4$ composites with various CD loadings (at 360 nm excitation). (b) PL decay curves of g-C$_3$N$_4$ and CDs@g-C$_3$N$_4$ powders with various CD loadings, observed at 510 nm (with 375 nm excitation).[257] (c) HRTEM image of a hollow carbon nanoparticle. Arrow indicates a hole in the graphite shells.[89] (d) ZMS-CQD synthesis and NH$_3$ detection processes; (e) Studies of ZMS-derived CQDs' fluorescence lifetime with time resolution.[258]

Self-assembled multi-functional CQDs were prepared using Ziziphus Mauritiana stone biomass (as a bioresource precursor) via the direct pyrolysis technique using MWs (*Figure 10*d).[258] The produced CQDs (spherical, avg. 2-4 nm in ⌀) emitted strong bluish-green light with stable dispersion and high photostability in the aqueous medium. The emission properties of CQDs were investigated by quenching them with NH$_3$ and other compounds in aqueous conditions, and the generated CQDs shown fluorescence for the selective/sensitive detection of NH$_3$, showing a detection limit of 10 nM. The UV-Vis absorption spectrum at 360 nm verifies the presence of π-π* and n-π*, which correspond to the C=O bond and conjugated C=C band transitions, respectively (Figure 10e).

The aqueous solution of citric acid and urea was irradiated with MWs for different times (power level 700 W for 0, 30, 60, 90, 120, 150, 165, 180, 210, 225, and 300 sec) to produce hydrophilic carbon quantum dots (CQDs).[259] The bandgap of the synthesized carbon QDs and their absorbance and photoluminescence, were shown to rise with an increase in sample exposure times up to 225 sec due to an increase in state density. Nevertheless, increasing the heating times causes PL saturation due to the saturation of the intrinsic density of states and a reduction in the band gap, which is most likely caused by CQD agglomerations. The CQDs synthesized after 165 sec have the most stable excited states but a lower PL than the other CQDs. Using citric acid as a carbon source, sulfur as a donor, and glutathione as a N source, N and S co-doped CQDs (N/S-CQDs) were synthesized employing a MW-heated oil bath.[260] At 350 nm excitation wavelength, the QY reached up to 76.5%. At the same time, N/S-CQDs exhibited good anti-photobleaching characteristics, and the concentration of NaCl had negligible effect on the fluorescence intensity of N/S-CQDs. N/S-CQDs may detect drug methimazole by fluorescence resonance energy transfer (FRET) with gold NPs, creating an elegant fluorescent probe. Luminescent graphene quantum dots (GQDs) with N, F, and S substitutions (avg. size ~ 2 nm) were produced by MW treatment of MWCNTs using an ionic liquid, {1-methyl-1-propylpiperidiniumbis (trifluoro methylsulfonyl)imide}, which works as an effective source of N, S, and F.[261] This method resulted in a high yield of ~85% for GQDs, as well as a 70% PLQY. In this study, CNTs were dispersed in ionic liquid and sonicated before being heated in a MW-1100 W oven for 15 minutes. The doping contents of N, F and S were reported to be 11.6%, 8.2% and 2.3%, respectively.

Hsin et al. reported a MW arc technique to synthesize graphitized core/shell iron/carbon NPs (Fe@CNPs) in toluene solutions containing $Fe(CO)_5$-$C_{60/70}$.[89] The reaction utilizes iron pentacarbonyl $(Fe(CO)_5)$ as a carbon source and Cu wire as an arc generating agent during MW irradiation. In the absence of $C_{60/70}$, the graphene shell structures were not satisfactory. The carbon source is thought to be a blend of $C_{60}$ and $C_{70}$, which can improve the structure of graphene shells encapsulating metal NPs (*Figure 10*c). In addition, pre-synthesized Co NPs were employed as templates to create graphene shells in toluene-$C_{60/70}$ solutions. Hollow carbon NPs could be obtained via acid etching and removing the central core Co NPs. The majority of the Co core NPs were etched away, leaving hollow carbon graphene shells, which have a hole (indicated by the arrow in *Figure 10*c). Guo et al. presented a MW-assisted carbonization technique for precise carbon coating of ZnO nanorods (NRs).[262] In the first stage, the surface of previously generated ZnO nanorods (NRs) was altered with amino groups via contact with (3-aminopropyl) triethoxysilane (APTES). The glucose molecules were then grafted onto NRs by coupling surface amino groups with glucose aldehyde groups. Finally, glucose-grafted ZnO NRs were exposed to MWs (100 °C, 30 min) to stimulate the conversion of glucose into carbon. The thickness of the carbon layer was 2 nm.

## 2.4 Preparation of graphene derivatives and their hybrids and MW triggered reduction techniques in solid-state and liquid-state

The first documented method for producing graphene involved the mechanical exfoliation of highly oriented pyrolytic graphite.[263] In recently years, several new techniques have been designed and implemented to prepare graphene and its derivatives via a top-down approach: exfoliation and reduction from graphite or graphite oxide[264, 265] or via adapting bottom-up processes: graphene sheet preparation via a CVD technique.[266-268,] Increasing the number of graphene layers in a graphene film is one way to make it highly conductive, mechanically strong, and chemically stable.[269] The sheet resistance of multilayer graphene (MLG), $R_{s,MLG}$ is found to be inversely proportional to the number of graphene layers ($n$) i.e. $R_{s,\,MLG} = R_{s,SL}/n$. Where $R_{s,SL}$ is the sheet resistance of a single layer. Since $R_{s,MLG}$ varies in direct proportion to the addition of $R_{s,SL}$,[270] increasing $n$ is not straightforward as each graphene layer may have different carrier concentration and mobility and these properties depend on the surface and interfacial conditions of each grapheme layer.[271, 272] The mobility and carrier inhomogeneities of graphene devices fabricated on single-crystal h-BN substrates utilizing exfoliated and mechanically transferred mono- and bilayer graphene are nearly an order of magnitude better than those of devices on SiO$_2$ substrate,[273] suggesting that the mobility and carrier concentration in layers close to the substrates are different in both cases. CVD on catalytic metals with high carbon solubility, such as Ni, is an easy way to produce MLG. But, direct fabrication of MLG using CVD does not guarantee homogeneity of the number of graphene layers.[274] Furthermore, in many cases, after CVD of graphene films on metal substrates, graphene needs to be extracted and transferred to a suitable substrate. In such situations, transferring atomic-layer thick graphene is a daunting task, limiting CVD graphene's practical applicability.[275]

The reduction of oxidized graphite is considered the most cost-effective approach. Graphite oxide is a crucial precursor for the synthesis of graphene. Because it has a significant number of oxygen-containing functional groups[276] and high solubility in water,[277] it is easy to disperse and exfoliate into graphene oxide flakes. These graphene oxide suspensions offer an easy way to process and deposit graphene materials. There are multiple models describing the molecular structure of graphene oxide.[278-280] The general consensus is that graphene oxide is rich in oxygen-containing groups such as carboxyl, hydroxyl, epoxy, carbonyl, and so on.[280] The probable arrangement of oxygen-containing groups in a graphene oxide sheet is that epoxides and alcohols reside on the surface of the basal plane, whereas carboxylic acids exist on the outer edges of the basal plane of the graphene oxide's graphitic platelets.[281] A study on the atomic and electronic structure of graphene oxide using composition sensitive annular dark field (ADF) imaging of single (1.6 nm-thick) and multilayer sheets and electron energy loss spectroscopy (EELS) reveals that the graphene oxide has an avg. surface roughness of 0.6 nm and the structure is semiamorphous due to distortions from sp$^3$ C−O bonds. These sheets contained around 40% sp$^3$ bonding and a measured *O/C* ratio of 1:5.[282]

The two most widely used techniques for reducing GO are chemical reduction (which uses hazardous and explosive compounds[283, 284]) and thermal reduction.[285, 286] Chemical approaches to producing rGO generally result in limited *SSA*, dismal $\sigma$, and sheet aggregation.[287] The thermal reduction of graphene oxide involves high-temperature processes that allow the elimination of oxygen groups. However, the reduction procedure may introduce defects in the lattice as a result of carbon evolution and the use of high temperatures, resulting in a higher production cost. In addition to the aforementioned conventional techniques of reduction, environmentally friendly non-chemical approaches such as electrochemical reduction[288-290] light,[291, 292] atmospheric-pressure glow discharge (AGD) plasma,[293] and solar irradiation-induced methods[294] have been utilized to reduce graphene oxide.

Reports on MW-assisted bottom-up approaches are scarce. Regardless, MW-assisted treatment is often used to exfoliate and reduce graphitic precursor materials. This includes the preparation of graphene via solid-state reduction of dry graphite oxide, reduction of graphite oxide in liquid state, or heating graphite intercalated compounds (GICs). It is also possible to simultaneously reduce and dope GO and fabricate hybrid materials. In liquid phases (aqueous media, ionic liquids, and organic media) and solid phases (with or without the presence of MW susceptors), MWs simultaneously reduced and exfoliated graphite oxide.[295-300] Due to their cost-effectiveness and short duration, MW-assisted reduction treatments using graphitic materials as internal and external susceptors have been extensively studied, and the next section reviews such studies. Table 3 summarizes the starting materials, MW-assisted thermal reduction types, conditions applied, rGO properties, and demonstrated applications.

PECVD-based approaches for carbon nanosheets (CNS) synthesis may yield vertically oriented graphene nanosheets (GNS, also called as graphene nanoflakes or carbon nanowalls) with a few atomic layers in thickness that stand perpendicular to the substrates. Graphene and CNSs have similar structures and properties, making them suitable for various applications such as emitters in field emission devices,[301, 302] LIBs,[303] supercapacitors,[304, 305] biosensors, photocatalysts, and catalytic oxygen reduction reaction[306], electrodes in sensors,[307, 308] etc.

*Table 3:* An overview of the experimental techniques, conditions, and properties of rGO derived from MW-assisted thermal reduction. List of abbreviations. $\sigma$ –electrical conductivity; $R_s$ –sheet resistance; *GNS* –graphene nanosheet; *SSA* –specific surface area; FLG –few-layer graphene. ExG –expanded graphite; EG – ethylene glycol.

| Materials used and methods | MW parameters: power, time & conditions | XPS C/O | Raman $I_D/I_G$ | $\sigma$, $R_s$, SSA, application tested etc. | Ref. |
|---|---|---|---|---|---|
| GO in DI water | 800 W, 120 min | 1.86 | | SSA = 7.2997 m$^2$/g; poor exfoliation yield | 309 |
| Graphite, IL2PF$_6$ or IL4PF$_6$ | 30 W, 30 min; exfoliation and reduction in liquid | 30 | 0.14 | Exfoliation yield: 93%; selectivity: 95%; Thickness of single-layer graphene: < 1 nm | 310 |
| Graphite oxide powders in C$_4$H$_6$O$_3$ | 700 W, 1 min; dry condition; simultaneous exfoliation and reduction | 0.79 for GO & 2.75 for the MEGO (rGO) | | $\sigma$ = 274 S/m; Ultracapacitor: $C_{sp}$ =191 F/g (in KOH); SSA =463 m$^2$/g; interlayer spacing ~ 0.36 nm | 296 |
| Graphene oxide, DMSO, EG, NMP, DMF & NMP | 800 W, 6 min; MW-solvothermal reduction | 7.8 | 1.22 (DMSO), 1.23 (EG), 1.25 (DMF), 1.30 (NMP) | 10$^4$ S/m | 311 |
| Free-standing graphite oxide (GO) film | 500 W, VFM (6.425 GHz); 2 sec; solid-state reduction | | 0.3 | 10$^4$ S/m for graphene sheets | 102 |

Preprint of https://doi.org/10.1016/j.est.2025.115315| | | | | | |
|---|---|---|---|---|---|
| GO sheets (0.36 to 1.69 nm thick); GO on a Si wafer | 500 W, 240 sec; substrate temperature of 155-175 °C; solid-state | | $I_D/I_G$: 0.87 | $R_s = 7.9 \times 10^4$ Ω/□ for MW-rGO on $SiO_2$/Si; $10^5$ - $10^9$ Ω/□ for PDMS substrate; optical transparency of 92.7% at ~547 nm | 312 |
| GO in epoxy resin | 1000 W, for 3 min, reduction in dry ambient conditions | 10.38 | | electrical percolation threshold ($\phi_c$) of composites: 1 wt% and 0.3 wt% for 2 and 3 min. irradiation samples | 313 |
| Highly oriented pyrolytic graphite | 42 W; 5 min in the air | 17.5 | 1.56 for FLG | $R_s = \sim 6 \times 10^3$ Ω/□ for FLG | 314 |
| Graphite oxide; GNS (10 wt%) as susceptor | 1600 W pulsed mode; 50 sec; solid-state under $H_2$ | 18.5 | 0.853, 0.842, & 0.785 resp. for GNS obtained under ambient air, Ar, and $H_2$/Ar | $\sigma = 1.25 \times 10^3$ S/m for GNS (1-4 layers) obtained under $H_2$/Ar; $SSA$ = 586 $m^2$/g for GNS produced under $H_2$; GNS flake lateral dimension of several μm; mean thickness 3.15 nm; $SSA$: 586 $m^2$/g | 297 |
| Graphite oxide & graphite flakes | 1000 W, few sec in ambient air | 19.4 | 0.88 | μm-sized rGO flakes; $\sigma$ = 53180 S/m; $SSA$= 886 $m^2$/g; Tested for LIB & SIB with reversible capacities of 2260 mAh/g and 460 mAh/g at 0.1 A/g | 99 |
| Graphene oxide; template of rGO as MW trigger template | 1000 W, 4-5 sec; Ar (95%)-$H_2$ (5%); Template assisted reduction | | 0.753 | $R_s$ = 36.5 Ω/□ | 315 |
| Graphene oxide; mildly rGO membrane as external susceptor | 2000 W, 30 sec; solid-state | 17.84 | 2.209 | $\sigma$ = 8.12 S/cm | 316 |
| ExG graphite | 1000 W, 60-120 sec; solid-state | 14.4 for FLG | 0.23 for FLG | $\sigma$ = 165 S/m for FLG; Avg. thickness of FLG = 1.8-2 nm | 317 |
| Mixed solvent; Graphene Oxide in N,N-dimethylacetamide and water (DMAc/$H_2O$) | 800 W, 1-10 min, under dry $N_2$ gas | 5.46 | 0.96 | $\sigma$ = 200 S/m for graphene paper; 0.015 S/m for Graphene Oxide paper; Thicknesses: GO (0.8 nm) & graphene (0.45 nm); | 318 |
| Graphene oxides (0.8-1 nm thick; ⌀ =1-5 μm), & [Bmim]$BF_4$ | MW-assisted ionothermal treatment at 600 W | 1.32 to 7.65 | | $\sigma$:193 S/m for rGO | 298 |
| GO powder, [EMIm][$NTf_2$] | 700 W, 15 sec; liq. state | > 3 | | Supercapacitor: *power density* of 246 kW/kg and *energy density* of 58 Wh/kg | 295 |
| Graphite flakes, sodium tungstate dihydrate, 1-ethyl-3-methylimidazolium tetrafluoroborate, nafion, $H_2O_2$, and isopropanol | 1200 W, 100 sec; liq. State | 9.6 | 1.1 for bilayered GNS | Supercapacitor: Capacitance: 219 F/g, *energy density*: 83.56 Wh/kg & *power density*: 15.29 kW/kg; cyclability: 3000; *SSA*: 1103.62 $m^2$/g | 319 |
| Graphene oxide; small amount of graphite as MW trigger | 1000 W, different exposure time | | 1.07 for 120 sec MW exposure | $R_s$ =4 Ω/□ for 20 sec MW exposure | 320 |
| rGO "paper" as a trigger & graphene oxide (35 to 45 μm) "paper" | Graphene-triggered MW reduction; 1-3 sec & 3-5 sec | | | $R_s$ of MW-rGO: 40 Ω/□ | 321 |
| Pre-reduction of GO using vitamin C & subsequent MW exposure | 2000 W, 60 sec | 13.79 | 2.078 | $\sigma$ = 4.7 S/cm | 322 |
| Thermally annealed (300 °C for 1 h) GO flakes; dry reduction | 1000 W, 1-2 sec under Ar | | $I_{2D}/I_G$ = 1, $I_G/I_D$ > 10 | Crystallite size ~ 180 ± 77 nm; Tested as FET channel: mobility = 1000 $cm^2$/V-s; | 323 |
| GO sheets embedded in the poly (vinyl alcohol) matrix | 800 W, 1 & 3 sec | 2.05 | 1.09 for 3 min | $\sigma$ = $2.13 \times 10^{-2}$ S/m for PVA/1 wt% GO | 324 |

| | | | exposure | | |
|---|---|---|---|---|---|
| Graphite oxide | Two intermittent 50 sec MW exposure | 19.0 | | Interlayer distance: 0.35 nm for rGO, $SSA$: 1333.7 m$^2$/g; No capacitance decay after 80,000 cycles at current density of 30 A/g | 325 |
| Graphite flakes (500 μm-thick), TBAClO$_4$, propylene carbonate, N-methyl-1-pyrrolidone | 700 W, 10 sec, electrochemical intercalation & MW-assisted expansion of graphite | | $I_D/I_G$ < 0.07 for Holey graphene nanosheets | 691 S/cm; $SSA$ of 29 m$^2$/g | 326 |
| ExG graphite, aqueous solution of ammonia | Exfoliation of ExG; 300 W, 120 & 200 °C for 60-120 min | | | FLG thickness: 4.4 nm; lateral sizes: 0.6 μm; $\sigma$: 2.5 S/m | 327 |
| ExG | Exfoliation of ExG; 20 sec | | 0.85 | Energy storage: 221 F/g capacitance after 5000 cycles; graphene sheet of 10-100 μm in size & 1-10 nm thick | 328 |
| GO; H$_2$O based, HT, and MW reduction in sequence | | | $I_D/I_G$: 0.67 | 43.78 S/m; $SSA$ 81.767 m$^2$/g | 329 |

Vertically aligned carbon NSs (CNSs) comprising bi- and trilayer graphene were deposited on different metal substrates from solid carbon sources using H$_2$ and Ar plasma irradiation.[330] Compared to earlier studies employing gas sources, the produced graphene structures showed more graphitization, larger diameters, and fewer layers. The findings indicate that the interplay between the plasma (of H$_2$ and Ar) and carbon sources promotes the formation of larger and finer nucleation cites in the early stages, with atomic H etching dominating throughout the entire finer CNS growth. The vertically aligned graphene layer may be transferred directly over patterned SiO$_2$/Si substrate, yielding a heterojunction solar cell with a power conversion efficiency of ~0.9%. The original CNSs (OCNSs) were generated on a Cu substrate with 2.45 GHz MPECVD (MW plasma-enhanced CVD) at 900 W. Then, CH$_4$ gas was used to deposit the CNSs on both the Cu substrate and the quartz ampoule wall. The CNSs on the quartz ampoule wall were then employed as a solid carbon source to produce regenerated CNSs (RCNSs). At this point, several metals (Cu, Ti, and NiFe alloys) were employed as substrates. Finally, due to the different internal stresses, OCNSs and RCNSs easily detached from the substrates.

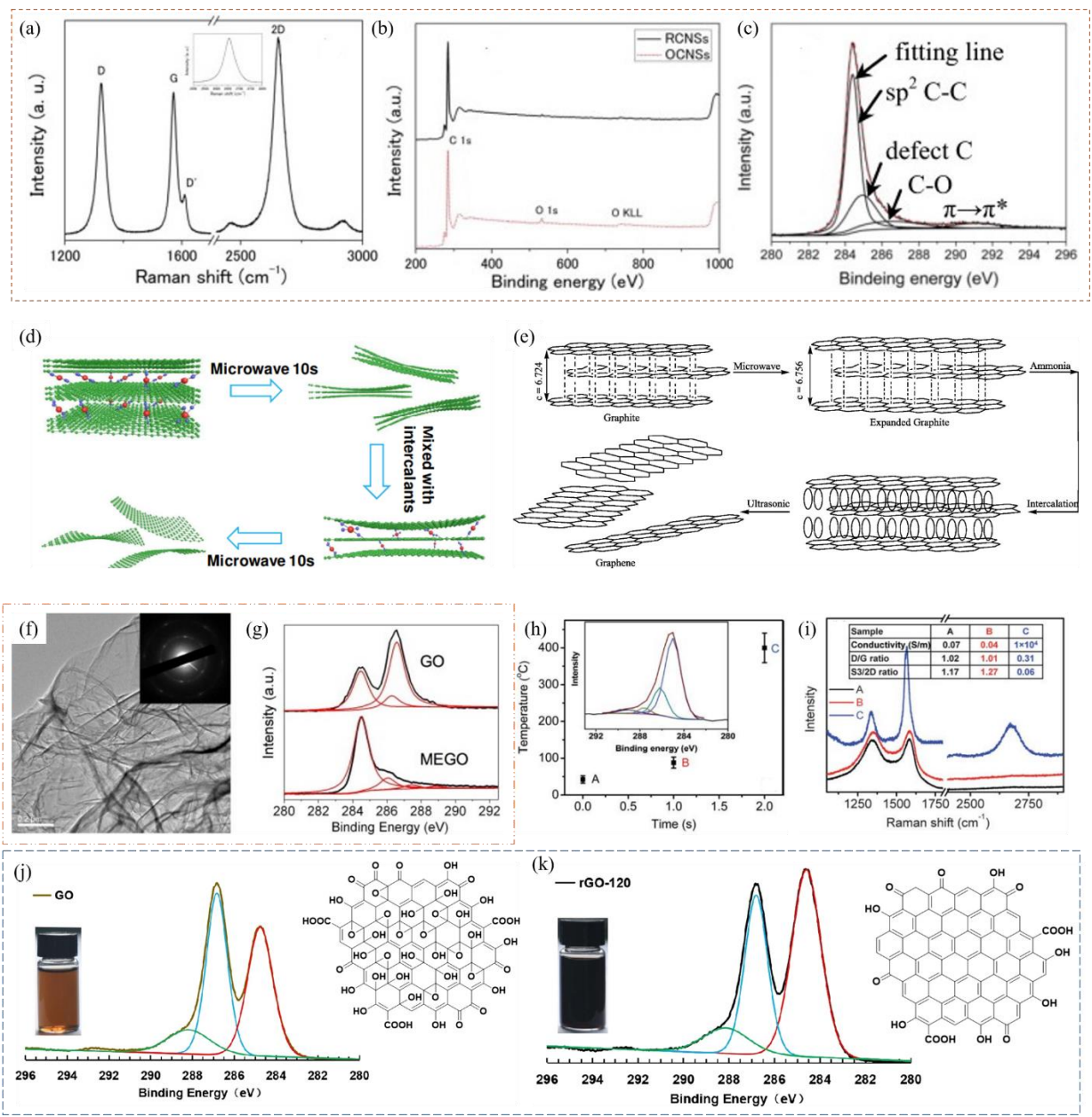

Figure 11: (a-c): Carbon NSs (CNSs). (a) Raman spectrum of the RCNSs. Inset: a single Lorentzian fitting of 2D peak. Laser: 633 nm; (b) XPS spectra for RCNSs and OCNSs; (c) XPS C 1s spectra of the RCNSs.[330] (d) Schematic of MGG fabrication process.[328] (e) Mechanism of MW-assisted intercalation exfoliation.[331] (f-g): MW exfoliated GO or MEGO. (f) TEM image of the MEGO (MW exfoliated GO) and associated electron diffraction pattern. (g) XPS C1s spectra of GO and MEGO.[296] (h-i): graphene sheets synthesis. (h) GO film temperature increment under 500 W MWs in 2 seconds. Inset: XPS data of rG; (i) Raman spectra of the GO sample prior to MW irradiation (A), imminent to transition state (B), after reduction to graphene (C), respectively. Table in inset: Comparison of $\sigma$, $D/G$ ratio, and S3/2D ratio for all samples at A, B, C stages.[102] (j-k) MW reduction of GO suspension. XPS spectra of (j) GO; (k) rGO prepared with 120 min of MW irradiation (rGO-120 sample).[309]

Figure 11a depicts the Raman spectra of RCNSs. As stated previously, the intensity ratio of D to G ($I_D/I_G$) correlates to the disorder degree and in-plane crystalline size $L_a$.[332] It should be noted that the following equation can be used to determine the size of crystalline domains, $L_a$, from Raman signals:[332]

$$L_a\,(nm) = (2.4 \times 10^{-10}) \times \lambda^4 \times \left(\frac{I_D}{I_G}\right)^{-1}$$

Where $\lambda$ (in nm), $I_D$ and $I_G$ are the wavelength of the Raman laser and the intensity of the D and G band respectively. The $I_D/I_G$ ratios of 3.62 and 1.98 were observed for the OCNSs and RCNS, respectively, indicating that the RCNSs had a higher $L_a$. The XPS measurements show a significant C 1s peak at 284.8 eV and a mild O 1s peak at 532.5 eV for both OCNSs and RCNSs (Figure 11b). The findings indicate that graphitic materials with a few oxygen species on their surfaces make up both CNSs. The growth of oxygen species may have been caused by the physical adsorption of oxygen and/or water vapor on the surface of the CNS when the samples were exposed to air. In the OCNSs, the atomic contents of C 1s and O 1s were 93.3% and 6.7%, respectively and for RCNSs, values for the same parameters were 99.6% and 0.4%, respectively. Figure 11c depicts the XPS C 1s spectra of the RCNSs. The OCNSs have around 52.9% non-oxygenated ring graphitic carbon (284.4 eV), whereas the RCNSs have 59.4%, indicating stronger graphitization in the RCNSs.

### 2.4.1 Exfoliation and reduction

FLG sheets ( > few μm-thick) were produced by exfoliating expanded graphite in an aqueous ammonia solution under MW irradiation.[327] The majority of the exfoliation product was made up of straight-edged mono-, bi-, and few-layer graphene sheets (<10 in number). The proposed exfoliation mechanism included the following steps: (i) permeation of the ammonia solution between the expanded graphite's graphene sheets; and (ii) spontaneous decomposition of the ammonia solution ($NH_4OH$) by MW radiation to form gaseous $NH_3$ and $H_2O$, which carried out the exfoliation. This technique's exfoliation effectiveness is proportional to the concentration of ammonia in the liquid medium, with higher ammonia concentrations producing larger graphene flakes. Comparison of the the C 1s peaks of samples produced using an aqueous solution of ammonia and pure water shows that graphene synthesized in ammonia solution contains comparatively few oxygenated species and is primarily made of graphitic carbon.

A chemical treatment in conjunction with MW-assisted heating for the synthesis of graphene NSs was demonstrated.[333] Natural graphite, hydrogen peroxide, and ammonium peroxy disulfate were mixed (in different wt. ratios) in a glass ampoule, ultrasonicated for 3 min, and directly exposed to MWs (500 W, 90 sec). Under MW irradiation, the precursors rapidly exfoliated, along with brilliant light. The authors propose that the oxidation and subsequent exfoliation include two stages: first, under MW irradiation, ammonium peroxidisulphate decomposes, producing oxide radicals that trigger radical-induced cutting along graphite NS edges and surface defects.[334] Second, as the process progresses, hydrogen peroxide decomposes and intercalates in the inter-layer graphene NSs, causing them to rapidly expand along the $c$-axis. The sheets had an $I_D/I_G$ ratio of 1.02 (correlative to an in-plane crystallite size of 3.7 nm) and an FWHM of 58/cm, suggesting strongly orientated and multi-layered graphene.[335] The exfoliation coefficient (volume of graphite, $V_{EG}$/volume of graphite intercalated compound, $V_{GIC}$[336]) was 150 and BET surface area of NPs was 590 m$^2$/g, which is comparable to the reported value.[337]

Expanded graphite was the starting material in a two-step process to make FLG (1.8-2 nm thick, 3-10 μm lateral size) with a low oxygen concentration.[317] After solvothermal treatment, expanded graphite was exposed to 1000 W MW radiation for 60-120 seconds. For comparison, GO was prepared from expanded graphite using a modified Hummer's approach. Exfoliated GO nanosheets were chemically reduced to graphene (called rGO) in the presence of hydrazine. The estimated sizes of sp$^2$ carbon domains in graphene oxide (GO) and FLG are 2-5 nm and 10-16 nm, respectively. The $\sigma$ of the as-synthesized FLG was determined to be 165 S/m, which is substantially higher than that of the GO (1.2 × 10$^{-4}$ S/m), perhaps because of the larger sp$^2$ carbon domain size, reduced oxygen concentration, and few structural defects. The projected $I_D/I_G$ values for expanded graphite, FLG, GO, and rGO are 0.14, 0.23, 0.86, and 1.0, respectively, showing that GO and rGO exhibit greater distortion. *C/O ratio of FLG is 14.4, which is greater than 2.2 of GO.*

Exfoliating expandable graphite repeatedly under MW irradiation resulted in large graphene sheets (10-100 μm in lateral dimension and 1-10 nm in thickness).[328] Figure 11d depicts the process of MW-assisted preparation of giant graphene sheets (MGG). The procedure begins with the exfoliation of commercial expanded graphite using 10 sec of MW irradiation in a typical MW oven. The volume of expanded graphite powders increases significantly when heated by MWs. In the second phase, the expanded graphite is combined with concentrated H$_2$SO$_4$ and KMnO$_4$. The optimal reactant ratio for KMnO$_4$ to expanded graphite is 2:1. In the final phase, the oxidized and intercalated expanded graphite is heated with MWs for 10 sec to produce a black and fluffy graphene powder. The MGG displays electrochemical capability for energy storage: 400 mAh/g in a coin-type Li-ion battery after 300 cycles (at 0.5 A/g) and $C_{sp}$ of 221 F/g after 5000 cycles (at 10 mV/s) in a symmetric capacitor type.

According to Matsumoto et al., when graphite suspended in molecularly engineered oligomeric ionic liquids (IL2PF$_6$ or IL4PF$_6$) exposed to MWs for 30 min results in single-layer graphene (thicknesses <1 nm) with high-efficiency exfoliation (93% yield), a high selectivity (95%), excellent structural integrity ($I_D/I_G$ ~0.14, C/O ~30) as compared to that of the precursor graphite ($I_D/I_G$ ~0.13, C/O ~32).[310] Subramanya et al. made FLG NSs by heating natural graphite, sodium tungstate, and hydrogen peroxide with MWs.[319] Sodium tungstate contributes to both oxidation and the formation of defects in graphite's π-structure. The prepared graphene is bilayered with a reduced domain size of 3.9 nm, which accounts for the FLG NSs' larger SSA (1103.62 m$^2$/g). Furthermore, XPS analysis of FLG NSs reveals a high C/O ratio (~ 9.6).

Using dibasic ester (DBE) as a solvent, Jiang et al. have demonstrated a technique to synthesize graphene through NH$_3$ molecule intercalation-exfoliation of graphite with MWs.[331] The prepared graphene has a few-layer structure and a size in excess of 3 μm. MW treatment increased graphite interlayer distance to 6.756 Å, overcoming weak Vander Waals interactions between layers. Consequently, the intercalator, such as ammonia, can easily penetrate the enlarged graphite interlayer (Figure 11e). Finally, expanded graphite is separated into monolayer and multilayer graphene and dispersed in the DBE solvent by ultrasonication.

Zhu et al. reported a method for concurrently exfoliating and reducing graphite oxide (MW exfoliated GO or MEGO, Figure 11f-g).[296] Within 1 min of processing graphite oxide powders in a commercial MW-700 W oven, reduced graphite oxide materials could be created. The materials as prepared were crumpled graphitic sheets (few layers) and electrically conductive. The $C_{sp}$ values of ~191 F/g was observed utilizing MW exfoliated graphite oxide as an electrode material in an ultracapacitor cell with a KOH electrolyte. Graphite oxide (GO) powders prepared by the modified Hummers method were treated in a MW oven at ambient temperatures for 1 minute. Upon MW irradiation, the GO granules expanded significantly, accompanied by brisk fuming. In addition to GO powders, MW treatment could be used to reduce graphene oxide suspensions in propylene carbonate, an organic solvent. The MEGO powders had a $SSA$ of 463 m$^2$/g. The XPS results of MEGO powders are shown in Figure 11g with the C1s spectrum and GO precursor for comparison. The peaks with binding energies higher than sp$^2$-bonded carbon (284.5 eV) are smaller for the MEGO than for the GO precursor. The peaks between 286-289 eV range are commonly associated with epoxide, hydroxyl, and carboxyl groups.[338] The XPS data indicate that MW irradiation considerably eliminated oxygen-containing compounds. The $C/O$ ratios for GO powders and for the MEGO were 0.79 and 2.75 respectively. The reduction of GO was further validated by the MEGO's improved $\sigma$, which was estimated to be ≈ 274 S/m, four orders of magnitude greater than graphite oxide.

rGO was prepared by a solid state MW treatment of a free-standing graphite oxide (GO) film in a variable-frequency MW (VFM, 6.425 GHz, 500 W).[102] The heat generated causes temperatures to rise at a rate of more than 200 °C/sec and the increased thermal impact of MW irradiation enables the rapid reduction of GO to graphene without the need of reducing agents or solvents. The surface temperature of the GO sheet during MW treatment was measured in situ (Figure 11h). Because graphitic materials absorb MW energy well, the GO's surface temperature rose to 400 °C in about two seconds. The MW treatment enhanced bulk $\sigma$ from 0.07 S/m to $1 \times 10^4$ S/m, equivalent to $\sigma$ of high-temperature (1050 °C) exfoliated graphene.[339] In Raman spectra, the G mode correlates to the vibration of sp$^2$-hybridized carbon, while the D mode corresponds to the conversion of sp$^2$-hybridized carbon to sp$^3$-hybridised carbon.[340] Following the MW reduction, the G band redshifts to 1565/cm, showing that the carbon atoms' hexagonal network has recovered.[341] Furthermore, the $D/G$ peak area ratio falls from 1 to 0.3. Because the $D/G$ ratio is proportional to the avg. size of the sp$^2$ domains, the drop under MW significantly enlarges the avg. size of the crystalline graphene domains. The S3/2D peak area ratio decreased considerably to 0.07 following reduction, indicating that the defect density was lowered (Figure 11i).[342] Xiang et al. demonstrated a MW-assisted reduction of graphene oxide (GO) suspension to produce rGO with a low reduction degree.[309] In this method, GO powder dispersed in deionized water was irradiated with MWs (800 W) for different time durations. It was proposed that the reduction began on the GO's surface and progressed to the basal plane's periphery. As the degree of reduction on the surface of the GO increased, so did the extent of conjugation, resulting in a structure that increasingly resembled graphene's conjugated structure. After 120 min of MW reduction, the intensity of the C-O and C = O peaks reduced (Figure 11j-k), confirming the removal of oxygen-containing groups, and the intensity of the C=C peak rose. The peak intensity ratios of C=C and C-O increased from 0.975 to 1.495, and the *C/O* ratio rose from 1.80 to 1.86. The *SSA* of GO is 2.6744 m$^2$/g, whereas the *SSA* of rGO-120 is 7.2997 m$^2$/g, which is lower than the values reported by the solid-sate MW reduction methods. The *SSA* of rGO significantly rises following MW treatment, indicating that water trapped between GO layers exfoliates GO layers under the action of MWs. MW reduction in dry condition may yield rGO with a high reduction degree along with a favorable exfoliation amount.

MWs were used to initiate solid-state deoxygenation of graphene oxide (GO) at 165 °C. In MW-rGO, the concentration of oxygen-containing functional groups: carboxyl, hydroxyl, and carbonyl were significantly reduced.[312] It was revealed that MWs can aid in the deoxygenation of GO at comparatively low temperatures. The defect level in GO falls rapidly throughout the isothermal solid-state MW-reduction process at low temperatures, showing that the thin graphene lattice structure has been successfully recovered. The MW-reduced GO on SiO$_2$/Si substrates has a low $R_s$ (~7.9 × 10$^4$ Ω/□) due to deoxygenation and defect-level reduction, resulting in an optical transparency of 92.7% at ~547 nm. After 240 sec, the $R_s$ of MW-reduced individual GO sheets reaches approximately 7.9 × 10$^4$/□ (Figure 12a). Flexible transparent conductive coatings on polydimethylsiloxane (PDMS) substrates are prepared via a low-temperature solid-state MW reduction. The transparency of these coatings ranges between 34% and 96%. The coating has $R_s$ of 10$^5$ to 10$^9$ Ω/□. A variable-frequency MW system (similar to the one stated above) fitted with a N$_2$ protection environment and an IR temperature sensor was used. The substrate temperature was regulated and maintained at 155-175 ºC, while the MW's power was regulated at 500 W.

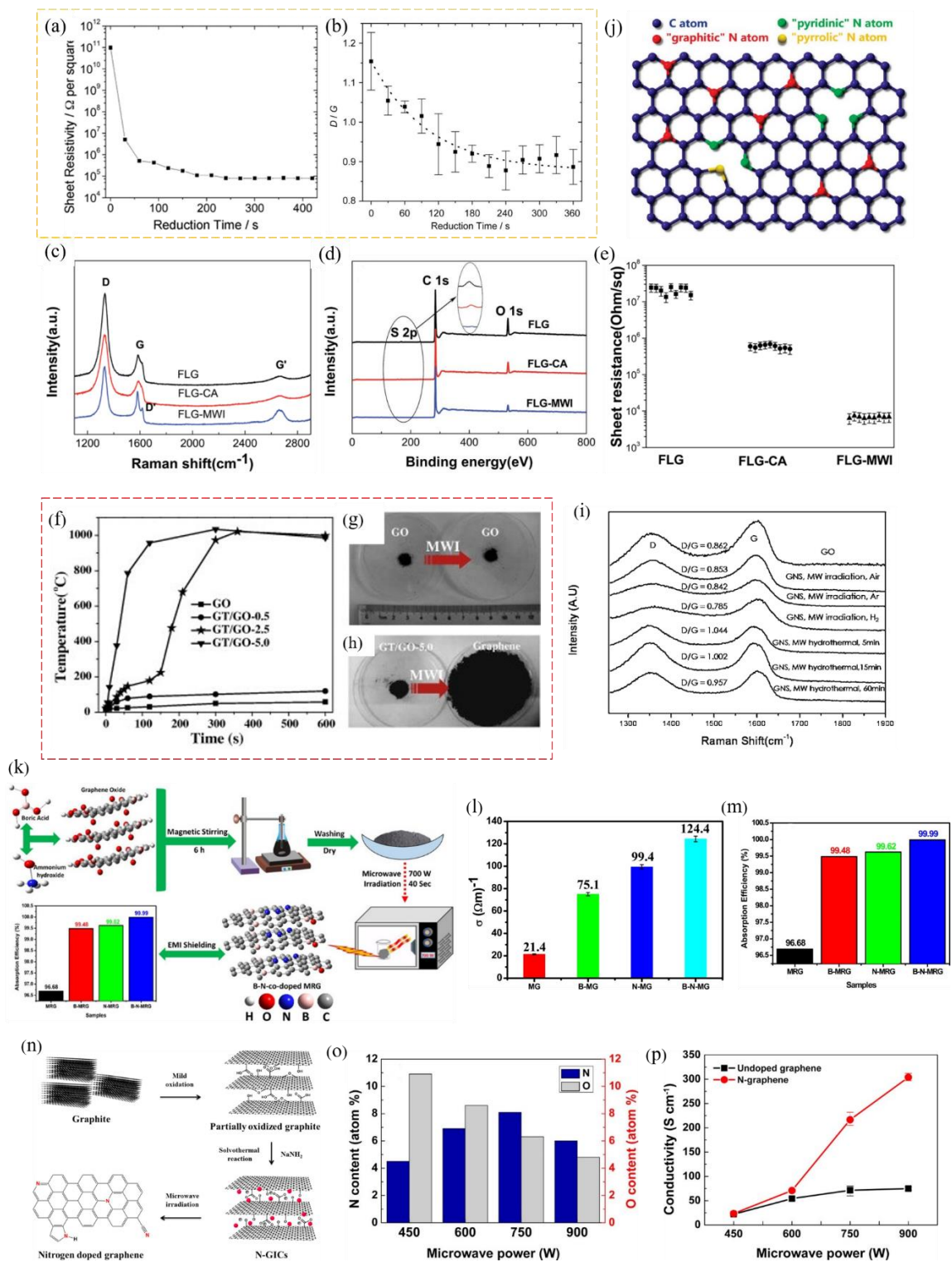

Figure 12: (a-b): rGO deposited on SiO$_2$/Si substrate. (a) $R_s$ of rGO oxide on a SiO$_2$-coated Si substrate. (b) $I_D/I_G$ ratio of the $D$ and $G$ bands of rGO vs. reduction time.[312] (c-e): Reduction of GO films. (c) Raman spectra (632 nm laser excitation) of FLG, FLG–CA and FLG–MWI films; (d) XPS spectra of three FLG films; (e) $R_s$ of FLG, FLG–CA and FLG–MWI films.[314] (f-h): GT/GO combination. (f) Temperature profiles of GO and GO/graphene from thermal expansion (GT) in different amount under MW irradiation (GO 0.2 g, GT 1-10 mg). Optical images of (g) GO & (h)

GT/GO-5.0 before and after MW irradiation, revealing insignificant response of GO and better results of GT/GO-5.0 to MW irradiation.[343] (i) Raman spectra of GO and GNS obtained by solid-state MW technique & MW-assisted HT technique under various settings.[297] (j) N-doped Gr by CVD. Blue, red, green, & yellow spheres represent the C, graphitic N, pyridinic N, & pyrrolic N atoms in the N-doped Gr, respectively.[268] (k-m) B & N co-doped rGO. (k) MW-assisted steps for the preparation of B–N co-doped rGO; (l) $\sigma$ of rGO, B-rGO, N-rGO and B–N-rGO at RT; (m) Absorption efficiency of rGO, B-rGO, N-rGO, and B–N-rGO.[344] (n-p) Exfoliation and doping of graphene with MWs from bulk graphite. (n) Schemes of graphene doping and exfoliation; (o) contents of N and oxygen atoms of N-graphenes prepared with different MW powers; (p) $\sigma$ of N-graphenes and undoped graphenes prepared with different MW power.[345]

As was previously discussed, fewer defects in the graphitic structures are typically indicated by a lower $D/G$ band intensity ratio in a Raman spectrum. The $D/G$ intensity ratios of MW-reduced GO studied as a function of reduction time (Figure 12b) indicate that as the MW-reduction period rises, the D/G ratio in the Raman spectra gradually falls from 1.16 (pre-reduction) to 0.87, signaling a decreasing defect density in MW-reduced GO and a restoration of the Gr lattice structure from $sp^2$-carbon bonds. Han et al. have investigated the MW irradiation effect on reduction of GO films.[314] The reduction process was accomplished with MW power < 42 W and temperatures below 250 °C in the air. The FLG sheets were produced using electrochemical exfoliation of highly oriented pyrolytic graphite (HOPG) and using dip coating, FLG films were coated on $SiO_2$/Si substrates. For MW irradiation (2.45 GHz, waveguide in TE103 mode),[346] each specimen was positioned at the center of the strongest $H$-field. The sample's temperature was recorded while it was being exposed to MW radiation. The MW irradiation treatment was conducted in the air for 5 min (at 42 W), resulting in FLG-MWI films. For comparison, FLG films were subjected to standard/conventional annealing at 250 °C for 30 min in an electric furnace in the ambient, resulting in FLG-CA film. Figure 12c displays the Raman spectra of the prepared graphene films (FLG), conventionally annealed graphene films (FLG–CA), and MW irradiated graphene films (FLG–MWI). The reported $I_D/I_G$ for FLG, FLG–CA and FLG–MWI is 2.89, 2.34 and 1.56 respectively. Following standard/conventional annealing or MW irradiation, the $I_D/I_G$ decreases, indicating that the oxygen functional groups were partially eliminated. The $C/O$ ratio goes up from 7.8 to 9.3 and 17.5 post /standard/conventional annealing or MW irradiation, demonstrating that both types of treatments may eliminate oxygen functional groups (Figure 12d). MWs have a greater impact on reduction than standard annealing ($C/O$ ratio of 17.5). FLG-MWI films have a decreased $R_s$ of $\sim 6 \times 10^3$ Ω/□ due to the elimination of oxygen functional groups triggered by MWs (Figure 12e, measured by four-point-probe technique[347]).

Pokharel et al. demonstrated a multi-step MW exfoliation of graphite oxide (GO) to highly reduced graphene sheets, followed by loading in epoxy resin to produce electrically conductive composites.[313] It was revealed that one-time MW irradiation (1000 W, 1 min) was insufficient to transform GO powder into highly conductive graphene sheets. As a result, the same sample was exposed to MWs twice for 1 minute. Approximately, 3 min of MW irradiation of GO in three phases was adequate to produce significantly reduced GO ($C/O \approx 10.38$). Then, using this material, a graphene/epoxy composite was prepared. The electrical percolation threshold ($\phi_e$)[348, 349] of composites was found to be 1 wt% and 0.3 wt% for 2 and 3 min. irradiation samples, respectively, indicating highly conducting percolation networks are formed.[349] Despite small 0.5 wt% loading of a 3 min irradiation sample in epoxy, the composite's glass transition temperature ($T_g$) enhanced by 10 °C, showing a substantial interfacial interaction between the graphene and the epoxy resin.[350]

Hu et al. showed that increased oxygen in graphite oxides (GO) significantly reduces its MW absorption capacity due to the size reduction of the π–π conjugated structure in these materials, and vice versa.[343] It is well known that graphene is a very efficient MW absorbent, however in GO with weak MW absorption capacity, the unoxidized graphitic area impurities in GO operate as MW absorbents to commence MW-induced deoxygenation. The inclusion of a minute quantity of graphene to GO causes an avalanche-like deoxygenation process of GO under MWs, resulting in graphene formation and the temperature profiles of reaction systems escalated rapidly to ~1000 °C in 180 seconds. To investigate the temperature profiles of GO and GO/varying quantities of graphene during MW exposure, thermal expansion (GT) was added to standard graphite oxide (GO) at amounts of 0.5%, 2.5%, and 5.0% (referred to as GT/GO-0.5, GT/GO-2.5, and GT/GO-5.0, Figure 12f). Both pure GO and GT/GO combinations were subjected to MW irradiation at 800 W. Pure GO showed negligible temperature variation and volume expansion (Figure 12g) throughout the MW irradiation, whereas GT/GO-0.5 ascended to 100 °C from ambient temperature after 600 sec of MW exposure. For the GT/GO-2.5, the reaction became intense, and the temperature climbed rapidly to ~1000 °C within 180 sec, along with a considerable volume growth (Figure 12h). An even faster reaction was shown by GT/GO-5.0, which ignited an avalanche-like reaction in < 10 seconds. The $C/O$ ratio of GO remained constant during the MW period, while only 2 min of MW removed the majority of the oxygen in GT/GO-5.0.

To increase the quality of graphene NSs (GNSs), a combination of graphite oxide (GO) and GNS in solid-state was irradiated with MWs under $H_2$ atmosphere.[297] Under solid-state MW irradiation conditions, GO couples with MWs poorly and does not get heated due to its partially unconnected π-system caused by multiple oxygen-containing functional groups. On the other hand, GNS is an efficient MW energy absorber. For this reason, GNS (10 wt%) was combined with GO as a MW susceptor to absorb MW energy, generate heat faster, and exfoliate GO during MW irradiation. For comparative study, GNS was produced using a MW-assisted HT treatment of GO. Figure 12i depicts the Raman spectra of GO produced via the Hummers method, GNS obtained by solid-state MW synthesis in various gas atmospheres, and MW-assisted HT synthesis. GNS produced by solid-state MW synthesis in ambient air, Ar, and $H_2$/Ar atmospheres, respectively had $I_D/I_G$ ratios of 0.853, 0.842, and 0.785 < 0.862 of GO−attributed to a self-healing or defect-healing action that ensues at high temperatures under the solid-state MWs exposure.[351] The lower $I_D/I_G$ ratio for GNS generated in an $H_2$/Ar environment than in an Ar atmosphere shows that the $H_2$ environment is very efficient in reducing the creation of vacancies/defects amidst the breaking of oxygen-containing functional groups in GO. For MW-assisted HT synthesis, the $I_D/I_G$ ratio of GNS was 1.044, 1.002, and 0.957 for reaction times of 5, 15, and 60 min, respectively. The GNS produced via solid-state MW synthesis in a $H_2$ atmosphere has a high $SSA$ of 586 m$^2$/g and a $C/O$ ratio of 18.5. For GNS obtained by solid state MW synthesis in $H_2$/Ar gas environment, $\sigma$ was 1.25 × 10$^3$ S/m; for GNS acquired by solid state MW synthesis in Ar gas atmosphere, 7.41 × 10$^2$ S/m; and for GNS obtained by MW-assisted HT synthesis, $2.77 \times 10^2$ S/m. For solid-state MW irradiation synthesis of GNS, the GO/GNS mix in a quartz vial was subjected to MW irradiation for 50 sec (1600 W, pulsed mode). For the HT procedure, the MW-precursor containers were heated in a MW oven and held at 200 °C for 5 to 60 min.

In a strategy to reduce graphene oxide (GO) embedded in polymers using MW irradiation, the MWs only reduced the GO sheets moderately without affecting the microstructures of the poly (vinyl alcohol) (PVA) matrix.[324] The hydrogen bonds produced between GO and the PVA matrix were removed after GO reduction, resulting in a decrease in interfacial bonding. However, the mechanical characteristics and glass transition temperature ($T_g$) of GO/PVA nanocomposites, particularly GO sheets with lateral dimensions (i.e., 0.1-5 µm), were significantly improved. There have been two hypothesized explanations for this phenomenon. First, the MW generated heat in GO welded GO and the neighboring PVA matrix, resulting in an interfacial region with a decreased free volume; second, the MW promoted the formation of covalent bonds between PVA and the GO sheets. To reduce GO with MW irradiation, a short duration of irradiation (1 & 3 min, respectively, at 800 W) from a commercial MW oven was applied to GO/PVA nanocomposites (with varying GO content) in ambient. The $C/O$ ratio elevated to 2.05, equating to an $I_D/I_G$ ratio of 1.09 for the sample that was exposed to MW irradiation for 3 minutes. The $I_D/I_G$ ratio increased with increased irradiation time, showing that as GO was reduced, the avg. size of sp$^2$ domains fell, while the new domains generated were smaller in size but larger in number.[352] The elimination of oxygen is a critical step during reduction because oxygen groups isolate sp$^2$ domains in GO.

Theoretical studies show that substitutional doping can change the band structure of graphene, leading to a metal-semiconductor transition.[353-355] Doped graphene has several exciting features and possible applications, including superconduction[356] and ferromagnetism.[357] N-doped graphene has been synthesized via the CVD method.[268] It was found that the structure is a few-layer graphene. Some single-layer graphene has also been found. Since in the CVD process doping occurs simultaneously with the recombination of carbon atoms to form graphene, N atoms can be inserted into the graphene lattice. Electrical measurements demonstrate that N-doped graphene has *n*-type behavior, demonstrating that substitutional doping can efficiently change graphene's electrical characteristics.[268] As shown in Figure 12j, in N-doped Gr, N atoms are arranged in three different bonding characteristics within the graphene network.

By permitting Li-ion interaction with active sites, heteroatom-doped graphene enhances cycle performance, first cycle charge/discharge capacity, and Li storage capacity.[11, 358] Heteroatom doping (N, B, F, S, or P) improves electrochemical performance because Li-ion storage is influenced not only by heteroatom contents but also by synergistic coupling effects among heteroatoms.[359-361] Recent theoretical studies have revealed that doped graphene is more effective at storing Li-ions owing to enhanced defects in the graphene plane.[362, 363] Doping rGO with *p*-type and *n*-type elements (e.g., B, N) improves its $\sigma$ and physical attributes.[364-367] B and N doping in graphene disrupts the ideal sp2 hybridization of carbon atoms, altering their electronic characteristics and chemical reactivity.[368-370] B,N-co-doped graphitic carbon exhibits good oxygen reduction reactions because of the beneficial interaction between B and N.[371] There are several reports on the fabrication of B- and N-doped Gr for different applications.[372-377] As a result, B and N can be ideal dopants for rGO due to their superior properties such as electron and hole transfer into carbon-based materials and abilities to change electrical and transport properties.[365, 378] For more information about the effect of heteroatom doping on different nanocarbon properties, readers should consult recently published articles.[11, 358]

Reduced graphene oxide (rGO), B-doped rGO (B-rGO), N-doped rGO (N-rGO), and B-N co-doped rGO (B-N-rGO) have been fabricated using MWs and studied for their effectiveness in electromagnetic interference (EMI) shielding applications in the *K*u-band frequency range (12.8-18 GHz).[344] B-N-rGO exhibited higher $\sigma$ in comparison to rGO, B-rGO, and N-rGO, thus making B-N-rGO well suited for EMI applications. The B and N co-doping significantly increases the $\sigma$ of rGO (21.4 to 124.4 S/m), because N contributes electrons and B introduces holes in the system potentially forming a nanojunction (Figure 12*l*). After doping with B and N, the spatial configuration of rGO improves space charge polarization, dielectric polarization, natural resonance, and trapping of EM waves by internal reflection, resulting in a high EMI shielding of −42 dB (∼99.99% attenuation) at a critical thickness of 1.2 mm compared to undoped rGO (−28 dB, Figure 12m). Graphene oxide (GO) was synthesized (using graphite flakes of 1-2 mm) using a chemical route[379] with some modifications (Figure 12k). GO was converted to rGO utilizing the MW-assisted thermal expansion process using MW irradiation (700 W, 40 sec). B-rGO was prepared using simple chemical mixing method employing boric acid solution. The end solid material was exfoliated using MWs (700 W, 40 sec). In a similar fashion, N-rGO was produced using an ammonia solution. B-N-rGO was produced by adding ammonia and boric acid solutions at the same time and repeating the methods described above.

In another case of co-doping, N and B co-doped graphene (NB-Gr) with a hierarchical framework was developed using a MW-assisted strategy.[372] The NB-Gr network provided various electron transport routes and was used to fabricate a sensitive $H_2O_2$ amperometric sensor. The NB-Gr modified electrochemical sensor had a linear range in 0.5 μM to 5 mM, a detection limit of 0.05 μM at a signal/noise ratio of 3. This performance was attributable to both the advantageous structure of NB-Gr and synergetic effects caused by the co-doping of N and B in Gr. The biosensor based on NB-Gr was also used to detect $H_2O_2$ in real time from living cells at the nanomolar level. To prepare NB-Gr, first graphene oxide (GO) was synthesized through the oxidative treatment of purified natural graphite.[380] Then GO aqueous dispersion mixed with cyanamide solution (N precursor) was treated in a commercial MW oven for 1 h (160 °C, 200 W) and dried under vacuum. The powder was then intimately mixed with $B_2O_3$ (B precursor) and subjected to pyrolysis in a tube furnace at 900 °C for 0.5 h under Ar.

Using sodium amide as a N source, N-doped GNSs were prepared via MWs[345] where graphite exfoliation and N doping happened simultaneously in a minute, with the N content of the doped graphene reaching ∼8.1%. It was also discovered that the N atom's binding configuration on the graphitic layer was made up of a variety of N-containing moieties, including pyrrolic-N, pyridine-N, and quaternary-N, and that their composition varied with irradiation power. To prepare N-graphene (Figure 12n), graphite was intercalated with a mixture of $H_2SO_4$ and $HNO_3$ to create graphite-intercalation compounds (GICs, 3 μm-thick). These GICs were then changed into N-GICs through a solvothermal reaction with $NaNH_2$ in benzene. N-graphene (1.2 nm-thick) was produced with MW-heating of N-GICs under $N_2$ environment. Figure 12o depicts how the overall N content in N-graphene varies with MW irradiation power. When the radiation power rises, intramolecular dehydration or decarbonylation occurs, resulting in thermally stable heterocyclic aromatic moieties such as pyridine, pyrrole, and quaternary type N sites. As a result, higher MW power would promote the formation of pyridinic N in graphene sheets, whereas lower MW power would result in more incorporation of N in the form of nitrile N and a greater proportion of pyrrolic N, regardless of MW power. N-graphene exhibits higher $\sigma$ (~305 S/cm) compared to undoped graphene (Figure 12p). This is due to increased $sp^2$ carbon networks, decreased oxygen concentration, and defects caused by the N atom. On the contrary hand, when the irradiation power was reduced, very low conductivity was observed, presumably due to insufficient restoration of $sp^2$ carbon networks from the oxidized graphitic layer generated during graphite's severely oxidative expansion. The N-doped graphene demonstrated a significantly increased specific capacitance of 200 F/g (current density of 0.5 A/g), which was attributed to the pseudocapacitive effect caused by the insertion of N atoms into the graphitic layer.

A moderate amount of graphite flakes can act as a catalyst, markedly promoting MW exfoliation and graphite oxide reduction in ambient air in just a few sec (Figure 13a).[99] This material, catalytic MW exfoliated graphite oxide (abbreviated as CMEGO), showed better quality as compared to conventionally MW-exfoliated graphite oxide (MEGO): larger exfoliation degree with fine graphene sheets, *SSA* (886 m$^2$/g vs. 466 m$^2$/g), a larger *C/O* ratio (19.4 vs. 6.3), better lattice crystallinity, and improved $\sigma$ (53180 S/m vs. 5140 S/m). The CMEGO is also tested as an anode material in LIBs and sodium-ion batteries (SIB), delivering (listed consecutively): reversible capacities of 2260 mAh/g and 460 mAh/g at 0.1 A/g; rate capabilities of 469 and 148 mAh/g at 30 A/g; improved cycling stability of 91.4% and 85.7%. According to Raman spectroscopy analysis, the MEGO has a lower reduction efficiency and is extremely disordered because of the lack of a recognizable 2D band and the existence of a sharp disordered D band in the Raman spectra ($I_D/I_G$ = 1.38). However, the CMEGO showed highly ordered Raman characteristics, including a distinct 2D peak and a faint D peak ($I_D/I_G$ = 0.88). It was discovered that MW energy absorption by flake graphite and the subsequent generation of an intense plasma triggered accelerated heating of the GO, resulting in the elimination of oxygen functional groups in one instance along with a major reordering of the graphene basal plane.

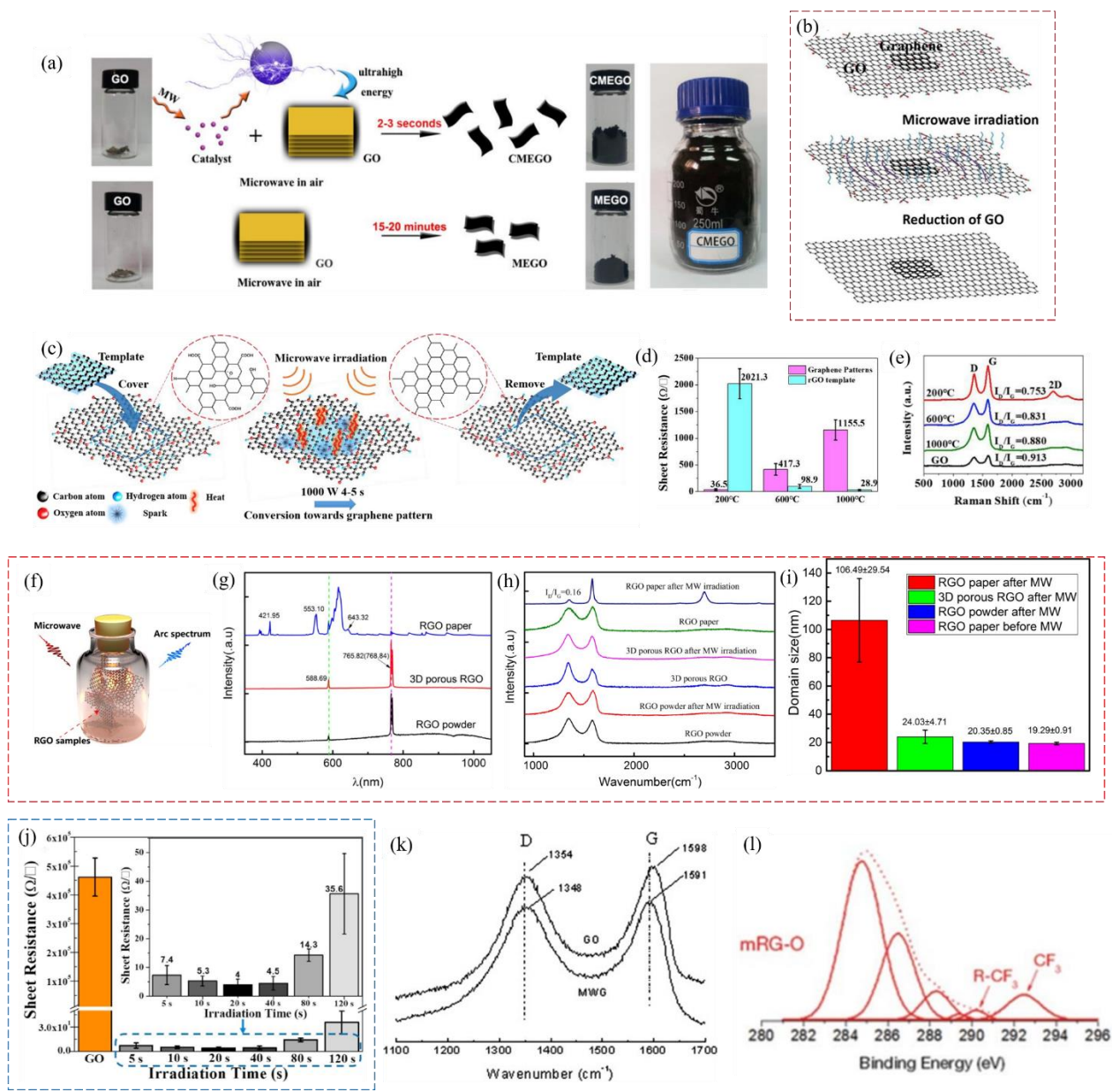

Figure 13: (a) The synthesis steps for MEGO and CMEGO.[99] (b) Schematic diagram of the graphene-triggered MW reduction.[321] (c-e): MW & template-assisted graphene oxide to graphene conversion. (c) Schematic representation of template-assisted MW conversion of GO to graphene patterns; (d) $R_s$ values for graphene patterns and rGO templates; (e) Raman spectra of GO thin film and graphene patterns.[315] (f-i) Arc spectra of various solid rGO samples. (f) Experimental set up; (g) Arc spectra of rGO powder, 3D porous rGO and rGO paper under MWs; (h) Raman spectra of rGO samples before & after MW exposure; (i) Domain size of several rGO samples acquired following MW irradiation.[381] (j) $R_s$ of GO and hierarchically porous graphene prepared under varied durations of MW irradiation.[320] (k) GO reduction in mixed solvents. Reduction using DMAc/$H_2O$ mixed solvents. Raman spectra of GO and MW thermally reduced graphene (MWG).[318] (l) Synthesis of RG-O using [EMIm][NTf$_2$]. The C 1s and N 1s XPS spectra of GO and mRG–O.[295]

rGO paper was used to trigger the MW arc discharge and reduction of graphene oxide paper in air. A considerable diffusion propagation process was seen during the MW reduction of graphene oxide (GO), resulting in graphene sheet.[321] Separately, paper clips or iron wire were used to prepare rGO via MWs. The paper clips or bent iron wires were combined with GO in a glass beaker. Following MW irradiation, graphene-triggered MW reduction yielded identical arc discharge and reduction outcomes. A small graphene (rGO) paper produced via thermal annealing or other reduction techniques was placed over a large GO paper in a glass vial and irradiated with MWs (Figure 13b). The thickness of the various GO papers (15 samples) were 35-45 μm. The arc discharge was noticed after 1-3 sec of MW irradiation, followed by another 3-5 sec of irradiation, resulting in MW reduced graphene oxide (MW-rGO) paper. In comparison to rGO, MW-rGO had a higher $I_{2D}/I_G$ ratio and larger graphene domain sizes. The authors proposed that MW reduction is not limited to a simple thermal reduction process, which could be caused by thermal effect or arc discharge. Furthermore, it was claimed that the arc discharge mechanism during MW reduction is crucial. It is well known that rGO is an effective MW absorber.[53] When MWs interact with rGO, delocalized π electrons can freely enter the junction (interfaces) between graphene and GO regions. The gathered electrons at the graphene/GO interfaces discharge, generating an intense arc that reduces the adjacent graphene oxide regions. This is coupled by the rapid creation of energy, which completely reduces GO in a brief period. In this event, oxygen atoms are extracted and released as a gas, while carbon atoms are rearranged.

In a somewhat similar approach, a template-assisted MW conversion of graphene oxide to graphene patterns was devised, in which MW irradiation was interacted with a chemical template of rGO to produce conductive graphene patterns on an insulating graphene oxide thin film via a reduction transfer process.[315] The properties of the resulting graphene patterns were discovered to be very similar to those of the templates utilized. Thermal annealing of GO films at 200, 600, and 1000 °C under Ar (95%)-$H_2$ (5%) flow was used to generate rGO templates. Then the prepared rGO template was positioned atop a GO thin film in a glass container and MW'ed at 1000 W for 4-5 sec under an atmosphere. Arc discharge was detected at the end of the reduction process. Following irradiation, the rGO template was recovered, yielding a GO sheet with graphene patterns (Figure 13c). Figure 13e depicts the Raman spectra of GO thin films and graphene patterns obtained at various annealing temperatures. The ratio, $I_D/I_G$, declines from 0.880 to 0.753 as the heat-treatment temperature of the rGO template is reduced, and all of them less intense than the GO. Further, the covered part transitioned from insulating to conducting, as confirmed by the $R_s$ measurement (Figure 13d). While the $R_s$ of the rGO template decreases with increasing the annealing temperature, the $R_s$ of the graphene pattern decreases with reducing the annealing temperature of the rGO template, indicating the inverse tendency. Annealing the rGO template at 200 °C resulted in a graphene pattern with a $R_s$ of 36.5 Ω/□.

A mildly reduced graphene oxide (MG) membrane was used as an external susceptor in the reduction of graphene oxide (GO) using hybrid MW heating.[316] The volumetric reduction degree of GO steadily increases as $R_{MG/GO}$ (mass ratios of MG to GO) increases, however the surface reduction degree falls, which should be attributed to decreased MW penetration into the surface of GO owing to the increased thickness of the covered MG. The initial stage involves conventional GO heating, followed by a two-way heating mode where MG heats the GO from the surface and MW heating from the GO itself. Mildly reduced graphene oxide (MG) was produced by annealing graphene oxide (GO) at 300 °C for 1 h in a horizontal furnace with nitrogen. GO was reduced using an industrial MW oven (2.45 GHz) in Ar. GO powder was covered with MG membrane in various $R_{MG/GO}$ ratios and heated by MW irradiation at 2000 W for 30 seconds. With 30 sec of 2000 W MW irradiation, the rGO sample's $\sigma$ reached 8.12 S/cm, with an $R_{MG/GO}$ ratio of 0.09. In XPS analysis, the rGO sample with an $R_{MG/GO}$ ratio of 0.045 had a highest $C/O$ ratio of 17.84, which corresponds to an $I_D/I_G$ ratio of 2.209.

The arc spectra generated during MW irradiation from rGO powder, 3D porous rGO, and rGO paper have been recorded and analyzed.[381] In the study, respective samples were placed in a glass vial and irradiated with MWs. The detector of the optical fiber spectrometer was set in front of the MW oven panel to gather and record the spectra. MW irradiation for 1-2 sec resulted in heat release and emissions, reduced rGO samples, and an arc. After completing the MW irradiation operation, the arc spectra were recorded for ~3 sec (Figure 13f). Based on the arc spectra analysis, it was proposed that the diverse spectral features may indicate different reaction process and mechanisms (Figure 13g). Carbon atoms can rearrange during MW-induced reduction.[99, 321, 323] Thus, the arc spectra of rGO paper under MW irradiation may be the consequence of carbon atom reorganization, and the reaction process of rGO paper under MW irradiation is the result of the heat effect of MW irradiation and carbon atom rearrangement. Further, the $R_s$ of the obtained rGO paper (16.87 Ω/□ from 2.11 kΩ/□) was lower than the 3D porous rGO (56.31 Ω/□ from 247.7 Ω/□) and rGO powder (1.47 kΩ/□ from 45.46 kΩ/□) after MW irradiation. The $C/O$ ratio of rGO paper (12.9) was higher than that of 3D porous rGO (7.9). Post MW irradiation, the graphene domain sizes of rGO paper, rGO powder, and 3D porous rGO were 106.49 ± 29.54 nm, 20.02 ± 1.30 nm, and 24.03 ± 4.71 nm respectively, indicating rGO paper has a better quality (Figure 13i). Post-MW irradiation, a smaller $I_D/I_G$ = 0.16 for rGO paper indicated that the decrease in defect level occurred due to the removal of partly oxygen functional groups (Figure 13h).

To prepare a hierarchically porous graphene, a small amount of graphite was used as a MW trigger to initiate the reduction of graphene oxide foam.[320] GO was entirely reduced in 5 sec under 1000 W MW irradiation. The rGO's sheet resistance steadily lowers with increasing the MW irradiation time (4 Ω/sq for 20 sec, Figure 13j).

### 2.4.2 Microwave- assisted reduction in mixed solvents

In addition to MW-assisted single-solvent synthesis, mixed solvents are often used to provide further control over the synthesis of nanostructures.[382] In mixed solvent reaction systems, the volume ratios of the different solvents can be adjusted, giving users greater control over the chemical composition, size, structure and morphology of the final product by manipulating the experimental parameters. A moderate amount of MW-absorbing ionic liquid can convert a nonpolar solvent into a MW-absorbing solvent. Leadbeater et al. showed that after MW heating for 10 sec (at 200 W), hexane with a little quantity of 1,3-dialkylimidazolium iodide could reach a temperature of 217 °C.[383] After 10 sec of MW heating at 200 W, the temperature was only 46 °C in the absence of an ionic liquid. Similar to this, as in the case of the $CHCl_3$-$CCl_4$ mixture, the heating rate of the entire combination increases when a polar solvent (with a high tan$\delta$) is coupled with a non-polar solvent.[40] In a different scenario, reactants can be extracted from one phase to the other in polar (water) and nonpolar (chloroform) solvent mixtures where the water phase can reach 100 °C and the chloroform phase can stay at a low temperature (below its BP of 61 °C).[384] The relaxation times ($\tau$) of the molecules—the amount of time needed to transition from aligned states in the presence of an $E$-field to random states in the temporary absence of an $E$-field—determine the tan$\delta$ of the solvents. A single relaxation process will typically be seen at an avg. position in a combination of liquids that are chemically identical and molecularly homogeneously mixed. But when the solvents are molecularly immiscible (such alcohols and ethers or bromides), two distinct $\tau$ are observed that are essentially the same as those of pure solvents. Since water has a tan$\delta$ of 0.123 at 2.45 GHz and 20 °C, ethanol has a greater tan$\delta$ of 0.941 at these frequencies, meaning that utilizing a combination of water and alcohols will enhance MW absorption more than using water alone.[38]

Graphene oxide (GO) was rapidly but moderately thermally reduced to graphene using MWs in a mixed solution of N,N-dimethylacetamide and water (DMAc/$H_2O$) in a few minutes.[318] The mixed solution acts as both a solvent for the resultant graphene and a medium for regulating the reactive system's temperature up to 165 ºC. The $\sigma$ of the GO and MW reduced graphene (MWG) papers was determined to be 0.015 S/m for GO paper and 200 S/m for MWG paper. The $C/O$ ratio increases from 2.09 to 5.46, showing that GO deoxygenates efficiently and graphene is formed. For the MW thermal reduction treatment, the produced GO solution in DMAc/$H_2O$ was placed in a MW oven and treated for 1-10 min at 800 W under dry nitrogen gas. Figure 13k displays the Raman spectra of GO and MW reduced graphene, indicating the presence of D and G bands in both. The G band for GO is 1598/cm, whereas the G band for decreased graphene has shifted to 1591/cm, which is near to the value of pure graphite, showing that GO was reduced during MW treatment. Simultaneously, the presence of the D band at 1354 and 1348/cm in both GO and MW treated GO suspensions indicates the sample defect as well as the size of the in-plane $sp^2$ domains.[385] The ratio, $I_D/I_G$ shifts from 0.95 to 0.96, showing a small increment in the avg. size of the $sp^2$ domain after GO reduction.[318]

The Ionic liquids (ILs) are effective MW absorbers at both low and high temperatures because of their ionic nature and strong polarizability. These organic salts, known as "green solvents," are nonflammable, thermally stable, and nonvolatile. They are ideal for replacing traditional organic solvents in chemical processes ascribed to their notable properties: a low melting point (<100 °C), low vapor pressure, low viscosity, high ionic conductivity, low interfacial tension, and adjustable polarity.[38] Tetrabutylphosphonium halides and other phosphonium ILs are ideal for dissolving S, Se, and Te at high temperatures.[38, 386] Under MW heating, the dissolved chalcogens readily react with several elemental metal powders to form micro/nanostructured metal chalcogenides. The phosphonium ILs' low binding strength enables for practically full elimination of organic residues with simple washing. Furthermore, ILs have excellent dissolving and metal cation stabilization properties, making them suitable as solvents, possible surfactants, or capping agents in inorganic synthesis.

The 1-ethyl-3-methyl imidazolium bis(trifluoromethylsulfonyl)imide [EMIm][$NTf_2$], an IL with high dielectric constant and polarizability,[52] efficiently absorbs MW energy and converts it into heat, resulting in rapid reductions of GO. Kim et al. presented a method for synthesizing reduced graphene oxide (RG-O) in 15 sec using [EMIm][$NTf_2$], in MW-assisted GO reduction.[295] The reduced graphene oxide (mRG-O) electrodes have a large $C_{sp}$ of ~135 F/g−mainly attributed to their open architecture packed with IL moieties. A supercapacitor constructed using mRG-O in an IL electrolyte operated at 3.5 V demonstrated a *power density* of 246 kW/kg and an *energy density* of 58 Wh/kg. In the study, the GO powder was combined with a predefined amount of [EMIm][NTf$_2$], and the sample in a glass vial was irradiated at 700 W (2.45 GHz) for 15 sec. MW irradiation of the GO/[EMIm][NTf$_2$] mixture for more than 30 sec leads to IL decomposition and the formation of a fuming gas. Therefore, the reaction duration was limited to < 30 sec. In the presence of IL, GO's MW reduction was accomplished in 15 sec, whereas pristine GO powders took over 60 sec under the same conditions. The C 1s XPS spectrum of mRG-O (Figure 13*l*) shows that treatment to MW radiation significantly eliminated oxygen-based functional groups as demonstrated by a significantly enhanced *C/O* ratio (from <1 to >3).

### 2.4.3 Graphene based hybrid material synthesis

Various types of composites, such as graphene derivative/metal oxides, graphene derivative /CNTs, graphene derivative/QDots, graphene derivative/metal NPs, graphene derivative/polymer, and ternary composite materials, have been fabricated using different routes, such as polyopl methods, mixed solvent methods, solvothermal (or HT), etc., all using MWs as energy sources and characterized for different applications.

Polyalcohols (polyols): ethylene glycol (EG), 1,4-butanediol, 1,3-propanediol, and glycerol ($C_3H_8O_3$) have high tan$\delta$s (*Table 1*), forming a family of strongly MW-absorbing solvents employed in MW-assisted synthesis of inorganic nanostructures. Polyols feature a variety of OH functional groups linked to their carbon backbone. Due to their high tan$\delta$ (relative to water), they are suited for MW heating. Polyols are readily miscible with water and their molecules can form intermolecular hydrogen bonds with each other and with water molecules in solution. Overheating polyol solvents increases their reducing ability, allowing them to easily reduce metal ions to metal in the zero state. With a tan$\delta$ of 1.350 at 20 °C (for 2.45 GHz) and a boiling point of ~198 °C, EG is an excellent MW-absorbing solvent and reducing agent. The polyol method makes use of polyol as both a reducing agent and a solvent.[387]

Sn-based oxide materials, including SnO$_2$, have been widely researched as potential substitutes for graphitic anode materials (theoretical capacity: 372 mA h/g), because to their high theoretical reversible specific capacity (e.g., 781 mA h/g). The main disadvantage of its industrial use is the poor cycling performance, which is due to the large change in specific volume caused by frequent charging and discharging of the battery. This leads to mechanical damage and loss of electrical contact at the anode.[388, 389] Morphological and structural modifications to SnO$_2$ anode materials are required to improve cycle performance, anodic capacity, and rate capacity. Carbon components added to SnO$_2$ improve the electrochemical performance of the resulting composite by volume expansion reduction.

The electrochemical properties of an anode composite, SnO$_2$ NPs (10-20 nm) embedded in an electrically conductive graphene matrix produced via the MW-assisted polyol technique are reported.[390] The composite was fabricated by MW reduction of poly acrylic acid (PAA) functionalized GO with a $SnO_2$ organic precursor (($C_6H_5$)$_2SnCl_2$) dispersed in EG. The PAA functionalization of GO inhibits the re-stacking of the graphene layers. In addition, PAA acts as a surfactant, promotes optimal metal dispersion and, after thermal decomposition, contributes to the formation of a carbon layer for better conductivity. In the final product, the $SnO_2$ NPs are evenly distributed throughout the rGO matrix. Graphene/$SnO_2$ nanocomposite electrodes produced with super-P carbon as a conducting additive and polyvinylidenedifluoride (PVDF) as a binder show good rate capability and cycle life testing. Beyond 140 cycles (at 500 mA/g), the electrodes maintain a steady specific capacity of ~430 mAh/g and a Coulombic efficiency near 100%. GO was synthesized using a modified Hummer's technique.[379] PAA-functionalized GO and a $SnO_2$ organic precursor dispersed in EG were exposed to 540 W MW radiation for 20 min, resulting in the partial reduction of GO and the formation of nanocomposite via a MW-assisted polyol process.[38, 391]

Similarly, for the synthesis of hybrid metal (PtRu, 2 nm)/metal oxide NPs ($SnO_2$, 2-3 nm) on graphene nanocomposites, a MW-assisted one-pot polyol synthesis strategy has been developed.[392] The combination of EG and water serves as both a solvent and a reactant. In the reaction system for the production of $SnO_2$/graphene, EG not only simply reduces graphene oxide (GO) to graphene, but it also helps to produce $SnO_2$ NPs, which is aided by the presence of a trace of water. In the PtRu/graphene nanocomposites reaction system, on the other hand, EG serves as both a reducing agent and a solvent, reducing graphene from graphene oxide and PtRu NPs from their precursors. The as-synthesized $SnO_2$/graphene hybrid composites have a substantially higher supercapacitance as compared to pure graphene, and the as-prepared PtRu/graphene has significantly superior electrocatalytic activity for methanol oxidation than commercial E-TEK PtRu/C electrocatalysts. In the experiment, 4 min of MW irradiation was delivered using a domestic MW-700 W oven. Gold NPs supported on rGO (Au/rGO) was synthesized at a MW power of 800 W within 30 min at two distinct reaction temperatures, 60 and 80 °C.[393] Using $N_2$ adsorption-desorption isotherm, a higher pore volume and *SSA* were noticed for the sample prepared at 60 °C. This suggests that the sample has more mesoporous volume and interconnections, which facilitates the mass transport process in proton exchange membrane fuel cells by allowing reactants easy access. The produced materials' stability was assessed by an accelerated electrochemical stability test with numerous cyclic voltammograms in the potential range of -0.35·0.8 and 0.35·1.0 V vs. Ag/AgCl. The *SSA* was estimated to be 244.3 $m^2$/g for Au/rGO-60 °C and 177.8 $m^2$/g for Au/rGO-80 °C.

Tin dioxide ($SnO_2$) is the most commonly investigated material for sensing applications due to its optimum carrier concentration, chemical and thermal stability, and high response to diverse gases and vapors.[394, 395] ALD-deposited *n*-type $SnO_2$ thin films (60 nm-thick with a crystalline size of ~2.9 nm) were annealed using MWs (2.45 GHz, 1 kW power, 5 min).[396] The carrier concentration was reported to be $1.50 \times 10^{20}/cm^3$, compared to $3.21 \times 10^{16}/cm^3$ in the non-annealed sample. During annealing at 500 °C, the formation of oxygen vacancies ($V_O^{**}$) outweighed the oxidation effect[397] in the $SnO_2$ lattice, resulting in more free electrons.[398] To achieve excellent chemical sensing capability, $SnO_2$ nanostructures with large *SSA*s are commonly utilized. $SnO_2$ must remain stable at high temperatures. To fully utilize its sensing capabilities and be dependable, it must be made operational at sufficiently high temperatures. To overcome this constraint, it is necessary to improve sensing capabilities by altering the microstructures of nanostructures and adding other additives. In this regard, scores of additives have been explored,[399] including graphene, which is recently garnered substantial interests in the gas sensor research community.[400, 401]

Graphene/SnO$_2$ nanocomposite, which is selective sensitivity to NO$_2$ gas was obtained by employing MWs.[402] The product had congregated structures of graphene and SnO$_2$ NPs (⌀ of 5 to 200 nm), with small secondary SnO$_x$ ($x \leq 2$) NPs settled on the surfaces. The SnO$_x$ ($x < 2$) phase resulted in a reduction in the total oxygen atomic ratio. Via the MW treatment of Gr/SnO$_2$ nanocomposites, with the graphene effectively promoting transmission of MW energy, facilitated the evaporation and redeposition of SnO$_x$ NPs. At an optimal temperature of 150 °C, a sensor response of 24.7 for 1 ppm of NO$_2$ gas was observed from the graphene/SnO$_2$ nanocomposites. The high sensitivity of the graphene/SnO$_2$ nanocomposites to NO$_2$ gas, according to the authors, is caused by the generation of SnO$_2$/SnO$_2$ homojunctions, SnO$_2$/SnO$_x$ ($x < 2$) heterojunctions, and SnO$_2$/Gr heterojunctions as a result of the formation of SnO$_x$ NPs and the SnO$_x$ phase in the matrix. Furthermore, the remarkable NO$_2$ sensing capability found can also be partially attributed to the growth of surface Sn interstitial defects. Graphene flakes were obtained using expandable graphite as a source material.[403] A combination of SnO$_2$ nanopowders with graphene (0.5 wt% graphene) was placed in an Al$_2$O$_3$ crucible and subjected to MW heating (1 kW, 2.45 GHz, 5 min, Figure 14a). According to the authors, when the MWs irradiate both SnO$_2$ powder and graphene sheets, graphene sheets reflects MWs because of its metallic nature. Because of this reflection, a large amount of MWs field exists around the source materials. As a result, a large number of MWs and an intense electric field will be applied to the nearby SnO$_2$ particles. Albeit SnO$_2$ contains no free carriers, it is feasible that oxygen vacancies or Sn interstitials may absorb kinetic energy from applied MWs and collide with atoms in metals or semiconductors, causing Joule heating. The evaporated SnO$_2$ vapors and remaining SnO$_2$ source powders deposit on the surface of the graphene sheets, resulting in the formation of SnO$_2$ NPs. Further, oxygen atoms are preferably vaporized, transforming the SnO$_2$ surface into SnO$_x$ ($1 < x < 2$).

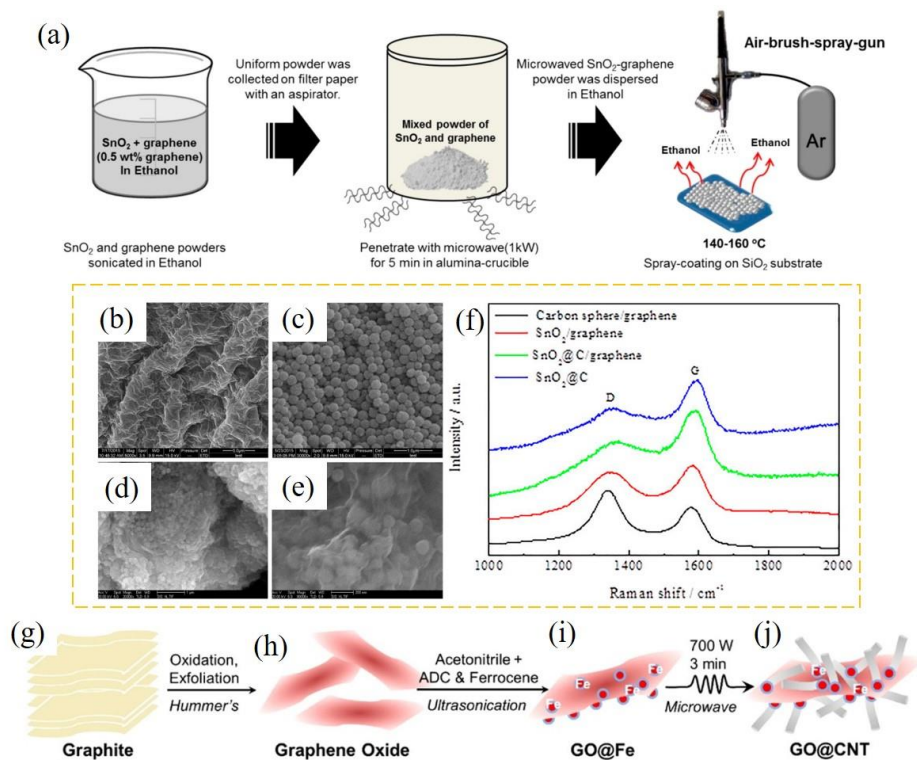

Figure 14: (a) Graphene/SnO$_2$ nanocomposite for NO$_2$ gas sensing. Illustration of Gr/SnO$_2$ nanocomposites synthesis using MWs.[402] (b-e): SnO$_2$@C/graphene ternary composite, SEM images: (b) GO, (c) SnO$_2$@C, and (d,e) SnO$_2$@C/Gr; (f) Raman spectra of carbon spheres/Gr, SnO$_2$/Gr, SnO$_2$@C, and SnO$_2$@C/Gr.[404] (g-j): GO@CNT hybrid, MW-assisted synthesis steps. (g) Pristine graphite with intrinsic stacking of layers; (h) GO NSs obtained by the modified Hummer's method; (i) After ultrasonication, dissolved ferrocene adsorbed onto GO NSs; (j) formation of GO@CNT hybrids by MW-assisted growth of CNTs at the Fe catalytic sites.[405]

A MW-assisted solvothermal technique was followed to prepare a porous SnO$_2$@C/graphene ternary composite material (Figure 14b-e).[404] The coating layer of SnO$_2$@C was found to be uniform and when tested as LIB anode material, the discharge and charge capacities of the initial cycle for SnO$_2$@C/graphene are 2210 mAh/g and 1285 mAh/g, resp. (at current density of 1000 mA/g). The Coulombic efficiency is 58.60%. Following 300 cycles, the SnO$_2$@C/graphene anode has a reversible specific capacity of 955 mAh/g. Despite a high current density of 5 A/g, the avg. reversible specific capacity is 572 mAh/g. Expanded graphite was produced from natural large graphite flakes.[406] The suspension of graphene oxide was prepared by modifying Hummer's method. The SnO$_2$@C was synthesized using SnCl$_4$ solution and carbon spheres via a HT treatment at 180 °C for 8 h. To prepare SnO$_2$@C/Gr composite, the prepared SnO$_2$@C material was mixed with GO aqueous solution and subjected to MW-assisted HT reaction (1200 W, 180° C, 60 min, 2 MPa). Figure 14f shows Raman spectra of carbon spheres/Gr, SnO$_2$/Gr, SnO$_2$@C, and SnO$_2$@C/Gr. The $I_D/I_G$ for SnO$_2$@C is 0.62 and $I_D/I_G$ = 0.83 for the SnO$_2$@C/Gr, showing that the disordered structure of the SnO$_2$@C/Gr is clearly increased.

SnO$_2$ crystals are produced via a pulse MW-assisted process and then uniformly implanted into rGO sheets to form SnO$_2$/rGO composites.[407] The $C_{sp}$ of electrochemical capacitors built with SnO$_2$/rGO composites can reach 348 F/g (at 50 mA/g), representing a 98% increase over that of a fresh rGO electrode. The SnO$_2$ crystals not only serve as spacers, providing more active sites on the rGO surface, but they also increase the proportion of hydrophilic surface available for the formation of the electric double layer, resulting in a higher capacitance. The SnO$_2$/rGO electrode maintains a maximum *energy density* of 32.2 Wh/kg and a *power density* of 1000 W/kg. The SnO$_2$/graphene composite is formed by preparing (i) GO sheets from a modified Hummers' method, (ii) SnO$_2$ nanocrystals using a pulse MW heating method (720 W, ~80 °C), and (iii) SnO$_2$/reduced GO (rGO) composites through a homogenizing process followed by MW reduction (450 °C, horizontal furnace under N$_2$ for 1 h). The SnO$_2$/rGO layer was deposited on carbon paper using a drop-coating technique, forming a flexible composite electrode.

With its wide oxygen functionalities, graphene oxide (GO), is an important study component in polymer composites. Proper exfoliation and controlled dispersion are crucial for transferring properties from GO NSs to host polymers. This is influenced by the competing evolution of GO-GO intersheet attractions and GO-polymer interfacial interactions. Figure 14g-j illustrates a MW-assisted synthetic technique to allow direct in situ development of CNTs on GO templates, resulting in minute-level 3D hierarchical nanohybrids (GO@CNT).[405] The nanohybrids were uniformly disseminated and appropriately exfoliated in racemic polylactide (PLA) blends in contrast to GO's microscale local aggregation caused by significant intersheet attractions. Although only present in trace amounts (0.05 wt%), GO@CNT significantly outperformed GO in terms of enhancing PLA stereocomplexation, yielding overall improvements in mechanical properties (young's modulus of ~2 GPa and tensile strength of ~70 MPa) and gas barrier performance (low oxygen permeability coefficient of 0.25 × 10$^{-15}$ cm$^3$ cm cm$^{-2}$ s$^{-1}$ Pa$^{-1}$) for stereocomplexed PLAs. During the preparation of PLA-based composites, nanofillers (GO and GO@CNT) were incorporated into racemic PLA via solution coagulation. The loadings of both GO and GO@CNT were set to 0.05 wt%. To synthesize GO@CNT hybrids, acetonitrile, azodicarbonamide (ADC), and ferrocene (C$_{10}$H$_{10}$Fe) were utilized.[408] With the help of MW irradiation, the ferrocene entities were efficiently separated and decomposed into Fe particles, which then functioned as catalytic sites to initiate CNT development—a mechanism akin to the results by Kumar et al.[409]

L-Ascorbic acid 6-palmitate (LAP) is used as the reducing agent in a MW reduction of solution-exfoliated graphene oxide to create carbon dot-modified reduced graphene oxides (LAP-rGO-CDs).[410] LAP-rGO-CDs have increased interlayer spacing and a quicker ion transfer rate due to the addition of carbon dots. LAP-rGO-CDs, used as a K-ion battery electrode, demonstrated a high specific capacity of 299 mAh/g at 1 A/g. The assembled LAP-rGO-CDs//AC full carbon-based K-ion capacitor cell (KIC) exhibits a maximum *energy density* of 119 Wh/kg and a *power density* of 5352 W/kg. LAP-rGO-CDs has a higher $I_D/I_G$ value (1.17) compared to rGO reduced by L-ascorbic acid without CDs (0.98) and GO (0.95), indicating an increase in disorder and structural defects. Kang et al. reported a composite based on MnO$_2$ and activated FLG (a-FLG).[411] For preparing a-FLG material, a KOH–assisted MW irradiation technique was followed. Using KOH as a chemical etchant effeciently activated the surface pores of graphene during MW irradiation-induced exfoliation. To coat $MnO_2$ on surface of a-FLG, a-FLG was dispersed in 0.1 M $NaMnO_4$/0.1 M $Na_2SO_4$ (neutral pH) solution and agitated at RT. The $MnO_2$/a-FLG shows good long-term stability, a $C_{sp}$ (265 F/g), and a high rate capability. Asymmetric supercapacitor devices using a-FLG as the negative electrode and $MnO_2$/a-FLG as the +ve electrode exhibit high *energy density* (40.8 Wh/kg) and *power density* (16.3 kW/kg), as well as 96% capacitance retention over 10000 cycles.

The electrochemical properties of $Fe_3O_4$(0.4 M)/rGO composites obtained via a MW-assisted reduction method with iron(II, III) oxide ($Fe_3O_4$) supported on the surface of graphene show that $Fe_3O_4$ plays an important role as a supplement to increase the electrical double-layer effect of graphene by adequately checking graphene re-agglomeration and also achieves a pseudocapacitance effect via a faradaic redox reaction.[412] The $Fe_3O_4$ (0.4 M)/rGO composite (where 0.4 M is the amount of $FeCl_3 \cdot 6H_2O$ precursor) exhibits a $C_{sp}$ of ~972 F/g at 1 A/g and a high capacitance of 265 F/g at 5 A/g. Furthermore, $Fe_3O_4$ (0.4 M)/rGO composites demonstrated extremely stable cycle performance, with 70% retention after 10000 cycles. To prepare the composite, graphite oxide in ethanol was mixed with different amounts of $FeCl_3 \cdot 6H_2O$ precursors, agitated at 90 °C for 1 h and 30 min, and heated in a domestic MW oven (800 W, several minutes).

Using MWs, NPs of binary $Mn_3O_4$-$Fe_2O_3$/$Fe_3O_4$ NPs (<100 nm) were uniformly decorated on the surfaces of rGO NSs to yield $Mn_3O_4$-$Fe_2O_3$/$Fe_3O_4$@rGO ternary hybrid as electrode materials for supercapacitors application.[413] In-situ exfoliation and reduction of graphite oxide to rGO NSs contributed to the surface decoration of the binary metal oxide (*Figure 15*a). The ternary hybrid in 1 M KOH electrolyte demonstrated a $C_{sp}$ of 590.7 F/g (0.5 mV/s scan rate) and retained 64.5% of original capacitance after 1000 continuous cycles (50 mV/s scan rate).

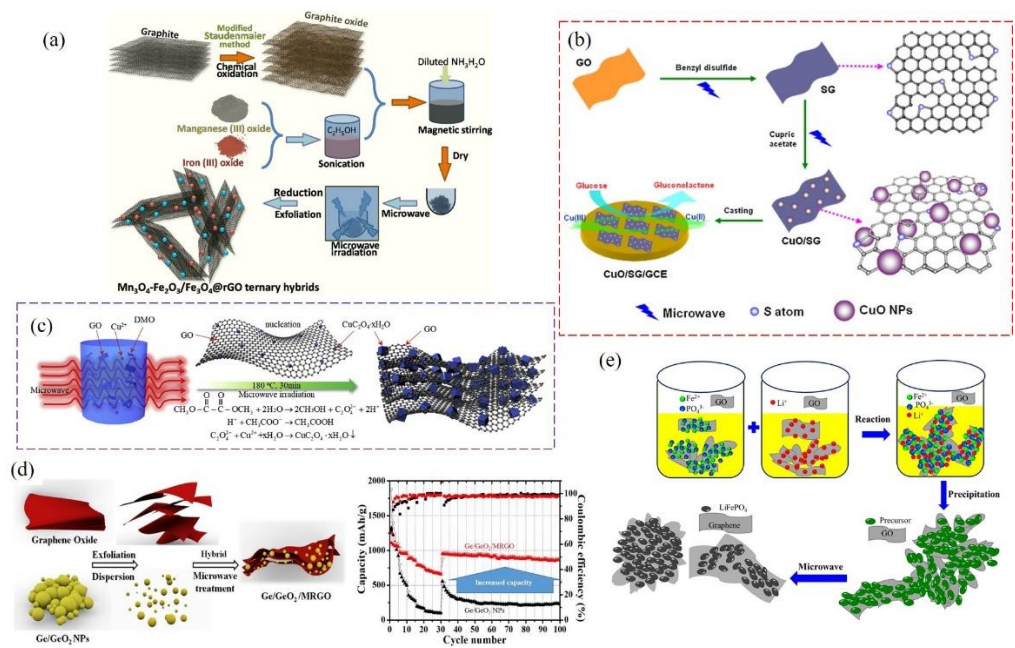

Figure 15: (a) Synthesis process of $Mn_3O_4$-$Fe_2O_3$/$Fe_3O_4$@rGO ternary hybrids.[413] (b) Illustration of the synthesis of CuO/SG and CuO/SG/GCE glucose.[414] (c) Synthesis steps to prepare $CuC_2O_4 \cdot xH_2O$/GO through MW- solvothermal reactions.[415] (d) Pictorial illustration of the synthesis process of Ge/$GeO_2$/MRGO & cycling performance and rate capability.[416] (e) Depiction of the assembly process of $LiFePO_4$/graphene composites.[417]

CuO NPs supported on S-doped graphene (CuO/SG) was synthesized by a MW-assisted approach (800 W, 10 min), where CuO NPs of ~13 nm were well-distributed and securely anchored to SG.[414] The CuO/SG/GCE-based nonenzymatic glucose detection sensor (Figure 15b) has a 2 sec response time, a linear range of 0.1-10.5 mM, a sensitivity of 1298.6 µA·mM$^{-1}$/cm$^2$, and a low detection limit of 80 nM. The working electrode was a pristine or modified glassy carbon electrode (GCE, ⌀: 0.3 mm). SG was prepared utilizing a MW-assisted solvothermal technique with GO as the graphene precursor and benzyl disulfide as the S dopant (Figure 15b). CuO/SG was produced using a modified precipitation process.[418, 419] The *C/O* atomic ratio in SG is 9.2 > 1.07 in GO, indicating that the majority of oxygen-containing functional groups are effectively eliminated during the S-doping process.

The MW-assisted mixed solvent approach was also used to generate graphene derivative/metal oxide hybrids suitable for energy storage as shown in the following studies.
Using a mixed solvent of EG and DI water with equal volume ratio, vanadium pentoxide ($V_2O_5$) NPs (20 nm) were grown on graphene.[420] The resultant $V_2O_5$/graphene composites were used to assemble a symmetrical supercapacitors, which exhibit $C_{sp}$ of 673.2 F/g (at 1 A/g) and 474.6 F/g (at 10 A/g), and 96.8% of capacitance retention after 10000 cycles at 1 A/g. Further, the fabricated devices showed better *energy density* and *power density* characteristics (46.8 Wh/kg at 499.4 W/kg and 32.9 Wh/kg at 4746.0 W/kg). The large theoretical capacity of $Co_3O_4$ (890 mAh/g) makes it a desirable anode material for next-generation LIBs.[421] In another case of employing mixed solvents, $Co_3O_4$ QDs/graphene composite was synthesized via MW irradiation.[422] A typical synthesis involved dispersing graphene NSs in a solvent mixture of DI water and ethyl alcohol. Then, $NH_3 \cdot H_2O$ and Co(CH$_3$COO)$_2$·4H$_2$O were added. The suspension was microwaved with reflux and then heated at 81 °C for 5 min using 500 W of MW irradiation. The Co$_3$O$_4$ QDs/graphene composites improve cycling performance (1785 mAh/g at 0.1 C after 90 cycles) and rate capability (485 mAh/g at 5 C) in LIB. Similarly, CuC$_2$O$_4$·xH$_2$O nanocubes (~0.8-1.2 μm size)/graphene composite was fabricated by a MW-assisted solvothermal process at 200 °C for 30 min using EG in distilled water with a ratio of 1:1.[415] As an anode material for LIB, a reversible discharge/charge capacity of 1043/1013 mAh/g after 100 cycles was observed. During the initial cycles, the capacity was noticed to be constantly increasing. Discharge/charge capacities of 952/949 (at 0.5 A/g), 719/720 (at 1 A/g), and 535/533 mAh/g (at 2.0 A/g) were observed. To prepare the composite, transparent graphene solution, chalcanthite, dimethyl oxalate lithium acetate and EG in distilled water with a ratio of 1:1 were used (Figure 15c). First, the attraction of the functional groups on the graphene surface causes Cu$^{2+}$ ions to disperse equally throughout the surface. When dimethyl oxalate decomposes, CuC$_2$O$_4$·xH$_2$O devitrifies. Within 30 min, the reaction was concluded to prevent CuC$_2$O$_4$·xH$_2$O from forming large conglomerates that can inhibit Li$^+$ formation and electron transport during electrochemical reactions.

A hybrid of Ge/GeO$_2$ and RGO composite was produced by reassembling fully exfoliated graphene oxide (GO) NSs with Ge/GeO$_2$ NPs, followed by MW treatment (Figure 15d).[416] To obtain the composite, the dry precursor, Ge/GeO$_2$/GO, was irradiated with MWs for 10 minutes. The $I_D$/$I_G$ ratio of Ge/GeO$_2$/MRGO was comparable to that of bare RGO, indicating that the GO in the as-prepared sample was completely reduced to RGO via MW treatment. When compared to the bare Ge/GeO$_2$ NPs anode, the current Ge/GeO$_2$/MRGO demonstrated higher electrochemical effectiveness with regard to rate capabilities, capacity retention, and cycle (Figure 15d).

With a high theoretical capacity of 170 mAh/g and a flat voltage plateau (3.45 V vs. Li/Li$^+$), olivine-phase lithium iron phosphate (LiFePO$_4$) is one of the well-studied cathode materials.[423] But, LiFePO$_4$ material has a lower $\sigma$ of ~10$^{-9}$-10$^{-10}$ S/cm and slow one-dimensional Li ion diffusion, which are not favorable to prepare high performance LIBs. A solid-state MW heating approach is used to produce micro/nanoscale LiFePO$_4$/graphene composites (1500 W, 10 min).[417] It is expected that the LiFePO$_4$ NPs, due to their nanosize can decrease the diffusion path of Li ions, increasing the Li-ion diffusion rate, whilst graphene sheets may offer a fast transport path for electrons, elevating the material's $\sigma$. The discharge capacities are reported to be 166.3, 156.1, 143.0, 132.4, and 120.9 mAh/g at 0.1, 1, 3, 5, and 10 C, respectively. In this study, LiFePO$_4$/C (sucrose as the carbon source) composites were also prepared and compared to LiFePO$_4$/graphene's properties. The $I_D$/$I_G$ ratio in LiFePO$_4$/graphene composites is 1.18, whereas in LiFePO$_4$/C composites it is 1.43, showing that graphene has a greater degree of graphitization than carbon derived from decomposed sucrose. The synthesis process of the LiFePO$_4$/graphene composites is shown in Figure 15e.

Zinc oxide (ZnO), a wide bandgap semiconductor, is considered a suitable candidate for supercapacitor applications due to its environmentally friendly properties.[424, 425] Due to its low $\sigma$ and accessible surface areas, ZnO typically has a low $C_{sp}$. Thus, ZnO is used as an addition to nanocarbon-based materials, particularly graphene.[426, 427] A one-step solid state MW irradiation synthesis of MW reduction graphene oxide (MRGO)/ZnO NPs (50-100 nm) for supercapacitor application is presented.[428] The composite-based supercapacitor electrode has a $C_{sp}$ of 201 F/g at 1 A/g and a rate

capability− far exceeding that of pure MRGO (77 F/g) and ZnO (5 F/g) at the same current density. Further, the $C_{sp}$ of the MRGO/ZnO electrode has excellent cycling stability (93% capacitance retention after 3000 cycles). In addition, MRGO/ZnO has an *energy density* of 4.37 Wh/kg; a *power density* of 8.33 kW/kg at 1 A/g −greater than that of pure MRGO electrode; and a large *SSA* of 109.5 m$^2$/g as compared to 9.2 m$^2$/g of pure ZnO and the published figure of G-ZnO (7.9 m$^2$/g).[429] To obtain MRGO/ZnO, GO and zinc acetate mixture was ground, desiccated and heated in a domestic MW oven for 5 min at 1000 W. Under the requisite wt. ratio of GO to $Zn(CH_3COO)_2·2H_2O$ of 2:1, the production of graphene/ZnO exhibited favorable repeatability and electrochemical performance. The $I_d/I_g$ ratio of MRGO/ZnO (0.95) > MRGO (0.38) because of the increased disorder of sp$^2$ caused by the existence of ZnO in the composite. The synthesis and characteristics of various nanocomposites have also been reported including MW-treated carbon cloths coated with rGO or $MnO_2$,[430] rGO/$TiO_2$ nanocomposite,[431] ZnO/G nanocomposite,[432] rGO/$Mn_3O_4$ (GM) composites.[433]

CNTs (20-50 nm) were grown on graphene NSs to prepare unique 3D carbon nanostructures utilizing defect-engineered processes involving an ionic liquid (carbon source for CNT growth), a palladium (Pd) catalyst, and MW radiation.[434] The one-pot approach involved three steps: creating defects on graphene nanoplatelets using MWs, attaching Pd NPs on these defects, and then growing CNTs using an ionic liquid. In a 1 M KOH electrolyte, capacitance measurement at 10 mV/s rate showed that for 3D G-CNT-Pd, it increased by 46% after 600 cycles as compared to the initial capacitance. In this process, mixture of expandable graphene platelets, ionic liquid EMIM $BF_4$, and a palladium (II) acetate catalytic precursor was exposed to MW irradiation (700 W, 10 min), yielding a fluffy powder of 3D G-CNT-Pd nanohybrid (*Figure 16*a-b).

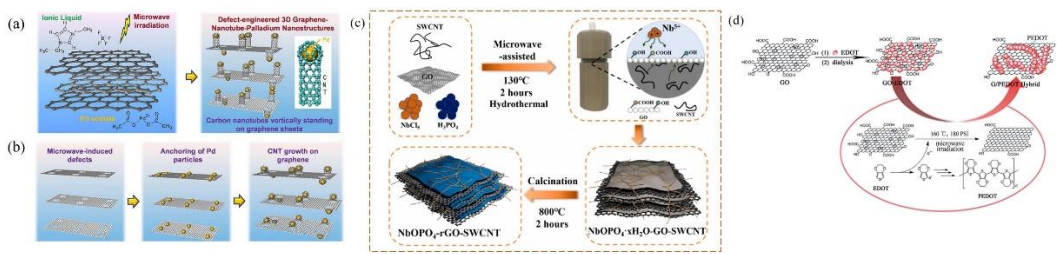

*Figure 16:* One-pot MW synthesis. 3D carbon hybrid nanostructures with vertically grown CNTs on graphene sheets. (a) scheme and (b) formation mechanism.[434] (c) illustration of NbOPO$_4$-rGO-SWCNT hybrid synthesis via a MW-assisted HT technique.[435] (d) The G/PEDOT hybrid synthesis illustration.[436]

Tao et al. presented a MW-assisted HT technique to produce NbOPO$_4$·$x$H$_2$O NSs that can grow in-situ on SWCNTs (8-250 nm) and rGO frameworks.[435] The rGO/SWCNT framework forms a hierarchically high-conductive network to manage the nucleation and growth of NbOPO$_4$ NSs (50 nm-thick). As a SIB anode material, because of the crystallographic structure of NbOPO$_4$ and enhanced $\sigma$ of rGO/SWCNT framework, NbOPO$_4$/rGO/SWCNT electrode delivers a reversible specific capacity (323.2 mAh/g at 0.1 A/g) with a capacity retention (114.6 mAh/g after 1700 cycles at 0.5 A/g, avg. capacity decay of 0.073 mAh/g/cycle) and rate capability of 147.9 mAh/g at 5 A/g compared to NbOPO$_4$ (143.1 mAh/g at 0.1 A/g and 43.5 mAh/g at 5 A/g). In this procedure, SWCNT that had undergone ultrasonic treatment was dissolved in DI water and subsequently adhered to a GO's surface (*Figure 16*c). The GO/SWCNT framework's carboxyl and hydroxyl functional groups act as an active site for $Nb^{5+}$ nucleation, interacting with the ion. The $NbOPO_4 \cdot xH_2O$ NSs can develop in-situ on GO/SWCNT surfaces using MW-assisted nucleation during a 2 h HT procedure at 130 °C. The powder was calcined at 800 °C in a $N_2$ environment to eliminate water.

     Polymer/nanocarbon composites have been tested as electrode materials for energy storage applications because of their ability to bear the high pressures caused by electrode volume expansion/contraction during charge and discharge cycles. Polymers provide additional advantages such as excellent conductivity and ease of processing.[437] Graphene/polypyrrole nanocomposite was obtained utilizing a sacrificial template polymerization approach.[438] Graphene/PPy nanocomposites with various combinations were created by altering the mass ratio of graphene and PPy, and each composite was then treated with MWs for 10 seconds. It was observed that the composite volume expanded when 320 W MW irradiation was applied. The MW-treated composite material had a BET area of 34.13 $m^2$/g. When tested as a material for supecapacitor, a maximum $C_{sp}$ of 240.4 F/g at 10 mV/s was observed for the sample with a graphene oxide to PPy mass ratio of 5:1.75. In another study, a simple two-step process was used to prepared 3D networked polypyrrole (PPy) nanotube/N-doped graphene (NDG).[439] The method involved in-situ polymerization using a $MoO_3$ template and MW-assisted HT method. The PPy nanotube production and N doping of reduced graphene oxide occur concurrently during the MW exposure. The PPy nanotube/NDG's suitable structure and good $\sigma$ result in a charge transfer resistance of 0.1 Ω, a low equivalent series resistance of 1.7 Ω, a $C_{sp}$ of 292 F/g at 5 mV/s, and a pore resistance of 0.1 Ω. $MoS_2$-decorated core-shelled $MoO_3$/PPy was also prepared using a MW-assisted HT process. This composite has exhibited a low charge transfer resistance of 0.5 Ω and a $C_{sp}$ of 527 F/g at 5 mV/s. An asymmetric supercapacitor with $MoS_2/MoO_3$/PPy as +ve electrode and a PPy nanotube/NDG as -ve electrode was developed, which achieves an *energy density* of 43.2 Wh/kg at *power density* of 674 W/kg and a capacitive retention of 126% after 5000 cycles.

     PEDOT, or poly(3,4-ethylenedioxythiophene) has garnered a lot of interest because of its superior environmental stability, $\sigma$, and transparency[437, 440, 441] and is being widely investigated for supercapacitor applications. By employing graphene oxide as an oxidant during the in situ polymerization of 3,4-ethylenedioxythiophene under MW heating, a graphene/poly(3,4-ethylenedioxythiophene) (G/PEDOT) hybrid was created.[436] During the charge-discharge process, the G/PEDOT hybrid supercapacitor electrode showed $C_{sp}$ of 270 F/g at 1 A/g of current density in 1M $H_2SO_4$ electrolyte solution. This value is higher than the published values under the same conditions for graphene (245 F/g),[442] PEDOT (130 F/g),[443] and G/PEDOT.[444, 445] The increased $C_{sp}$ of the G/PEDOT hybrid can be due to the adherence of the PEDOT on the surface of the graphene sheets− significantly reducing the van der Waals force among the graphene sheets and consequently blocks them from aggregating. Furthermore, the increased distance between graphene sheets increases the contact surface with the electrolyte solution, which improves ion transport in the hybrid (*Figure 16*d). The material showed good cycling stability (93% retention after 10000 cycles at a current density of 5 A/g). In addition, the *energy density* reached 34 Wh/kg at a *power density* of 25 kW/kg. The reaction was held at 160 ºC, 200 W power, and 180 psi for 20 min in a commercial MW oven. Table *4* lists the

distinct nanocarbon material types prepared via MW-assisted treatments, synthesis conditions, their attributes, and the outcomes of various applications with their use.

Table 4: Synthesis conditions and attributes of distinct nanocarbon materials generated by MW-assisted treatments, as well as outcome of varied applications with their use. Abbreviations: $\sigma$ –electrical conductivity; MW –microwave; HT –hydrothermal; ⌀ –diameter; *SSA* –specific surface area; NPs –nanoparticles.; ExG –Expanded graphite; $R_s$ –sheet resistance.

| Obtained material type | Materials & reagents used | Synthesis conditions, time, MW power | Dimension & other properties | Application tested & relevant parameters | Ref. 444 |
|---|---|---|---|---|---|
| CNTs | Granulated polystyrene & Ni NPs | 600W under $N_2$; 800 ºC for 10 min | ⌀: 25-100 nm; length: 2 μm | - | 94 |
| CNTs | Graphite, ferrocene & C fiber (length: 8–8.5 mm & wt.: ~2.5 mg) | 5 sec; power at different %s of 1800 W | Lengths & ⌀: 1.4 ± 0.7 μm & 17 ± 10 nm; $I_D/I_G$ = 0.8 | - | 193 |
| CNTs@ Graphene oxide (GO) | GO, acetonitrile, azodicarbonamide & ferrocene | 700 W (150−250 °C), 3 min | ⌀ of CNT: 10-20 nm & length: ~100 nm to ~1 μm; Crystal plane distance in GO: 0.36 nm | Gas barrier performance (oxygen permeability coefficient: $0.25 \times 10^{-15}$ $cm^3$ cm $cm^{-2}$ $s^{-1}$ $Pa^{-1}$ ) when racemic polylactide blends added later | 405 |
| Carbon platelets & μm-sized carbon spheres | Graphene oxide (GO, heating element) & glucose (1:800 wt. ratio) | MW-assisted HT carbonization; 1250 W, 1 min | ⌀ of carbon sphere: ~1-4 μm; Low GO loading yielded, carbon platelets (tens of nm thick); at high mass loading, free-standing carbon monoliths; *C/O* :4 | - | 242 |
| Carbon nanospheres with ZnO NPs inside | C microparticles mixed with 10 wt% ZnO microparticles | MW-assisted HT reaction; 340 W; 3.5 min | Carbon nanosphere avg. ⌀: < 35 nm with ZnO NPs (4 nm) inside them | - | 251 |
| rGO/ZnO | GO & $Zn(CH_3COO)_2$ $\cdot 2H_2O$ (2:1 wt. ratio) | 1000 W, 5 min | ⌀ of ZnO NPs in rGO/ZnO: ~20 nm; $I_D/I_G$: 0.95 | Supercapacitor electrode: $C_{sp}$ of 201 F/g in 1 M $Na_2SO_4$ (at 1 A/g) | 428 |
| rGO/ZnO | Zinc acetate + graphene oxide + NaOH | 450 W, 20 min | $I_D/I_G$: 0.90; ⌀ of ZnO ≈ 10–20 nm | Glucose sensor: detection limit of 0.2 nM; sensitivity of 39.78 mA $cm^{-2}$ m/M; Supercapacitor: specific capacitance of 635 F/g | 446 |

| N-doped CQDs (N-CQDs) | Xylan | 200 W; 200 °C for 10 min | Avg. $\varnothing$: 7.89 nm; $I_D/I_G$: ~0.4 | Detection of tetracycline: detection limit of 6.49 nM | 245 |
|---|---|---|---|---|---|
| Graphene/SnO$_2$ nanocomposites | SnO$_2$ nanopowders (< 100 nm) & expandable graphite | 5 min; 1 kW | $\varnothing$ of SnO$_2$ particle: 5-200 nm; Interplanar distance in graphene sheet: 0.21 nm | Sensor response to NO$_2$ gas: 24.7 for 1 ppm of | 402 |
| Graphene NSs/SnO$_2$ composite | Poly acrylic acid functionalized graphene oxide & SnO$_2$ in ethylene glycol | 20 min & 540W | $\varnothing$ of SnO$_2$ particle: 10-20 nm | Anode material for LIBs Super-P carbon as conducting additive & polyvinylidenedifluoride as binder; Specific capacity: about 430 mAh/g; Coulombic efficiency: 100% | 390 |
| Porous graphene material | Graphene oxide | 1 min in Ar; 250 W | Single layer thickness: 0.857 nm; $C/O$: 11.36; $I_D/I_G$: 0.8205 | $SSA$: 461.6 m$^2$/g; H$_2$ & CO$_2$ adsorption: H$_2$ uptake of 0.52 wt% (77 K & 1 atm); CO$_2$ adsorption (273 K & 1 atm) was 7.1 wt% | 447 |
| SnO$_2$@C/graphene ternary composite | Graphite Oxide & Graphene Oxide | MW-assisted HT; 1200 W, 180° C, 2.0 MPa & 60 min | $\varnothing$ of SnO$_2$@C: 100-120 nm | Energy storage Coulombic efficiency: 58.60%; Discharge & charge capacities of initial cycle: 2210 mAh/g & 1285 mAh/g, resp., at current density of 1000 mA/g; Reversible specific capacity of anode: 955 mAh/g after 300 cycles | 404 |
| MoTe$_2$/graphene | Mo(CO)$_6$, Te-powder, graphene | 90 sec, 1250 W | - | Supercapacitive behavior: symmetric MoTe$_2$/graphene//MoTe$_2$/graphene configuration had cyclic stability to 10000 cycles; max. *energy density* of 43.2 Wh/kg at high *power density* of 3000 W/kg in 1 M Na$_2$SO$_4$ electrolyte | 448 |
| NiS/rGO hybrid | Graphene oxide (GO); nickel nitrate & thiourea | MW-assisted HT; 700 W for 4 min for MW & 130 °C for HT | NiS NPs size: ~80 nm; $I_D/I_G$ = 1.06-1.31 for different amounts of GO | Electrode material: $C_{sp}$ ~1745.67 F/g (1 A/g); Symmetric solid-state supercapacitor: *energy density* of 7.1 Wh/kg & *power density* of 1836 W/kg | 449 |
| rGO/NiCo$_2$S$_4$ | Graphene oxide (GO), ethylene glycol, Ni(NO$_3$)$_2$·6H$_2$O, Co(NO$_3$)$_2$·6H$_2$O, CS$_2$ | At 150 °C, 15 min, 1000 W | ⌀ of NiCo$_2$S$_4$ nanoneedle: several nm | As electrode: $C_{sp}$ (1320 F/g at 1.5 A/g), high rate capability (70.2% retention from 1 to 9 A/g), & cycling stability (≥ 96% retention after 2000 cycles) | 450 |
| Graphene/CuO | GO, cupric acetate, DMF, | 10 min at 800 W | CuO monoclinic NPs of 13 nm lodged on graphene | Glucose detection: 80 nM detection limit, Sensitivity of 1298.6 µA · mM$^{-1}$ · cm$^{-2}$ | 414 |
| MWCNT/RuO$_2$ | RuCl$_3$, MWCNT (Length: tens of nm; 10–50 nm in ⌀) | 30 sec at 700 W | Avg. ⌀ of RuO$_2$ NPs: 1-2 nm; Ru/O ratio = 0.5 | Tested for $C_{sp}$: 493.9 F/g | 224 |
| rGO/Fe$_3$O$_4$ | Graphite oxide, FeCl$_3$·6H$_2$O, C$_2$H$_5$OH, NH$_3$ | 700 W, 30 sec | $d$-spacing: 0.25 nm & 0.34 nm; $I_D/I_G$: 0.22 for rGO nanosheets, 0.61 for rGO/Fe$_3$O$_4$ | Specific capacitance of 471 F/g at 10 mV/s; 96.5% of capacitance retention post 1000 cycles | 451 |
| MoS$_2$/rGO (10%) | graphite powder, MoCl$_5$, thiourea | 700 W, 35, 45, and 45 sec | $I_D/I_G$: 1.033; µm size particles | Supercapacitor: specific capacity of 1023 F/g at 1 A/g; 82% retention after 1000 cycles | 452 |

## 3 Summary, challenges, and outlook

In this review, recent developments in the fabrication of nanostructured carbon materials and their modifications using MW irradiation as an energy source were examined. MW-assisted techniques have the advantages of being rapid and energy efficient, with certain reactions taking place in the air (MW arcing approaches). The MW arc technique has good selectivity for heating targets and a fast reaction rate. This technique is used to produce various nanocarbons in liquid and solid states. Many carbonaceous materials, metals, and oxides absorb MWs well, therefore they can be employed as internal or exterior susceptors, making MW energy delivery effective. It is very clear that one component of the reaction mixture—either the solvent, reagents, a susceptor, or a catalyst—must be a strong MW absorbent (high tan$\delta$). If the specimen holder itself is made of a MW absorbing material (e.g., CuO, SiC, or solid carbon), neither an internal susceptor nor a high tan$\delta$ solvent is required. Despite the well-known advantages of MWs in materials processing, several key limitations of MW-assisted nanocarbon processing make the technology uncompetitive for practical application compared to many conventional methods. Recent trends in this research discipline suggest that MW application in nanocarbon processing is promising provided its shortcomings are remedied.

To enable uniform heating, the addition of carbon absorbers (internal susceptors) and reactants is critical. Conversely, the separation of the carbon absorber and the nanocarbon end product may require additional time-consuming tasks. Further, the end product's properties may not meet expectations. The thermal uniformity of MW absorbers is determined by the frequency of the MWs as well as the depth of penetration into the interior. Using a variable-frequency MW waveguide is an effective approach for increasing penetration depth, resulting in uniform heating and higher product quality. The current obstacles to MW-initiated nanocarbonization are the nanocarbons' producing scale is insufficient, and their size and shape consistency is still rather poor.

Carbon dots with narrow emission, a high QY, a long lifetime, and distinct absorption are necessary. These qualities are dependent on the reaction conditions and carbon sources, making it difficult to optimize all factors for a single application, especially using MWs, as multiple synthesis variables must be optimized. Regardless, the MW-assisted liquid-state synthesis of quantum dots and NPs of a certain shape and size appears promising. This is quite similar to traditional solution-based synthesis, with the exception that MW energy delivery is different. This technique retains some of the disadvantages of solution-based synthesis, particularly extraction and purification. A majority of reports on MW-assisted synthesis of carbon dots show that the obtained carbon dots emit only in the blue-green wavelength range after exposure to UV excitation, and long-wavelength emission (red) is rarely reported. Therefore, it is very important to study effect of synthesis conditions on carbon dots' fluorescence emission properties. The size of the carbon dot may be different at the early stages of the exposure as compared to lengthy exposure for the same MW operating power. X-ray or TEM characterization may provide some clues on the evaluation of carbon dot sizes during different periods of MW exposure. In general, there is a need to exploit the potential of carbon dots as temperature, pH and heavy metal detection sensors. One example is the modification of surface functions to achieve selectivity and sensitivity to different analytes. Possible applications of carbon dot-based nanosensors will be the simultaneous detection of multiple analytes and the quantitative determination of all analytes without influencing each other. Carbon microspheres may be prepared quickly using MW-assisted HT methods.

By applying MW heating, the possibility of introducing defects into the sidewalls of CNTs during acid reflux cleaning can be reduced. It is shown that metals can be successfully removed from CNTs by first selectively oxidizing the amorphous carbon and then refluxing the CNTs in concentrated $HNO_3$. In the MW synthesis of CNTs and CNFs from biomass feedstock, cellulose is recognized to be the component responsible for CNT production. It would be beneficial to directly grow CNT or CNF arrays on substrates and employ the combination for various applications where the nancarbon's quality is not a major concern. The ratio of carbon source, other precursors, and MW operating power must optimized to get higher CNT yield. But, post-synthesis processes are required to remove metal wires and metal particles using hazardous chemicals such as aqua regia, which necessitates extra stages using water.

MW exfoliation of graphene oxide offers several advantages including purity, high product yield and fast processing time. However, the $\sigma$ of the rGO is low as compared to graphite. The metal oxide nanoparticles can offer high capacity with higher potential (2-3 times that of graphite) due to their size, structure and crystallinity. Because of the volume expansion (~300%) of the oxide particles during charge/discharge, poor cycling performance occurs, resulting in the loss of electrical connection between active material and the current collectors. Graphene/metal oxides nanocomposites offer the combined salient features of both constituents towards materials for electrochemical energy storage. Graphene facilitates chemical functionality to pave the way for easy processing of oxide component. The metal oxide particle deposited on GNSs prevents GNS agglomeration and restacking o the sheets (which may lead to low energy density), in addition to providing specific surface area for GNSs. As a support for the metal oxide nanoparticle, graphene can assist in nucleation, growth of metal oxide nanoparticle and graphene related microstructures with required functionalities−increasing the electronic conductivity and ion transport due to conducive interfaces. In graphene/metal oxide composites, the elastic buffer helps to alleviate the problems associated with the volume expansion of the metal oxide particles during the Li insertion and extraction process. Therefore, it is very important to study the microstructural properties of the composite and especially the metal particle size after several cyclic stability tests using TEM images. The MW method is ideal for preparing 3D structures. It was found that MW irradiation time and operating power are important and practical parameters for adjusting the $C/O$ ratio of graphene derivatives. Since the review discussed the use of graphene sheets as susceptors for the reduction of graphene oxide, the resulting graphene derivative may not exhibit the homogeneous conductivity property. Such techniques may need to be improved to obtain large-scale graphene derivatives with uniform properties. The local superheating effects created by hotspots and arc plasmas that enable or accelerate certain chemical reactions need to be thoroughly investigated. For graphene, the exact relationship between the amount of Li storage with respect to the number of layers, the number of defects, the dimensions of the graphene, and oxygen-containing functional groups is not well understood.

As mentioned earlier, ILs are ideal for dissolving some metals at high temperatures. It will be interesting to see how such mixtures can help in the deposition of metal nanoparticles, especially semiconductors on graphene or other nanocarbons using MWs, and what properties such composites will have. The MW-assisted synthesis of various nanocarbon-based composites uses polyols, a family of strong MW-absorbing solvents. The polyol method makes use of polyol as both a reducing agent and a solvent. The disadvantage is that polyol techniques make it difficult to control reactions in large-scale sample preparation; batch variation in morphology and characteristics is possible. Furthermore, separation and purification of the final product are difficult. Similarly, mixed solvent methods have been used in the thermal reduction of graphene oxide and the fabrication of different hybrids. Mixed solvent technique expands solvent options in nanocarbon synthesis while providing acceptable alternatives with low cost and toxicity. A mixed solvent's composition, however, varies with heat because the solvents evaporate at different rates. Open reaction systems with mixed solvents allow only low temperature chemical reactions (e.g. water/alcohols), depending on the lowest BP of a constituent liquid. Therefore, closed-system solvothermal processes are necessary for reasonably high-temperature MW-assisted synthesis.

A major disadvantage of MW heated reactions in a domestic MW oven is the lack of tracking of the reaction temperature, including local temperature. Therefore, it is not possible to track the exact temperature during MW heating, although automatic power adjustment is available. This problem can be solved in MW-assisted solvothermal/HT reactions, where the autoclave can be equipped with a temperature and pressure measurement system. Since HT/solvothermal reactions are very sensitive to pressure and temperature in the pressure chamber, accurate recording of temperature, pressure, MW power and duration of exposure is very important to produce the product consistently and reliably. When using domestic MW ovens, it is vital to check for the emissions of gases and take precautions to avoid inhalation. Running the MW oven in a fume hood is recommended. In many circumstances, spontaneous emissions and even minor explosions might occur. Building a customized MW reactor or modifying an existing MW oven is the best option.

The lab-scale synthesis and modification of nanocarbons are straightforward. However, because MW reactors cannot be scaled arbitrarily, scaling these techniques for large-scale synthesis is challenging. A continuous MW process for nanocarbon synthesis (in liquid state) in larger batches will pave the way for large-scale, commercially viable products for potential novel composites, catalytic, energy, and other applications with good yield and reproducibility. Only a small proportion of reported studies provided repeatability of the outcome of the MW-assisted synthesis or modification process. Since MW-assisted heating is a complex process with many factors to consider, it is very important to keep track of each variable in case the same material needs to be prepared. Some research debates whether MWs are energy efficient as a heat source. Magnetrons, for example, convert 65% of electrical energy into electromagnetic radiation. As we have observed, certain metals absorb $E$-field while others absorb $H$-filed. This factor made it difficult to use standard household MW ovens. Furthermore, in polar solvents, there is less conversion of electromagnetic radiation into heat. It is apparent that depending on the synthesis variables and the type of the solvents (or materials) used in a certain procedure, the outcome is extremely beneficial, but in some circumstances, it is not satisfactory. As a result, it is critical to group the elements that are useful in any MW-assisted synthesis. For example, many studies report that MW heated oil bath is very useful instead of direct MW irradiation.

To use nanocarbon in high energy density supercapacitors, more work needs to be done to tune the surface area and pore size distribution of carbon materials using MW methods. It is very important to note that nature has bestowed us with numerous natural resources that can have special structures for energy storage. One such material is a biomass. MWs can be used to produce electrode material (activated carbon, microporous carbon) from a variety of biomasses, including agricultural crop residues, forest residues, and wood industry by-products, considering their low cost, availability, and environmental friendliness. Any successful venture can have a significant positive impact on the global energy market.

Conflicts of interest: No competing interests.

Preprint of https://doi.org/10.1016/j.est.2025.115315

78Preprint of https://doi.org/10.1016/j.est.2025.115315127. Mao, Y.; Zhang, X.; Zhou, Y.; Chu, W., Microwave-assisted synthesis of porous nano-sized $Na_3V_2(PO_4)_2F_3$@C nanospheres for sodium ion batteries with enhanced stability. *Scr. Mater.* **2020,** *181*, 92-96.

128. Ferrari, A. C.; Robertson, J., Interpretation of Raman spectra of disordered and amorphous carbon. *Physical review B* **2000,** *61* (20), 14095.

129. Wu, T.; Wang, G.; Zhang, X.; Chen, C.; Zhang, Y.; Zhao, H., Transforming chitosan into N-doped graphitic carbon electrocatalysts. *Chem. Commun.* **2015,** *51* (7), 1334-1337.

130. Zhang, H.; Wang, Y.; Wang, D.; Li, Y.; Liu, X.; Liu, P.; Yang, H.; An, T.; Tang, Z.; Zhao, H., Hydrothermal transformation of dried grass into graphitic carbon-based high performance electrocatalyst for oxygen reduction reaction. *Small* **2014,** *10* (16), 3371-3378.

131. Bai, J.; Xi, B.; Mao, H.; Lin, Y.; Ma, X.; Feng, J.; Xiong, S., One-step construction of N, P-codoped porous carbon sheets/CoP hybrids with enhanced lithium and potassium storage. *Adv. Mater.* **2018,** *30* (35), 1802310.

132. Pan, Z.; Ren, J.; Guan, G.; Fang, X.; Wang, B.; Doo, S. G.; Son, I. H.; Huang, X.; Peng, H., Synthesizing Nitrogen-Doped Core-Sheath Carbon Nanotube Films for Flexible Lithium Ion Batteries. *Adv. Energy Mater.* **2016,** *6* (11).

133. Xu, Y.; Zhang, C.; Zhou, M.; Fu, Q.; Zhao, C.; Wu, M.; Lei, Y., Highly nitrogen doped carbon nanofibers with superior rate capability and cyclability for potassium ion batteries. *Nat. Commun.* **2018,** *9* (1), 1720.

134. Luan, Y.; Hu, R.; Fang, Y.; Zhu, K.; Cheng, K.; Yan, J.; Ye, K.; Wang, G.; Cao, D., Nitrogen and phosphorus dual-doped multilayer graphene as universal anode for full carbon-based lithium and potassium ion capacitors. *Nano Micro Lett.* **2019,** *11*, 1-13.

135. Niu, J.; Liang, J.; Shao, R.; Liu, M.; Dou, M.; Li, Z.; Huang, Y.; Wang, F., Tremella-like N, O-codoped hierarchically porous carbon nanosheets as high-performance anode materials for high energy and ultrafast Na-ion capacitors. *Nano Energy* **2017,** *41*, 285-292.

136. Chen, M.; Cao, Y.; Ma, C.; Yang, H., AN/S/O-tridoped hard carbon network anode from mercaptan/polyurethane-acrylate resin for potassium-ion batteries. *Nano Energy* **2021,** *81*, 105640.

137. Cao, M.; Jin, X.; Zhao, J.; Wang, X., Triggering hollow carbon nanotubes via dual doping for fast pseudocapacitive potassium-ion storage. *Applied Materials Today* **2023,** *30*, 101694.

138. Mo, F.; Wu, X., MgO template-assisted synthesis of hierarchical porous carbon with high content heteroatoms for supercapacitor. *J. Energy Storage* **2022,** *54*, 105287.

139. Liu, S.-H.; Tang, Y.-H., Hierarchically porous biocarbons prepared by microwave-aided carbonization and activation for capacitive deionization. *J. Electroanal. Chem.* **2020,** *878*, 114587.

140. Ando, Y.; Zhao, X., Synthesis of carbon nanotubes by arc-discharge method. *New diamond and frontier carbon technology* **2006,** *16* (3), 123-138.

141. Moisala, A.; Nasibulin, A. G.; Kauppinen, E. I., The role of metal nanoparticles in the catalytic production of single-walled carbon nanotubes—a review. *J. Phys.: Condens. Matter* **2003,** *15* (42), S3011.

142. He, D.; Li, H.; Li, W.; Haghi-Ashtiani, P.; Lejay, P.; Bai, J., Growth of carbon nanotubes in six orthogonal directions on spherical alumina microparticles. *Carbon* **2011,** *49* (7), 2273-2286.

143. Dorji, P.; Kim, D. I.; Hong, S.; Phuntsho, S.; Shon, H. K., Pilot-scale membrane capacitive deionisation for effective bromide removal and high water recovery in seawater desalination. *Desalination* **2020,** *479*, 114309.

144. Kim, N.; Lee, J.; Hong, S. P.; Lee, C.; Kim, C.; Yoon, J., Performance analysis of the multi-channel membrane capacitive deionization with porous carbon electrode stacks. *Desalination* **2020,** *479*, 114315.
78

Preprint of https://doi.org/10.1016/j.est.2025.115315355. Calandra, M.; Mauri, F., Electronic structure of heavily doped graphene: The role of foreign atom states. *Physical Review B—Condensed Matter and Materials Physics* **2007,** *76* (16), 161406.

356. Uchoa, B.; Castro Neto, A., Superconducting states of pure and doped graphene. *Phys. Rev. Lett.* **2007,** *98* (14), 146801.

357. Peres, N.; Guinea, F.; Castro Neto, A., Coulomb interactions and ferromagnetism in pure and doped graphene. *Physical Review B—Condensed Matter and Materials Physics* **2005,** *72* (17), 174406.

358. Naraprawatphong, R.; Chokradjaroen, C.; Thiangtham, S.; Yang, L.; Saito, N., Nanoscale advanced carbons as an anode for lithium-ion battery. *Mater. Today Adv.* **2022,** *16*, 100290.

359. Ma, X.; Ning, G.; Qi, C.; Xu, C.; Gao, J., Phosphorus and nitrogen dual-doped few-layered porous graphene: a high-performance anode material for lithium-ion batteries. *ACS Appl. Mater. Interfaces* **2014,** *6* (16), 14415-14422.

360. Zheng, Y.; Jiao, Y.; Ge, L.; Jaroniec, M.; Qiao, S. Z., Two-step boron and nitrogen doping in graphene for enhanced synergistic catalysis. *Angew. Chem. Int. Ed.* **2013,** *52* (11).

361. Wu, Z.-S.; Ren, W.; Xu, L.; Li, F.; Cheng, H.-M., Doped graphene sheets as anode materials with superhigh rate and large capacity for lithium ion batteries. *ACS nano* **2011,** *5* (7), 5463-5471.

362. Ma, C.; Shao, X.; Cao, D., Nitrogen-doped graphene nanosheets as anode materials for lithium ion batteries: a first-principles study. *Journal of Materials Chemistry* **2012,** *22* (18), 8911-8915.

363. Wang, X.; Weng, Q.; Liu, X.; Wang, X.; Tang, D.-M.; Tian, W.; Zhang, C.; Yi, W.; Liu, D.; Bando, Y., Atomistic origins of high rate capability and capacity of N-doped graphene for lithium storage. *Nano letters* **2014,** *14* (3), 1164-1171.

364. Wang, X.; Li, X.; Zhang, L.; Yoon, Y.; Weber, P. K.; Wang, H.; Guo, J.; Dai, H., N-doping of graphene through electrothermal reactions with ammonia. *science* **2009,** *324* (5928), 768-771.

365. Yu, S.; Zheng, W.; Wang, C.; Jiang, Q., Nitrogen/boron doping position dependence of the electronic properties of a triangular graphene. *Acs Nano* **2010,** *4* (12), 7619-7629.

366. Cruz-Silva, E.; Barnett, Z.; Sumpter, B. G.; Meunier, V., Structural, magnetic, and transport properties of substitutionally doped graphene nanoribbons from first principles. *Physical Review B—Condensed Matter and Materials Physics* **2011,** *83* (15), 155445.

367. Chauhan, S. S.; Srivastava, P.; Shrivastava, A. K., Electronic and transport properties of boron and nitrogen doped graphene nanoribbons: an ab initio approach. *Appl. Nanosci.* **2014,** *4*, 461-467.

368. Panchakarla, L. S.; Subrahmanyam, K. S.; Saha, S.; Govindaraj, A.; Krishnamurthy, H. R.; Waghmare, U. V.; Rao, C., Synthesis, structure, and properties of boron-and nitrogen-doped graphene. *Adv. Mater.* **2009,** *21* (46), 4726-4730.

369. Dai, J.; Yuan, J.; Giannozzi, P., Gas adsorption on graphene doped with B, N, Al, and S: A theoretical study. *Appl. Phys. Lett.* **2009,** *95* (23).

370. Meyer, J. C.; Kurasch, S.; Park, H. J.; Skakalova, V.; Künzel, D.; Groß, A.; Chuvilin, A.; Algara-Siller, G.; Roth, S.; Iwasaki, T., Experimental analysis of charge redistribution due to chemical bonding by high-resolution transmission electron microscopy. *Nat. Mater.* **2011,** *10* (3), 209-215.

371. Wang, L.; Yu, P.; Zhao, L.; Tian, C.; Zhao, D.; Zhou, W.; Yin, J.; Wang, R.; Fu, H., B and N isolate-doped graphitic carbon nanosheets from nitrogen-containing ion-exchanged resins for enhanced oxygen reduction. *Sci. Rep.* **2014,** *4* (1), 5184.
91